# Reference Physics Design for 1 GeV Injector Linac and Accumulator Ring for Indian Spallation Neutron Source

by

***Amalendu Sharma, Arup Ratan Jana, Chirag Bhai Patidar, Mukesh Kumar Pal, Nita Kulkarni, Pradeep Kumar Goyal, Prasanta Kumar Jana, Rahul Gaur, Ram Prakash, Rinky Dhingra, Urmila Singh and Vinit Kumar***

*Raja Ramanna Centre for Advanced Technology, Indore*

## ABSTRACT

As a part of the ongoing XII$^{th}$ plan project titled "R&D activities for high energy proton linac based spallation neutron source", the work on physics design of various subsystems of the injector linac and accumulator ring has been taken up at RRCAT, Indore. For the 1 GeV $H^-$ injector linac, physics design studies of individual systems have been completed, and the end to end beam dynamics simulation studies have been performed to ensure that the stringent beam dynamics criteria are satisfied for the optimized lattice. Physics design studies to optimize the linear lattice of the accumulator ring have also been completed. The design studies for the beam transport lines from the injector linac to the accumulator ring, and from the accumulator ring to target are currently in progress. This report describes the physics design of various systems of the injector linac and the accumulator ring.

# Table of contents

# 1. INTRODUCTION

A 1 GeV, 1 MW average power, pulsed accelerator is being designed for the proposed Indian Spallation Neutron Source (ISNS) at RRCAT. In this report, we discuss the reference physics design of the injector linac and the accumulator ring. Two of its features are important to mention. First, this will use superconducting accelerators after 3 MeV, unlike SNS at Oak Ridge [1], which uses such accelerators only after 185 MeV. Second, we propose to inject 2 ms long pulses to accumulator ring, compared to 1ms for SNS. This will help us in reducing the required macropulse beam current to 10 mA, such that the requirement of peak RF power is reduced, enabling us to use indigenous Solid State RF power sources.

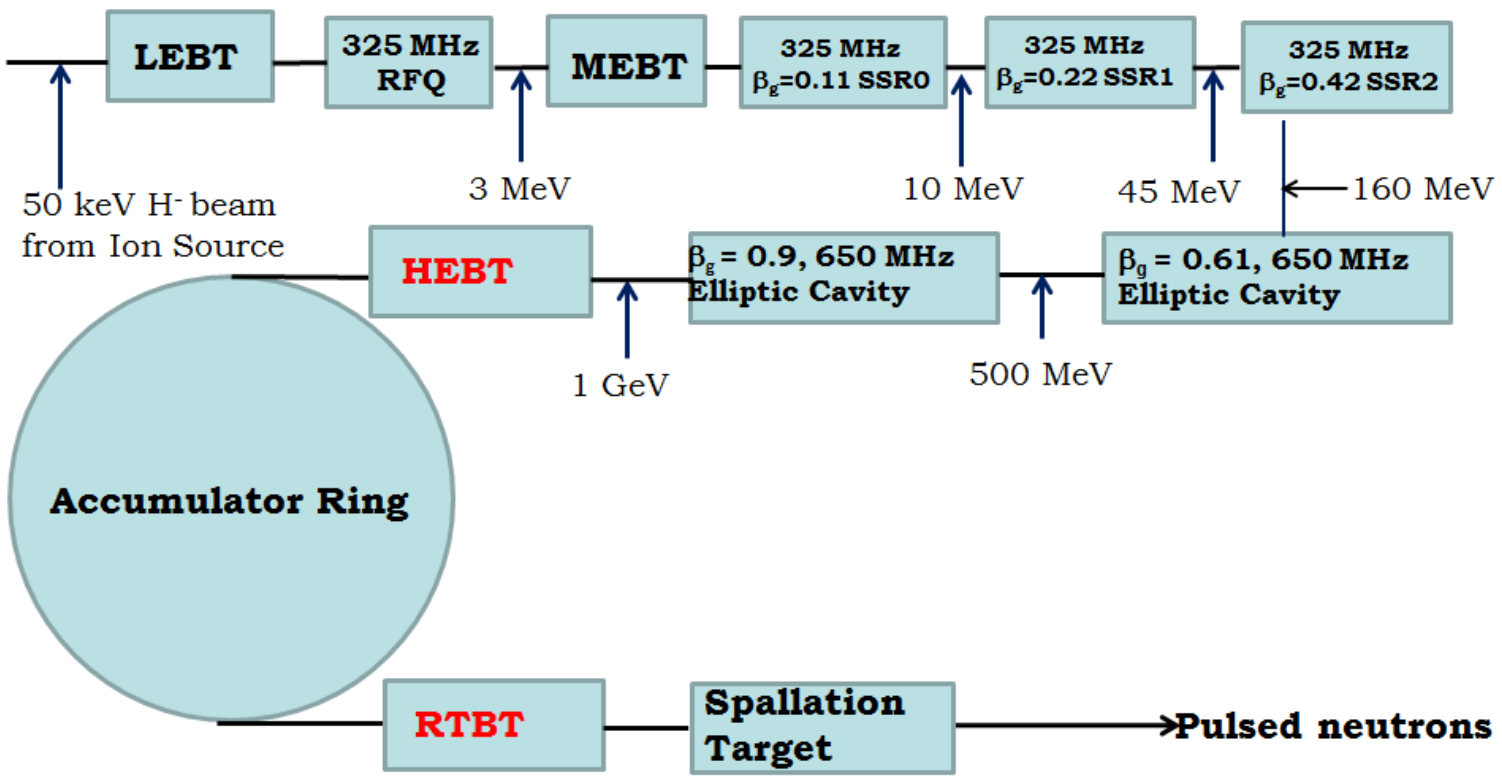

*Fig. 1: Schematic of the accelerator for ISNS.*

Figure 1 shows the schematic of the accelerator for the ISNS. The accelerator will be pulsed and a brief description of the pulse structure in the injector linac is shown in Fig. 2. There will be 2 ms long $H^-$ pulses, called macropulses, which repeat at the rate of 50 Hz. Average current within a macropulse is 10 mA. Each macropulse has midi-pulses, which are ~ 0.65 μs long, and repeat at the rate of 1 MHz, for 2 ms duration. The average current in a midi-pulse is 15 mA. Each midi-pulse has micropulses, which repeat after every 3.1 ns, corresponding to the RF frequency of 325 MHz of the front-end of the linac. In the next paragraphs, a brief introduction to each section of the accelerator is provided.

The Low Energy Beam Transport (LEBT) line will transport, and match the 50 keV, 15 mA $H^-$ beam from an RF antenna based ion source, to the entrance of Radio Frequency Quadrupole (RFQ) accelerator. Physics design studies have been performed for a 1.9 m long magnetic LEBT, having two solenoid magnets, and two steering coil magnets, along with a beam chopper. In order to reduce the growth of emittance due to space charge, space charge compensation technique will be employed, on which detailed studies have been started.

After LEBT, the 50 keV beam will be accelerated to 3 MeV in an RFQ structure. A 3.49 m long, 325 MHz RFQ has been designed, and it has been ensured in the design that beam transmission efficiency is > 95%. All the required geometrical tolerances and design details, along with the data for heat generation have been obtained. In order to ensure that unwanted dipole modes do not affect the performance of RFQ, dipole stabilization rods have been designed. This helps in keeping the frequency of dipole modes sufficiently away from the operating quadrupole mode. Co-axial type loop coupler has been designed to couple the RF power into RFQ structure. Multipacting studies for the power coupler are currently in progress.

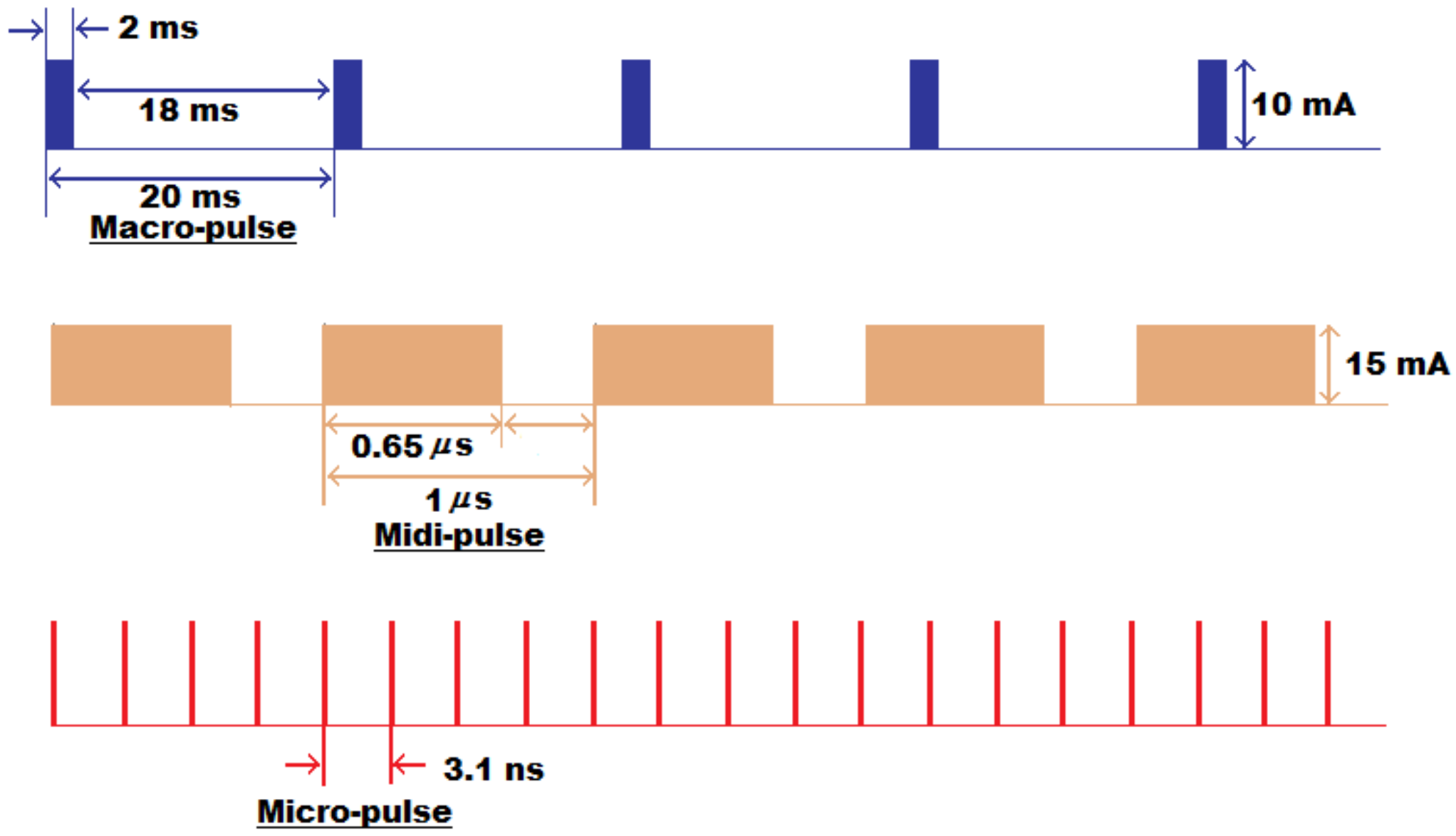

*Fig.2: Pulse structure in the injector linac.*

Beam from RFQ will be injected to superconducting section of the linac, the first of which will be the Single Spoke Resonator–0 (SSR0) section. The acceptable phase space for the injected beam at the entrance of SSR0 is different from the one available at the RFQ exit. The Medium Energy Beam Transport (MEBT) line will transport the beam from RFQ exit, and will match it to SSR0. Here, beam matching needs to be achieved in transverse as well as longitudinal planes in phase space. Matching in transverse planes will be achieved with the help of an optimized set of quadrupole magnets. Similarly, matching in the longitudinal plane will be achieved with the help of an optimized set of buncher cavities. In addition, MEBT has a chopper section to chop nearly ~35% of beam at the rate of 1 MHz, which is needed for injecting the beam into accumulator ring later. Note that the beam from the ion source has pulse width of 2 ms, which is chopped at 1 MHz in two stages – pre-chopping in LEBT as discussed earlier, and then final chopping in the MEBT. The chopped pulses are called midipulses. Micropulses are formed within the midipulses, while the beam traverses through the RFQ and the superconducting linac section later. Chopping is performed to avoid beam loss at the time of injection into the accumulator ring.

A first order design of a 3.86 m long MEBT, having eleven quadrupole magnets and three buncher cavities, to achieve the beam matching has been performed. Design of chopper, which will be installed inside the quadrupole magnets is in progress.

Next, the 10 mA beam will be accelerated in superconducting sections, consisting of low beta section having SSRs, medium beta section having beta = 0.61 elliptic cavities, and high beta section consisting of beta = 0.9 elliptic cavities. The low beta section will consist of SSR0 section having beta = 0.11, SSR1 section having beta = 0.22 and SSR2 section having beta = 0.42. Electromagnetic design study for SSRs and elliptic cavities have been completed, and geometry has been optimized to achieve maximum acceleration gradient. The total length of linac is ~ 254 m, and it has 20 nos. of SSR0 cavities, 28 nos. of SSR1 cavities, 48 nos. of SSR2 cavities, 54 nos. of beta = 0.61 cavities and 48 nos. of beta = 0.9 cavities. Multipacting studies have also been performed and accordingly required refinement in the geometry has been made. In order to keep the beam focused in the transverse and longitudinal directions, a suitable periodic arrangement (known as lattice) of solenoid magnets, drift space and SSR cavities in the low beta section, and quadrupole doublets, drift spaces and elliptic cavities in the medium and high beta section has been worked out. Beam dynamics studies have been carried out with the optimized lattice, and it was found that all the required criteria for minimum beam loss are achieved. First-order design of fundamental power couplers for these cavities has been worked out. Design of higher order mode (HOM) couplers, and multipacting studies in couplers are currently in progress. The end to end beam dynamics simulations, starting from the ion source exit, up to the end of beta = 0.9 section have been performed, using the code TRACEWIN [2].

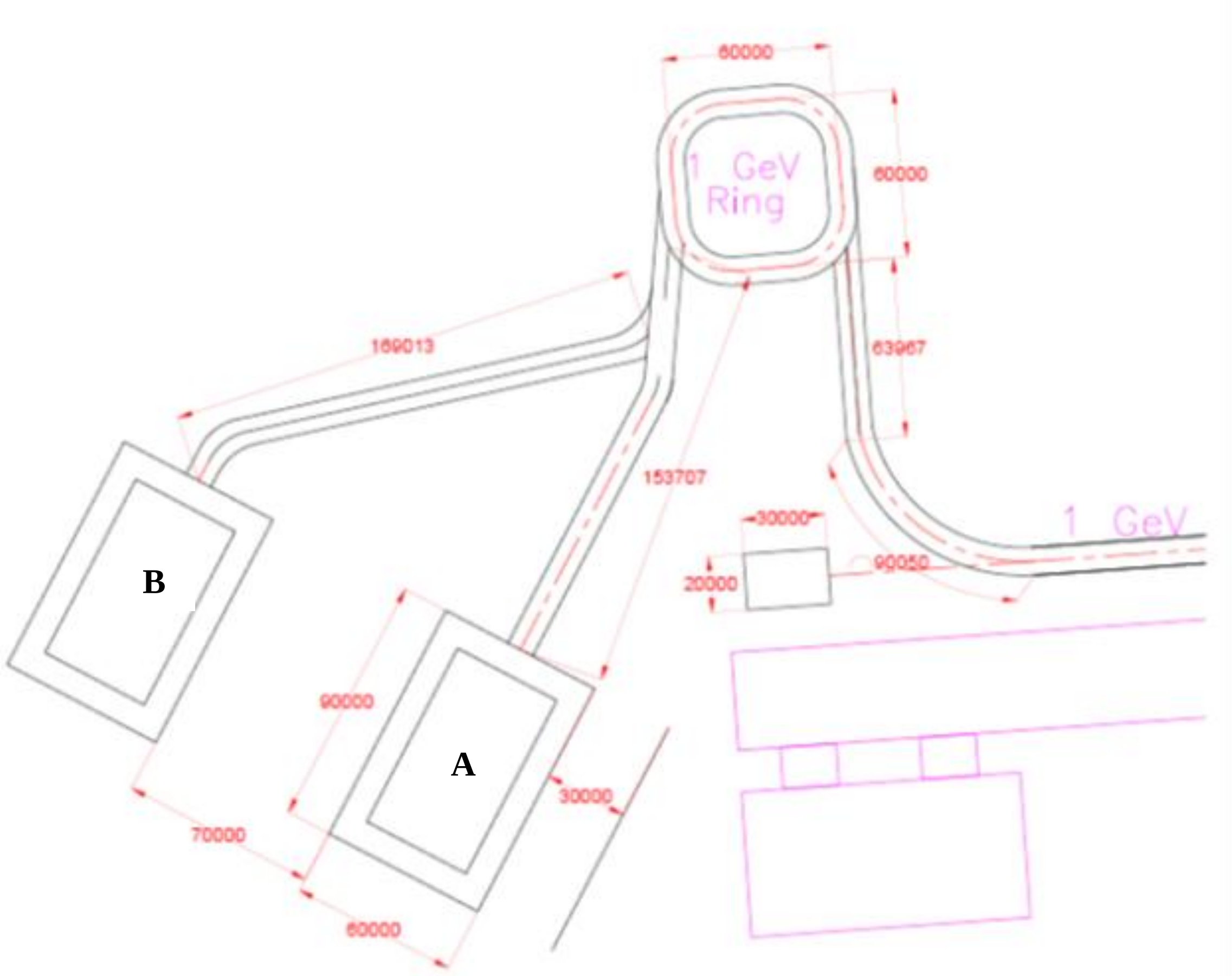

*Fig. 3: A schematic layout of the HEBT, Accumulator Ring and RTBT for ISNS. Note that two target stations A and B are shown. Design studies are currently being pursued for RTBT for Target A. All dimensions are shown in mm.*

The 1 GeV beam from the linac will be injected into the accumulator ring, for which a high energy beam transport (HEBT) line will be required. In the accumulator ring, the subsequent midipulses from the linac override each other, and we have ~ 0.65 μs long proton pulses repeating @ 50 Hz, after the beam is extracted from the accumulator ring. The extracted beam is transported to the target through Ring to Target Beam Transport (RTBT) system. The layout of HEBT and RTBT is currently being evolved. It is proposed that the HEBT will have an energy corrector cavity to make final correction for jitter in phase and energy before injecting into accumulator ring, for which design study is in progress. Fig. 3 shows a schematic of the layout showing the HEBT, Acuumulator Ring and RTBT.

First-order lattice design of the 1 GeV accumulator ring having a circumference of ~ 262 m has been worked out. Two types of lattices – FODO and hybrid have been considered in the initial design, which are used worldwide [1,3]. As the design will evolve, one of these lattices will be finalized subsequently. Detailed specifications of lattice magnets, along with corrector magnets and trim coils have been evolved. Closed orbit distortion (COD) studies, along with other linear optics errors and correction schemes have been worked out for the FODO lattice, as well as the hybrid lattice. Scheme of sextupoles to correct for the chromaticity has also been worked out, based on which some of the operating tune points of the machine have been selected.

*Table 1: Indian Spallation Neutron Source Primary Parameters*

| | |
|---|---|
| Proton beam power on target | 1.0 MW |
| Proton beam kinetic energy on target | 1.0 GeV |
| Average beam current on target | 1.0 mA |
| Pulse repetition rate | 50 Hz |
| Protons per pulse on target | $1.25 \times 10^{14}$ |
| Charge per pulse on target | 20 μC |
| Energy per pulse on target | 20 kJ |
| Proton pulse length on target | 680 ns |
| Ion type (Front end, Linac, HEBT) | H minus |
| Average linac macropulse current | 11 mA |
| Linac beam macropulse duty factor | 10% |
| Front end length | 9.1 m |
| Linac length | 254 m |
| HEBT length | 170 m |
| Ring circumference | 262 m |
| RTBT length | 150 m |
| Ion type (Ring, RTBT, Target) | proton |
| Ring filling time | 2.0 ms |
| Ring revolution frequency | 1.0 MHz |
| Number of injected turns | 2000 |
| Ring filling fraction | 68% |
| Ring extraction beam gap | 250 ns |
| Maximum uncontrolled beam loss | 1 W/m |
| Target material | Hg/Pb-Bi |

Beam injection studies are in progress. Currently, the carbon foil based injection scheme has been considered and studies on foil heating have been performed to find the maximum value of injected pulse width. In order to achieve an injected pulse width of 2 ms, studies on laser stripping based scheme will be taken up. First-order design of beam extraction scheme has been completed, and specifications of kicker magnets have been evolved. Design of beam collimators to reduce the uncontrolled beam loss to the required level is going on.

Studies on longitudinal beam dynamics in the accumulator ring are currently in progress, where the requirement of RF power and its temporal profile is being evolved for the RF cavities inside the accumulator ring, after taking the effect of beam loading into account. This required some code development work, which has been completed.

Studies on beam instabilities in the accumulator ring, e.g., electron cloud instability and the effect of wakefield and impedances will be performed in the later phase of the design of accumulator ring. Table 1 gives the primary parameters of ISNS.

In the next sections, we discuss in detail the physics design of various systems in a sequential manner.

## 2. Low Energy Beam Transport (LEBT)

Low Energy Beam Transport (LEBT) line is the first component of the front end of injector linac. It transports the 15 mA, 2ms beam at a rep. rate of 50 Hz. from the ion source exit to the entrance of the RFQ. The beam parameters at the exit of the RF antenna based ion source, are given in Table 2. Based on the beam dynamics studies, the required beam parameters at the RFQ entrance have been calculated, which are given in Table 2.

*Table 2. Beam parameters at the ion source exit and RFQ entrance*

| Parameter | Value at the ion source exit | Value at the RFQ entrance |
|---|---|---|
| $\alpha_{x,y}$ | -1 to -2 | 1.3 |
| $\beta_{x,y}$ | 0.12 to 0.24 m/rad | 0.041 m/rad |
| $\varepsilon_{x,y,n,rms}$ | 0.36 mm-mrad | 0.39 mm-mrad |

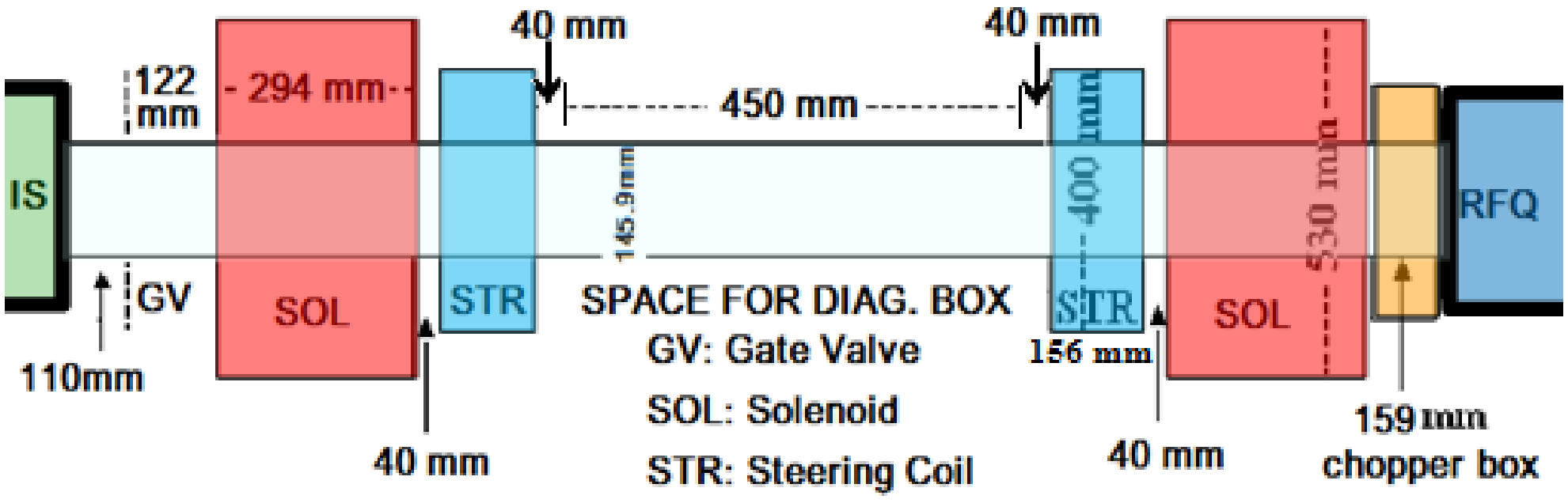

*Fig.4: Schematic layout of the LEBT.*

A schematic layout of the LEBT is shown in Fig. 4 [4]. Total length of LEBT is 1901 mm, starting from the end of first vacuum flange, till the end of RFQ flange. In addition to transporting and matching the beam to the requirements of RFQ, it has provision for beam diagnostics, and pre-chopping of beam. The specifications of solenoids and steers are given in Tables 3 and 4 respectively.

*Table 3. Specifications of solenoids*

| Parameters | Value |
|---|---|
| Number of solenoids | 2 |
| Aperture diameter | ≥ 160 mm |
| Peak central field ($B_z$) | 1620 – 3170 G |
| Integrated field ($\int B_z\, dz$) | 460 – 910 G-m |
| Effective length | 286 mm |
| Physical length | 294 mm |
| Coefficient of spherical aberration ($C_1 = \frac{\int \{B'(z)\}^2 dz}{2\int B_z^2 dz}$) | < 30 per m$^2$ |
| Positioning accuracy | < 300 μm |
| Alignment accuracy | < 5 mrad |

*Table 4. Specifications of steering coils*

| Parameters | Value |
|---|---|
| Number of steering coils | 2 |
| Aperture diameter | ≥ 160 mm |
| Magnetic field ($B_x$, $B_y$) | 0 – 105 G |
| Integrated field ($\int B_{x,y}\, dz$) | 0 – 17 G-m |
| Physical length | 156 mm |
| Good field region | ± 4 cm |
| Field uniformity (ΔBL/BL) | < 1% |
| Kick angle for 50 keV beam | up to 50 mrad |

Beam dynamics studies have been performed assuming 80% space charge compensation everywhere in the LEBT, except in the chopper box. The beam distribution is assumed to be 4D Waterbag in phase space. Beam size and normalized particle density at different locations inside the LEBT are shown in Figs. 5 and 6 respectively. The emittance growth has been found to be less than the maximum acceptable value for the range of input beam parameters specified in Table 2.

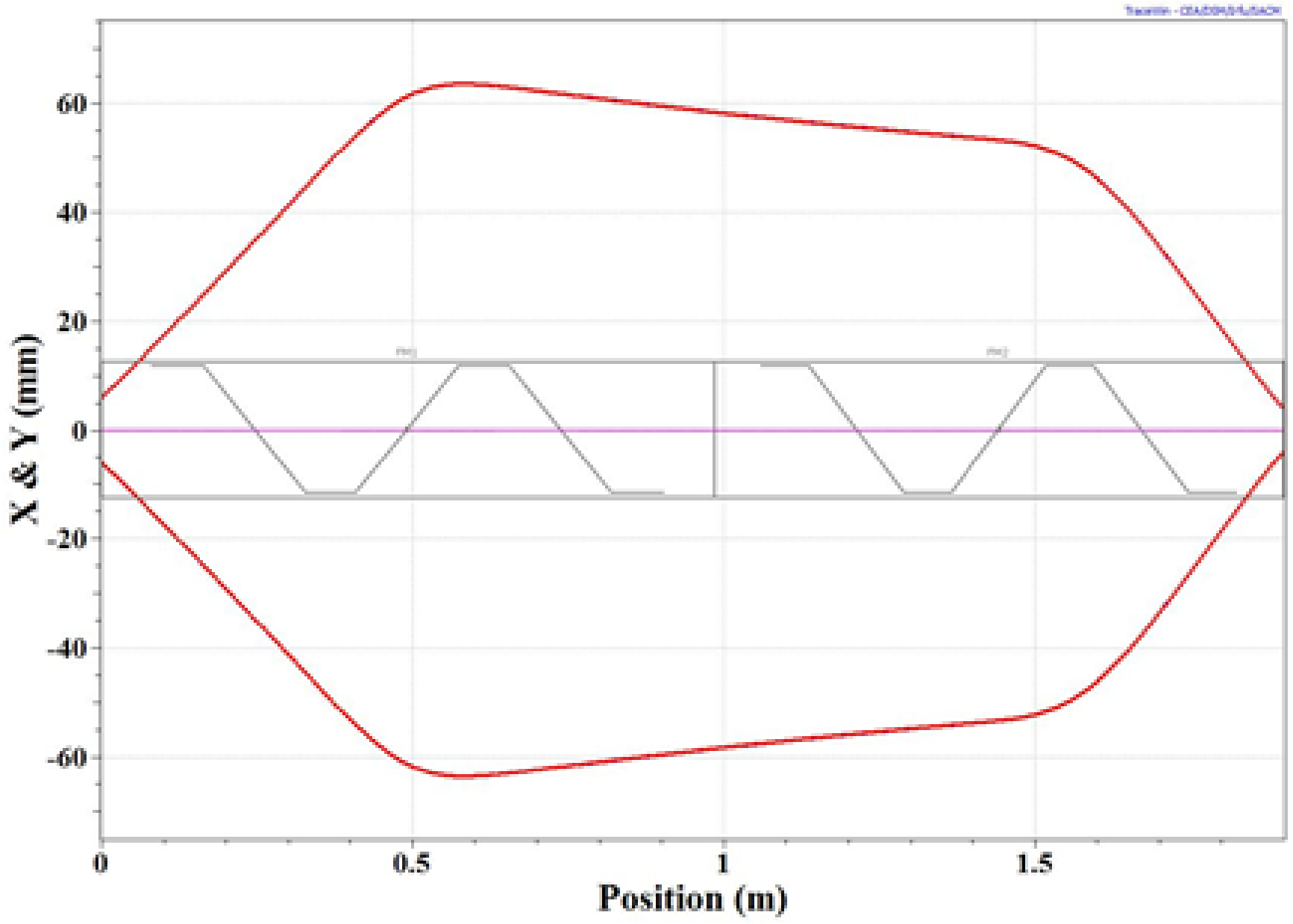

*Fig.5: Beam size (3σ) along the length of LEBT, assuming α=-2 and β=0.12 m/rad at input.*

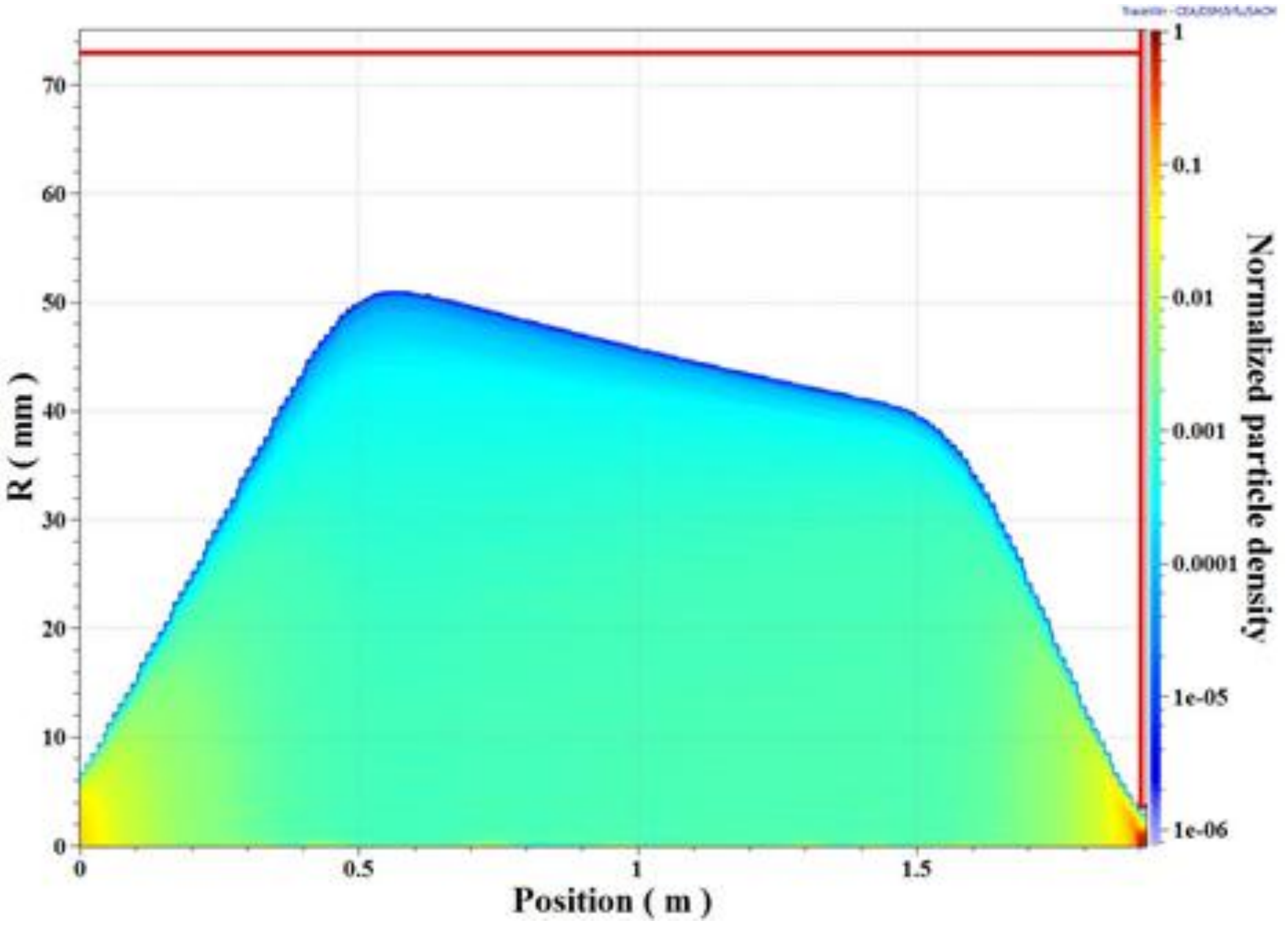

*Fig. 6: Normalized particle density along the length of LEBT. Bold red line shows the inner boundary of vacuum chamber.*

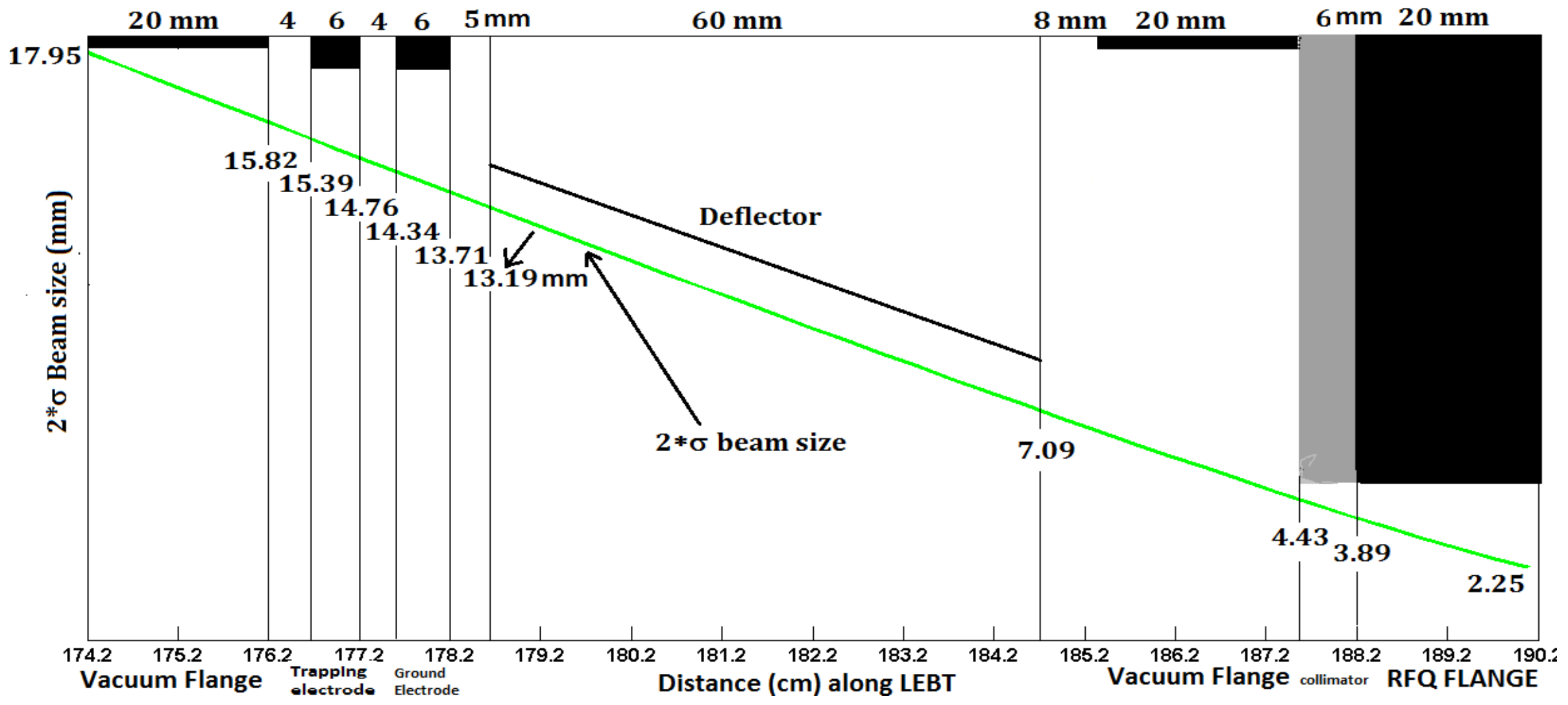

*Fig. 7: Schematic showing the beam profile, along the components of chopper box in LEBT.*

Preliminary studies of beam chopper have been performed. A 60 mm long inclined plane deflector, similar to the one used for Chinese Spallation Neutron Source (CSNS) [5] has been assumed, as shown in Fig. 7. Required deflector voltage has been calculated as 8.6 kV, and the gap between the plates varies between 31.6 mm to 17 mm, which gives a kick of ~ 163 mrad, and ~ 9 mm of beam deflection at the RFQ flange. Deflected beam will be dumped at the RFQ flange. Deflector voltage will be applied for 350 ns at the rate of 1 MHz, and the rise time of the voltage pulse will be around 20 ns. This will perform the preliminary beam chopping.

Analytical calculations have been performed on space charge compensation studies. A suitable gas can be injected with a controlled pressure to achieve the required space charge compensation. Calculation of the critical pressure required for this has been performed for different gases such as hydrogen, nitrogen, krypton, xenon etc. For xenon, the required critical pressure is around $3 \times 10^{-6}$ mbar. Stripping loss due to the presence of xenon is estimated to be 3%. It has also been estimated that the two stream instability will not be excited for this value of pressure.

## 3. Radio Frequency Quadrupole (RFQ) Linac

A 3.49 m long, four vane type RFQ, has been designed for an operating frequency of 325 MHz [6]. The RFQ will be built in three sections, as shown in Fig. 8. The vacuum ports, RF power coupling ports and tuner ports are also shown in the figure.

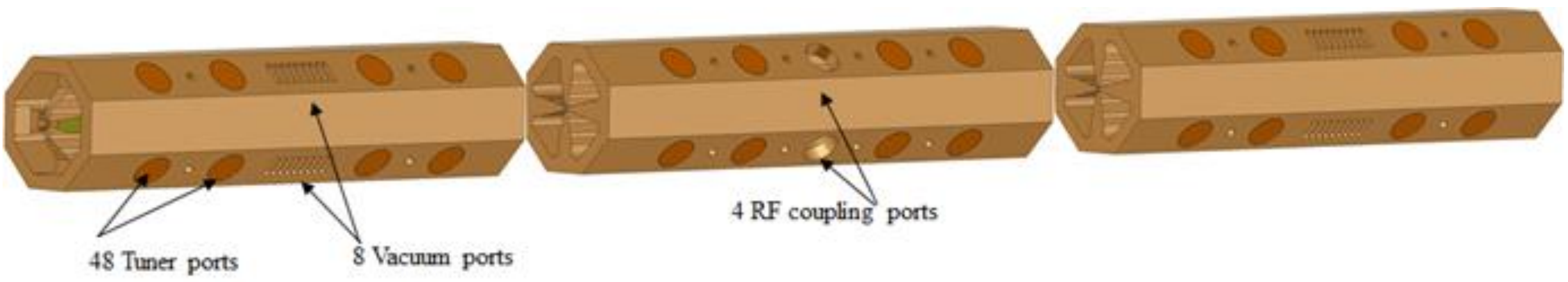

*Fig. 8: Computer model of RFQ structure to be built in three sections.*

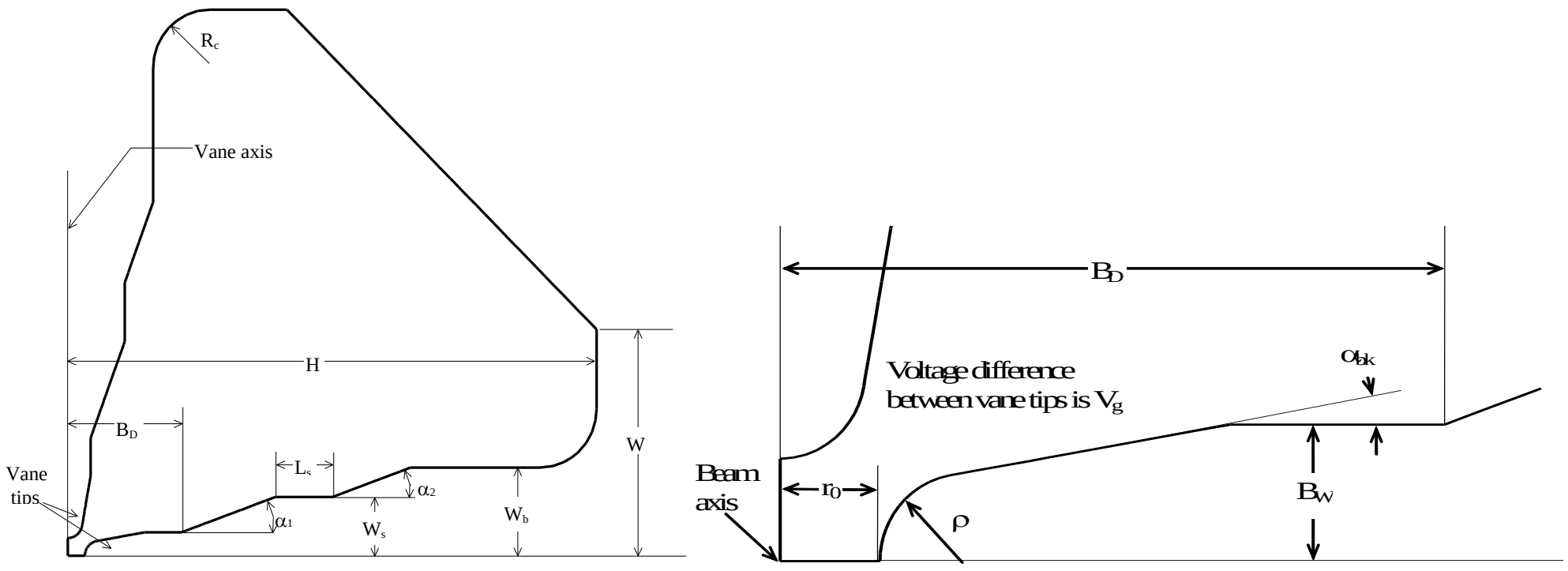

*Fig. 9: Cross-section of one quadrant of RFQ* (This fig. is taken from 'RFQ design codes, LA-UR-96-1836' by K.R. Crandall et al).

*Table 5: Geometrical parameters of RFQ quadrant*

| **Parameter** | **Value** |
|---|---|
| Average aperture radius, $r_0$ | 3.56 mm |
| Vane-tip transverse radius of curvature, $\rho$ | 2.67 mm |
| Breakout Angle, $\alpha_{bk}$ | $20^0$ |
| Vane-Blank Half Width, $B_w$ | 8 mm |
| Vane-Blank Depth, $B_D$ | 30 mm |
| Vane Shoulder Half Width, $W_S$ | 15 mm |
| Vane Base Half Width, $W_b$ | 20 mm |
| Vane angle 1, $\alpha_1$ | $20^0$ |
| Vane angle 2, $\alpha_2$ | $20^0$ |
| Vane shoulder length, $L_s$ | 10 mm |
| Corner Radius, $R_c$ | 10 mm |
| Vane height, $H$ | 103.38 mm |
| Vane half width, $W$ | 42.82 mm |

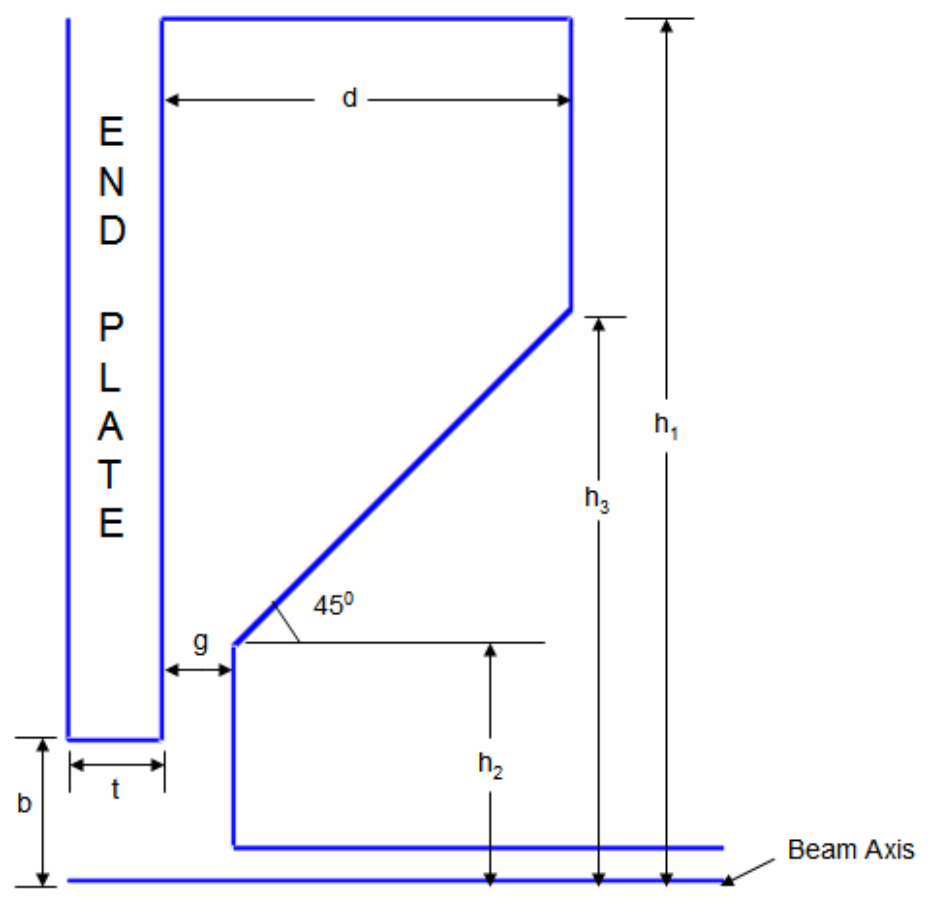

*Fig. 10: Schematic showing the details of vane cut-back.*

Cross section of RFQ has four-fold symmetry. The schematic of one quadrant of the cross section of RFQ is shown in Fig. 9, and all the geometrical parameters are tabulated in Table 5. At the two ends of the RFQ, the structure will be closed with vane cut-backs in such a manner that the entire structure along with appropriately tuned geometrical parameters will resonate at the same frequency as the infinitely long structure with 2D cross section described in Table 5. The geometry of vane-cut back is explained in Fig. 10 and the optimized parameters are described in Table 6.

*Table 6: Geometrical parameters of vane cut back*

| **Parameters** | **Value (entrance)** | **Value (exit)** |
|---|---|---|
| *g* | 7.09 mm | 4.93 mm |
| $h_1$ | 103.38 mm | 103.38 mm |
| $h_2$ | 30.00 mm | 30.00 mm |
| $h_3$ | 71.23 mm | 66.85 mm |
| *d* | 48.32 mm | 41.78 mm |
| *t* | 10.00 mm | 10.00 mm |
| *b* | 20.00 mm | 20.00 mm |

Based on the electromagnetic simulations, further details of RFQ are given in Table 7.

*Table 7: Design parameters of RFQ*

| | |
|---|---|
| Quad. mode frequency (with tuner insertion = 11 mm) | 325 MHz |
| Structure power loss | 385 kW |
| Total RF power | 415 kW |
| No. of sections | 3 |
| Coupling between sections | Direct, gap = 0.1 mm |
| No. of ports | 4 (RF), 8(Vacuum),48(Tuners) |
| Tuning Range | 316 – 344 MHz |
| Dipole mode cut-off frequency | 315.8 MHz |
| Mode stabilization scheme | Dipole rods |

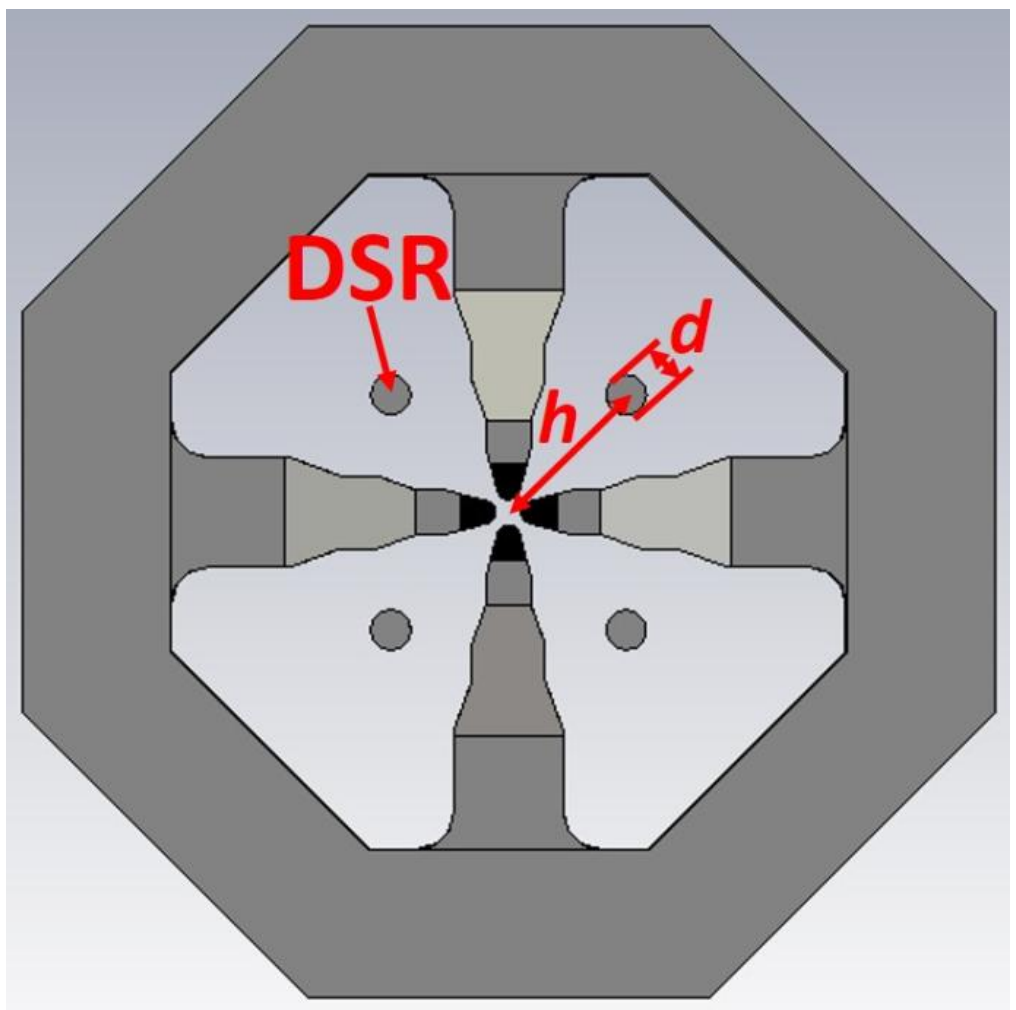

*Fig. 11: Cross-section of the RFQ showing the location of dipole rods.*

Studies have been performed on dipole rod stabilization scheme to ensure that the nearest dipole modes are sufficiently shifted from the operating quadrupole mode such that the unwanted dipole modes are not excited due to fabrication errors. Schematic showing the location of dipole stabilization rods is shown in Fig. 11. The optimized parameters specifying the location and the length of the dipole rods are given in Table 8.

*Table 8: Optimized parameters for dipole rods*

| **Parameter** | **Value** |
|---|---|
| *h* (entrance side) | 61.16 mm |
| *h* (exit side) | 60.81 mm |
| *d* | 14 mm |
| Length | 155 mm |
| Frequency shift from the operating mode, for nearest dipole modes | -4.2 MHz, +4.3 MHz |

Diameters of all the ports are given in Table 9. RF power will be coupled to the RFQ structure using 3-inch co-axial transmission line having inner diameter as 32.15 mm and outer diameter as 76.2 mm with loop type coupler [7]. The required value of cavity coupling coefficient for critical coupling $\beta$ in the presence of beam is calculated as 1.15. Two nos. of RF coupling ports will be used, and therefore, the required value of $\beta$ for each port is 0.575. This is achieved by using a co-axial coupling loop with surface area 290 mm$^2$ and insertion depth of 8.14 mm. The value of $\beta$ can be varied by rotating the loop. A schematic showing the two co-axial RF power couplers for RFQ is shown in Fig. 12. Frequency shift as a function of tuner insertion is shown in Fig. 13.

*Table 9: Diameter of different ports of RFQ*

| **Name of the port** | **Diameter** |
|---|---|
| Power coupler port | 80 mm |
| Tuner port | 80 mm |
| Sampling loop port | 20 mm |
| Vacuum port | 9 nos. of slots (12 mm × 54 mm), separation between slots = 6 mm |

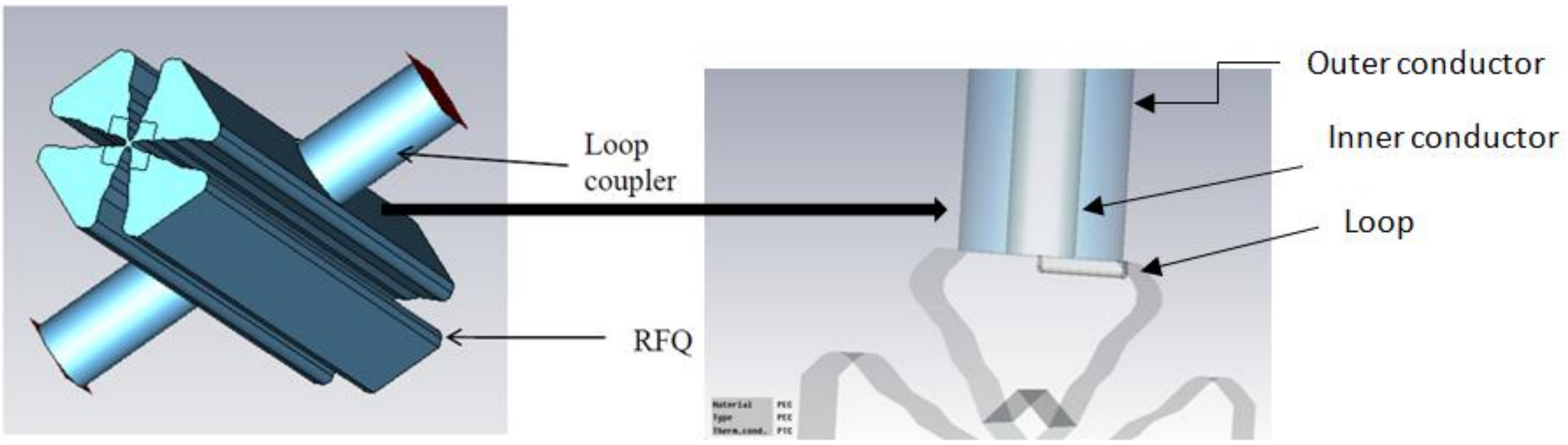

*Fig. 12: Schematic showing the two co-axial loop type RF power couplers for the RFQ.*

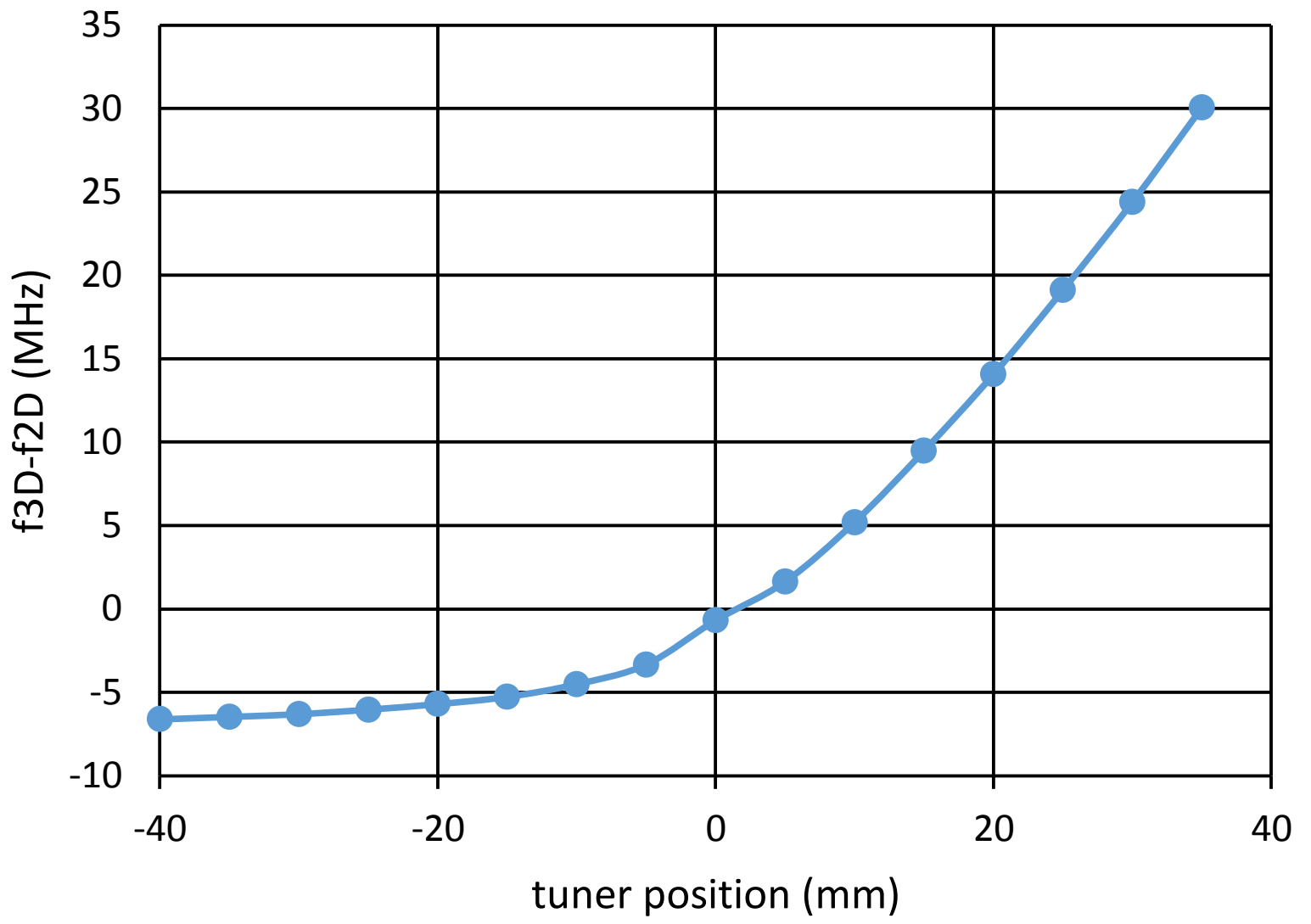

*Fig. 13: Frequency shift as a function of tuner position.*

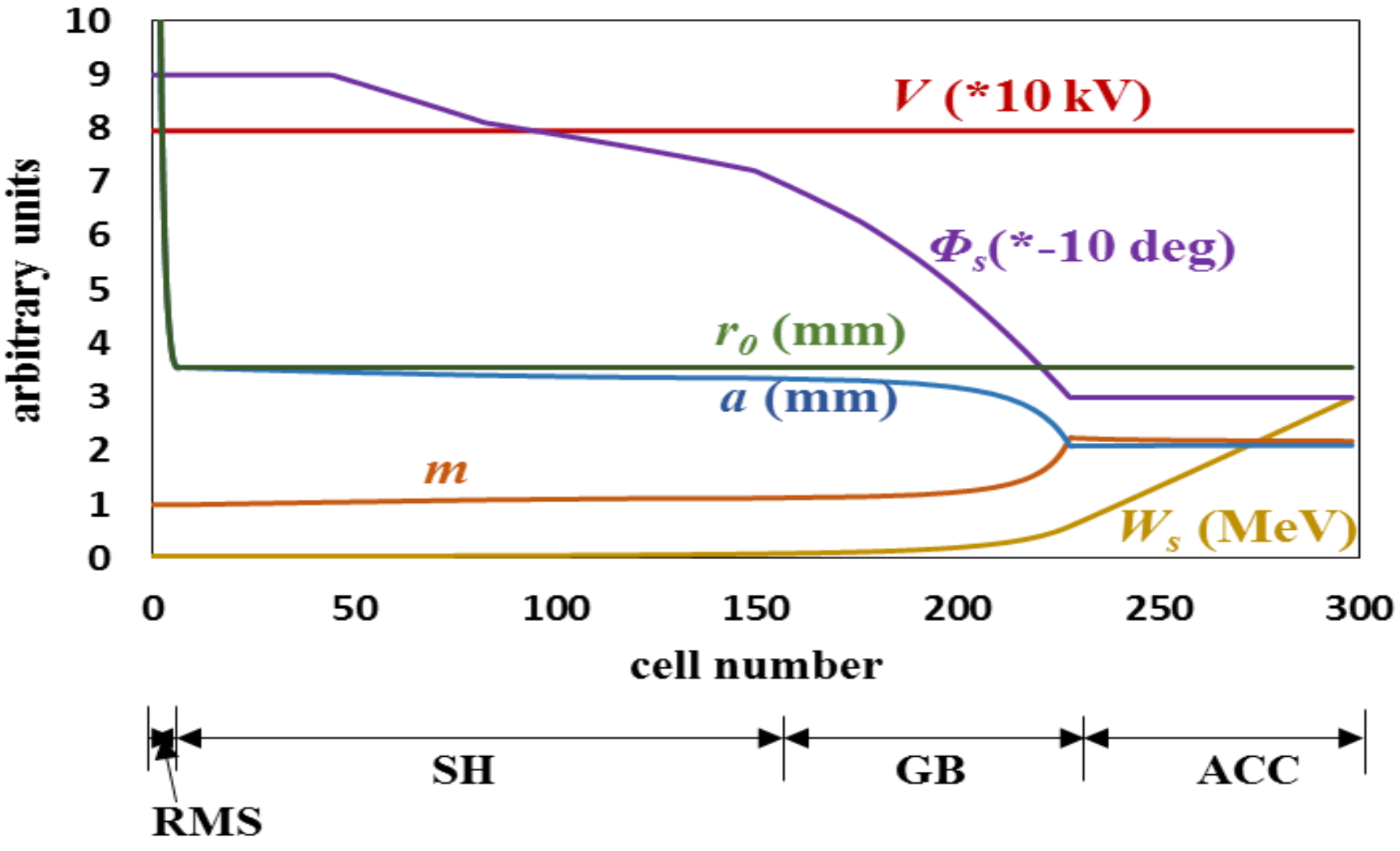

*Fig. 14: Variation of beam dynamics parameters along the RFQ*

Beam dynamics studies have been performed for the RFQ [6], and variation of beam dynamics parameters, along with variation in aperture radius and vane modulation parameter for the optimized case are shown in Fig. 14. Here, $V$ is the inter_vane voltage, $a$ is the aperture radius, $r_0$ is the average aperture radius, $m$ is the vane modulation parameter, $\phi_s$ is the synchronous phase and $W_s$ is the synchronous energy. Here, RMS denotes the radial matching section, SH denotes the shaper section, GB denotes the gentle buncher section, and ACC denotes the accelerating section. Based on the beam dynamics studies, the important RFQ design parameters are given in Table 10.

*Table 10: Important design parameters of RFQ*

| Parameter | Value |
|---|---|
| Operating frequency | 325 MHz |
| Peak beam current | 15 mA |
| Input energy | 50 keV |
| Output energy | 3 MeV |
| $\varepsilon_{x,y,rms,n}$ (input) | 0.39 mm-mrad |
| $\varepsilon_{x,rms,n}$ (output) | 0.39 mm-mrad |
| $\varepsilon_{y,rms,n}$ (output) | 0.39 mm-mrad |
| $\varepsilon_{z,rms,\,n}$ (output) | 0.44 mm-mrad |
| Inter_vane voltage | 80 kV |
| Maximum surface electric field, $E_s$ | 1.72 Kilpatrick |
| Total length | 348.53 cm |
| Transmission | 95 % |
| Beam power | 42.75 kW |
| Total beam power loss | 359 W |

Plot of beam envelopes in RFQ is shown in Fig. 15. Combined beam profile in the LEBT and RFQ is shown in Fig. 16. Evolution of halo parameter is shown in Fig. 17. Note that the beam loss in RFQ occurs at an energy below2.5 MeV, which is the threshold energy for neutron production in copper. This justifies our choice of 3 MeV as the energy of RFQ.

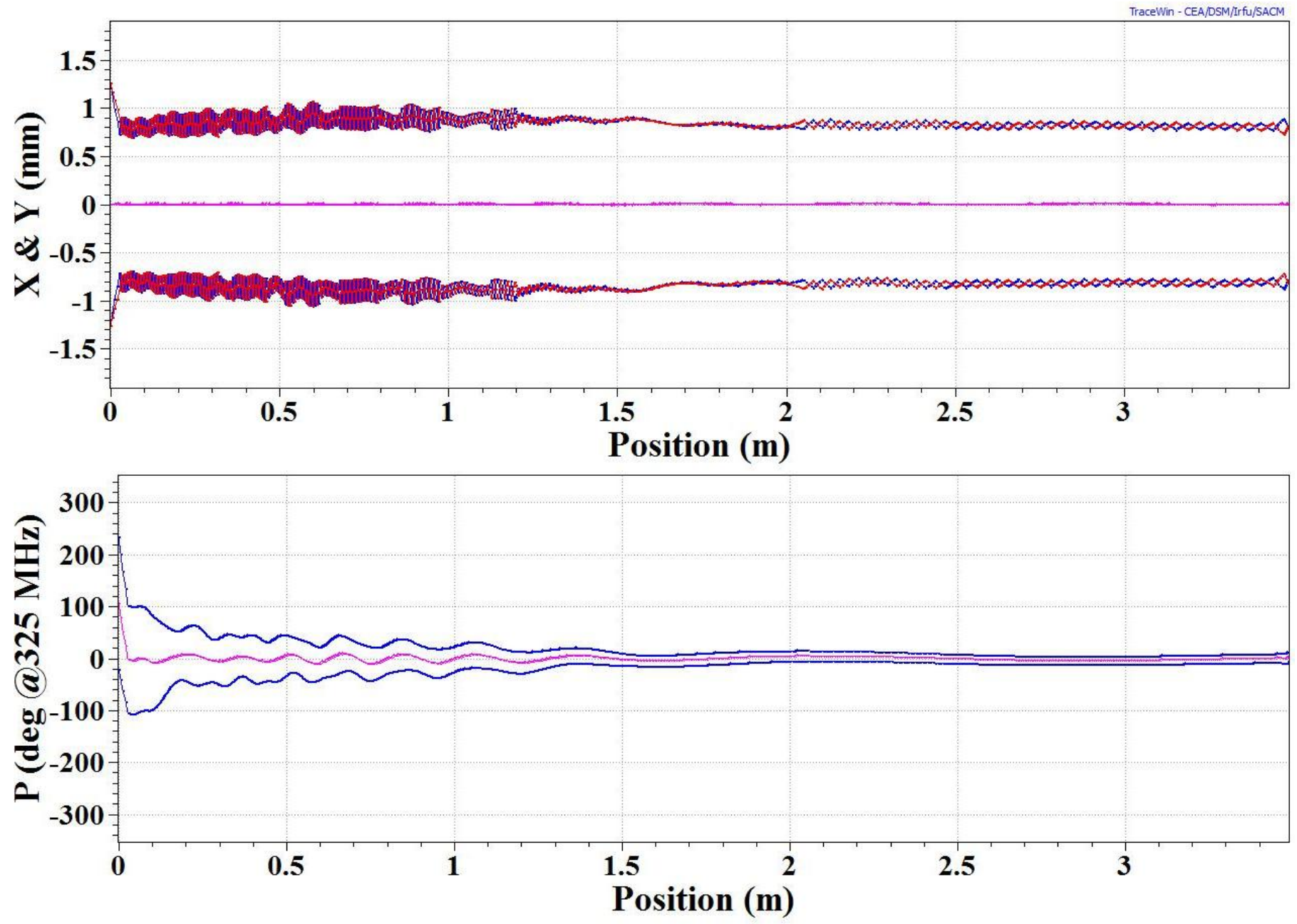

*Fig. 15: RMS beam envelope in transverse (upper) and longitudinal (lower) planes along RFQ.*

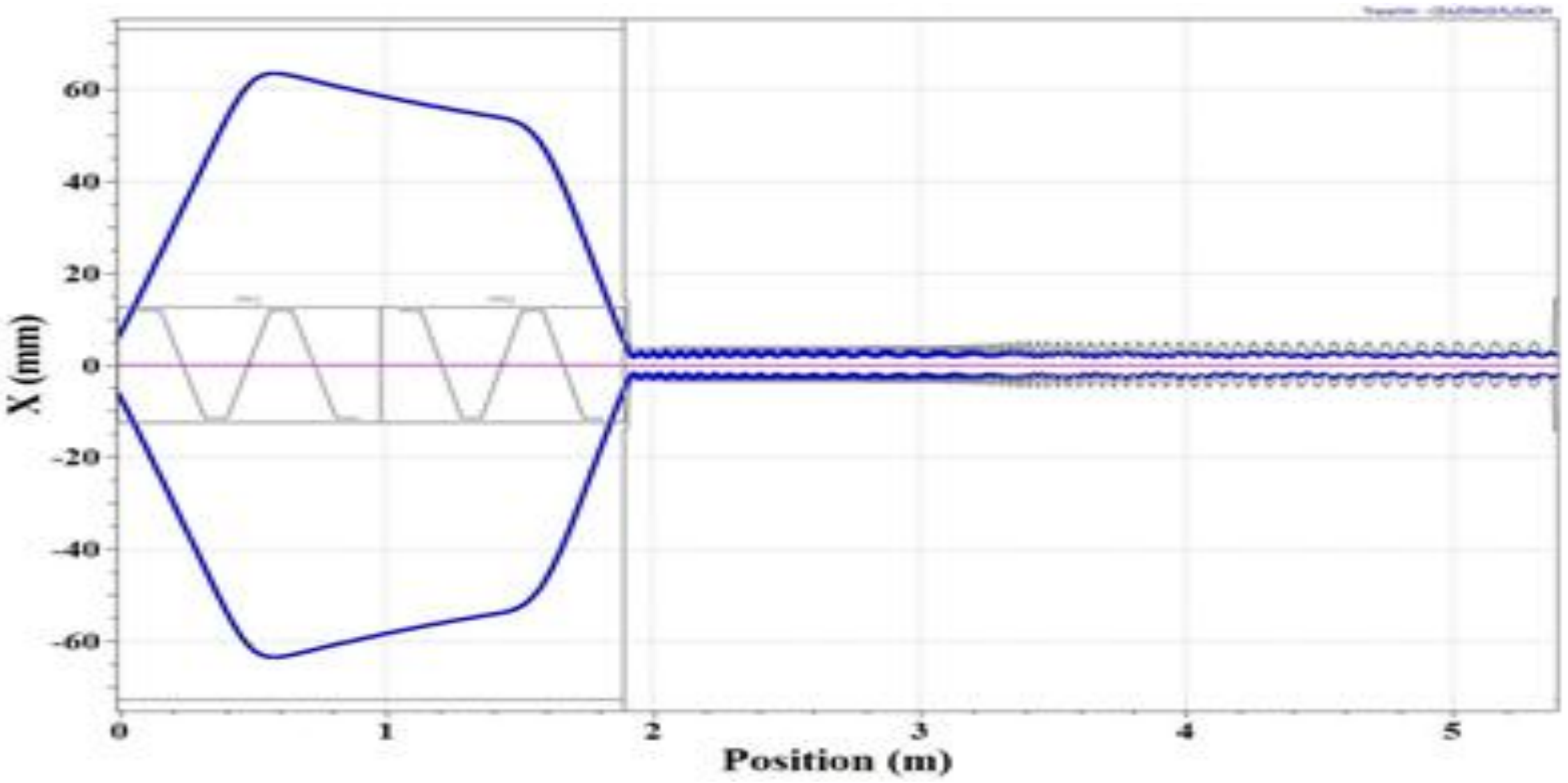

*Fig. 16: Beam envelope (3$\sigma$) along LEBT and RFQ.*

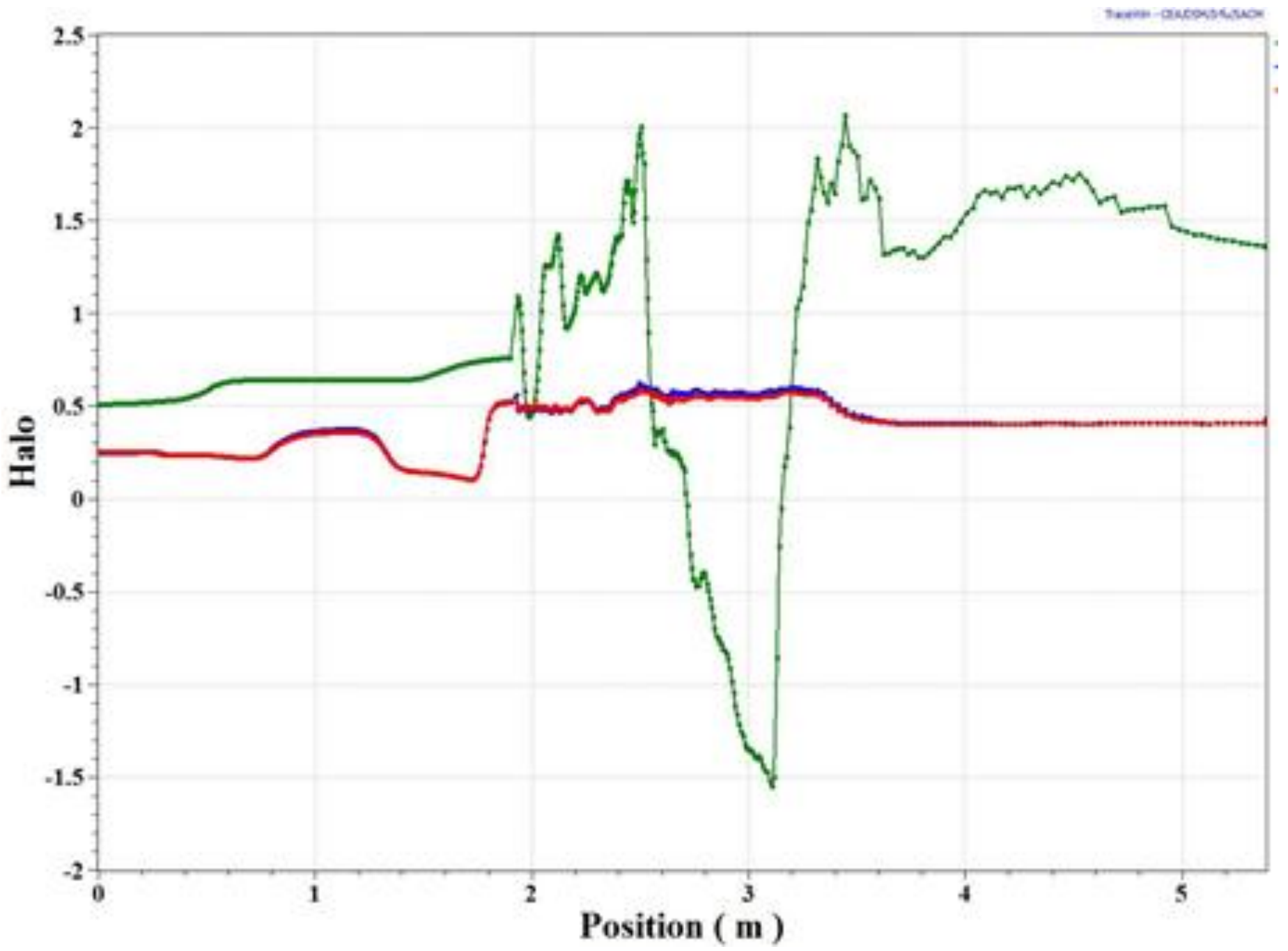

*Fig. 17: Halo parameter in longitudinal (green), vertical (blue) and horizontal (red) plane along RFQ.*

## 4. Medium Energy Beam Transport (MEBT)

Beam from RFQ will be further accelerated in SSSRs. The matched beam parameters at the input of SSR are different from the optimized value of beam parameters at the output of RFQ. In order to match the beam from RFQ to the SSR section, and also to accomplish the final beam chopping @ 1 MHz, with the width of chopped pulse around 650 ns and the rise time of few ns, a dedicated MEBT section is required. The matched beam parameters at the RFQ exit, and SSR0 entrance are given in Table 11 and Table 12 respectively.

*Table 11: Matched beam parameters at the RFQ exit*

| Parameter | Value |
|---|---|
| $\alpha_x$ | 1.2887 |
| $\beta_x$ | 0.1321 m/rad |
| $\alpha_y$ | -1.3502 |
| $\beta_y$ | 0.1360 m/rad |
| $\alpha_\varphi$ | 0.0266 |
| $\beta_\varphi$ | 0.4375 deg/keV |
| $\varepsilon_{xx',rms,n}$ | 0.3973 mm-mrad |
| $\varepsilon_{yy',rms,n}$ | 0.3996 mm-mrad |
| $\varepsilon_{\Delta\varphi\Delta w,rms}$ | 163.7 deg-keV |

*Table 12: Matched beam parameters at the SSR0 entrance*

| Parameter | Value |
|---|---|
| $\alpha_x$ | -1.0399 |
| $\beta_x$ | 0.5493 m/rad |
| $\alpha_y$ | -1.1050 |
| $\beta_y$ | 0.5420 m/rad |
| $\alpha_\varphi$ | -0.3008 |
| $\beta_\varphi$ | 0.3652 deg/keV |

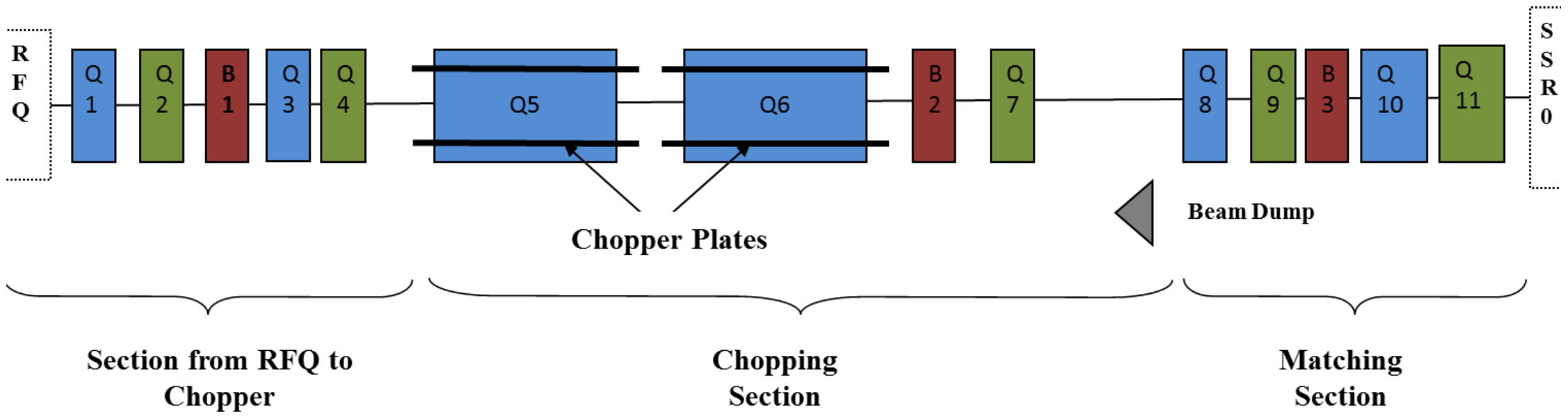

*Fig. 18: Schematic of the layout of MEBT.*

The schematic of the layout of MEBT is shown in Fig. 18. It consists of three sections. The first section consists of four quadrupoles and one buncher cavity, which are tuned to achieve the optimized parameters at the beginning of the second section, which performs beam chopping. The second section consists of three quadrupoles, one buncher cavity and two beam choppers, and a beam dump. Meander type traveling wave beam chopper is proposed to be installed in the aperture of two long quadrupoles as shown in Fig. 18. The third section consists of four quadrupoles, and a buncher cavity, which are tuned to match the beam to SSR0 section. Details of optimized parameter of each transport line element are given in Table 13.

For the optimized MEBT lattice, beam dynamics calculations were performed and plot of beam envelopes in transverse as well as longitudinal directions are shown in Fig. 19 [8]. Note that the beam profiles shown are the total beam sizes in each case, where the total beam size is √5 times the respective rms value. The length of complete MEBT lattice is 3.860 m. Phase space of the input and output beams are shown in Figs. 20 and 21 respectively. Variations of rms emittances along the length of the MEBT with space charge are shown in Fig. 22.

*Table 13: Optimized parameters of MEBT elements for operation with 15 mA beam current*

| Element | Parameters |
|---|---|
| Drift D1 | L=120 mm |
| Quad Q1 | G=30 T/m, L=50 mm |
| Drift D2 | L=120 mm |
| Quad Q2 | G=-25 T/m, L=50 mm |
| Drift D3 | L=130 mm |
| Buncher B1 | V=94 kV,$\phi_s$=-90$^0$ |
| Drift D4 | L=130 mm |
| Quad Q3 | G=26.5 T/m, L=50 mm |
| Drift D5 | L=80 mm |
| Quad Q4 | G=-23.8 T/m, L=50mm |
| Drift D6 | L=150 mm |
| Quad Q5 | G=0.685 T/m, L=255 mm |
| Drift D7 | L=115 mm |
| | L=115 mm |
| Quad Q6 | G=1.20 T/m, L=255 mm |
| Drift D8 | L=300 mm |
| Buncher B2 | V=75.19 kV, $\phi_s$=-90$^0$ |
| Drift D9 | L=220 mm |
| Quad Q7 | G=-1.3 T/m, L=155 mm |
| Drift D10 | L=355 mm |
| | L=215 mm |
| Quad Q8 | G=13.98 T/m, L=63.00 mm |
| Drift D11 | L=120 mm |
| Quad Q9 | G=-9.67 T/m, L=143 mm |
| Drift D12 | L=118 mm |
| Buncher B3 | V=101.57 kV, $\phi_s$=-90$^0$ |
| Drift D13 | L=120 mm |
| Quad Q10 | G=23.11 T/m, L=56 mm |
| Drift D14 | L=144 mm |
| Quad Q11 | G=-31.30 T/m, L=58 mm |
| Drift D15 | L=123 mm |

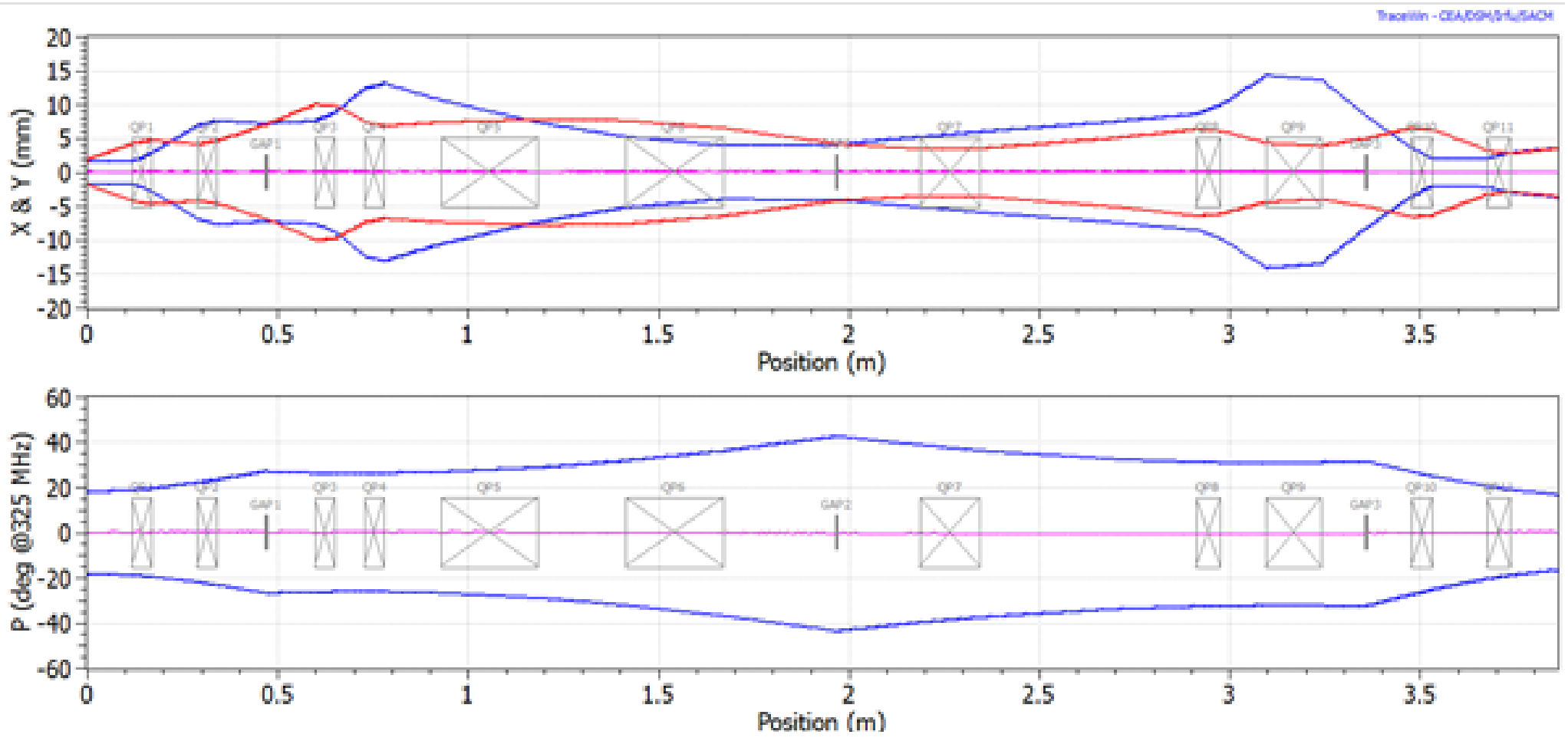

*Fig. 19: Horizontal (top blue), vertical (top red) and long. (bottom, blue) beam size along MEBT.*

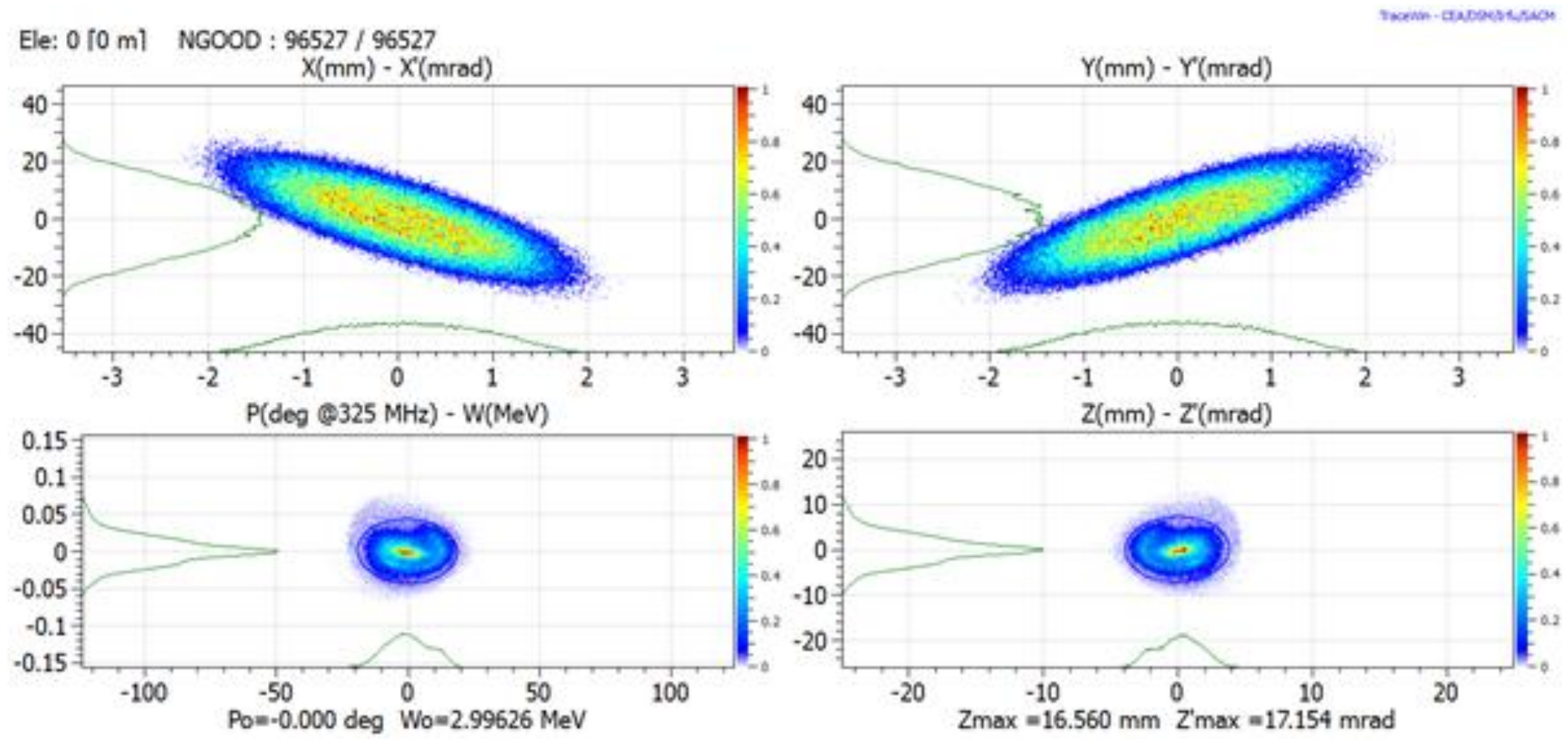

*Fig. 20: Phase space of the beam at the entrance of MEBT*

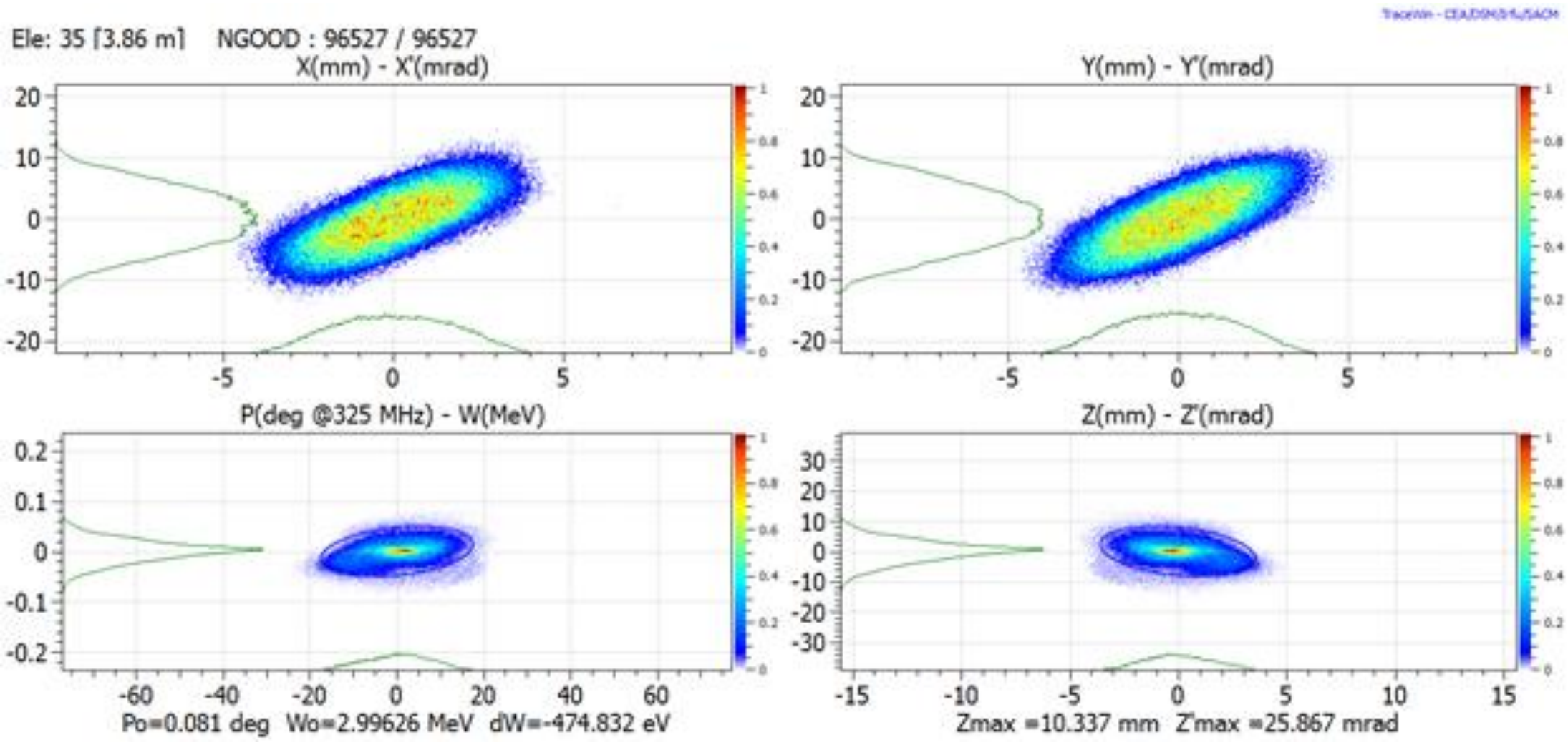

*Fig. 21: Phase space of the beam at the output of MEBT.*

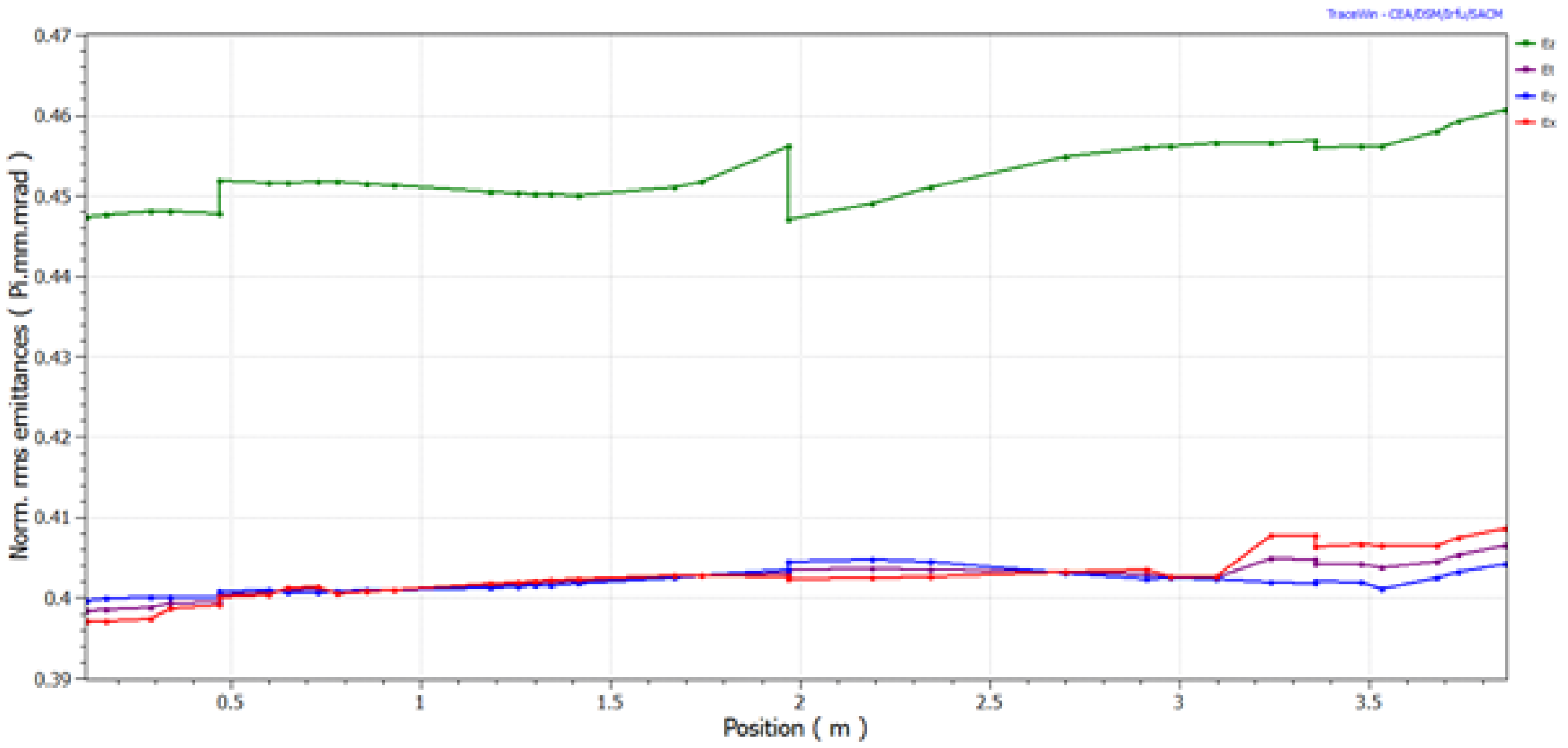

*Fig. 22: Variation of longitudinal (green), vertical(red) and horizontal (blue) emittances, along MEBT.*

Detailed studies on the design of meander type traveling wave chopper, having a fast rise time ~ 2ns, are going on. Meander type chopper structure is a slow wave structure that is chosen to make the phase velocity of transverse electric field same as that of ion velocity for efficient chopping. We briefly discuss here the initial results of the design calculations. The chopper structure is similar to that of the CERN-SPL project [9]. Metallic meander is itched on a dielectric substrate (alumina, relative permittivity=9.9), which is present on a metallic ground plane. Similar plate is present on the other side of the beam axis. The beam passes along the beam axis through the vacuum present in between the two plates. The voltage applied on the chopper plates will be around ±600 V. The on-time of the voltage pulse is equal to 350 ns, which is the time interval during which we want to chop the beam. The length of each chopper plate is around 40 cm. As the distance between chopper plates is 20 mm, the deflecting transverse field is ~ 60 kV/m. As the required rise time ($t_r$) of the voltage pulse is 2 ns, the bandwidth of chopper is given by $1/3t_r$, which comes out to be around 200 MHz. Thus, the frequency band of interest is 0-200 MHz, though time domain simulations have been carried out for a frequency band of 0-1000 MHz.

Power is fed from one end of meander geometry and taken out from the other through multi-pin connectors. A six period meander geometry has been simulated with the dimensions shown in Fig. 23. Straight section of 5 mm each has been considered on either side for connecting the coaxial connectors. The thickness of the metallic meander and ground plate is taken to be 50 microns each. The ground plane below the dielectric is not visible in Fig. 23. Dimensions of the triangular cut are shown in Fig. 24. Total number of periods in the meander structure is 66.

Figure 25 gives a plot of reflection coefficient $S_{11}$ versus frequency. It is seen that $S_{11}< -15$ dB. For almost zero reflection, it is desired to bring $S_{11}< -20$ dB. Simulations are ongoing to improve the results. Conductor losses are taken into account by considering the conductivity of copper in the simulations. Figure 26 gives a plot of magnitude of transmission coefficient $|S_{21}|$ versus frequency, which clearly shows that the attenuation is higher at higher frequency.

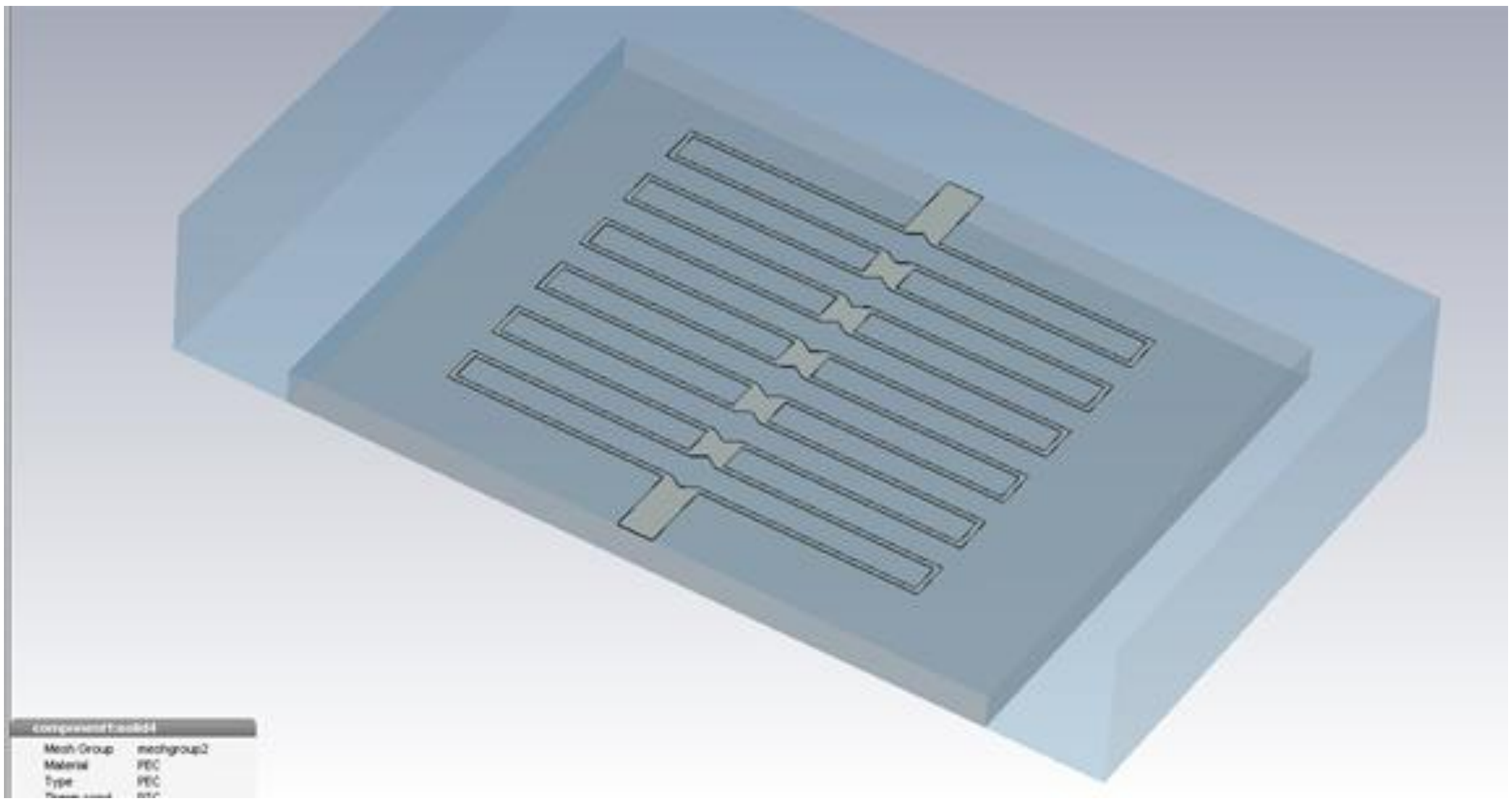

*Fig.23: Six period meander geometry*

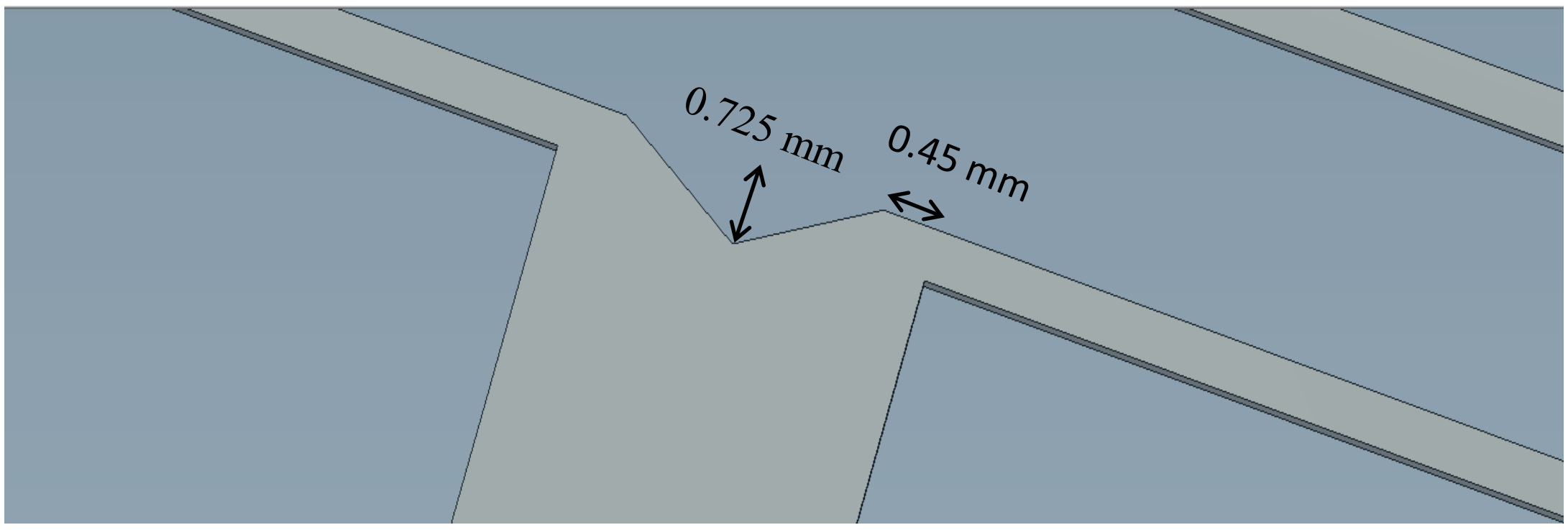

*Fig. 24: Dimensions of the triangular cut.*

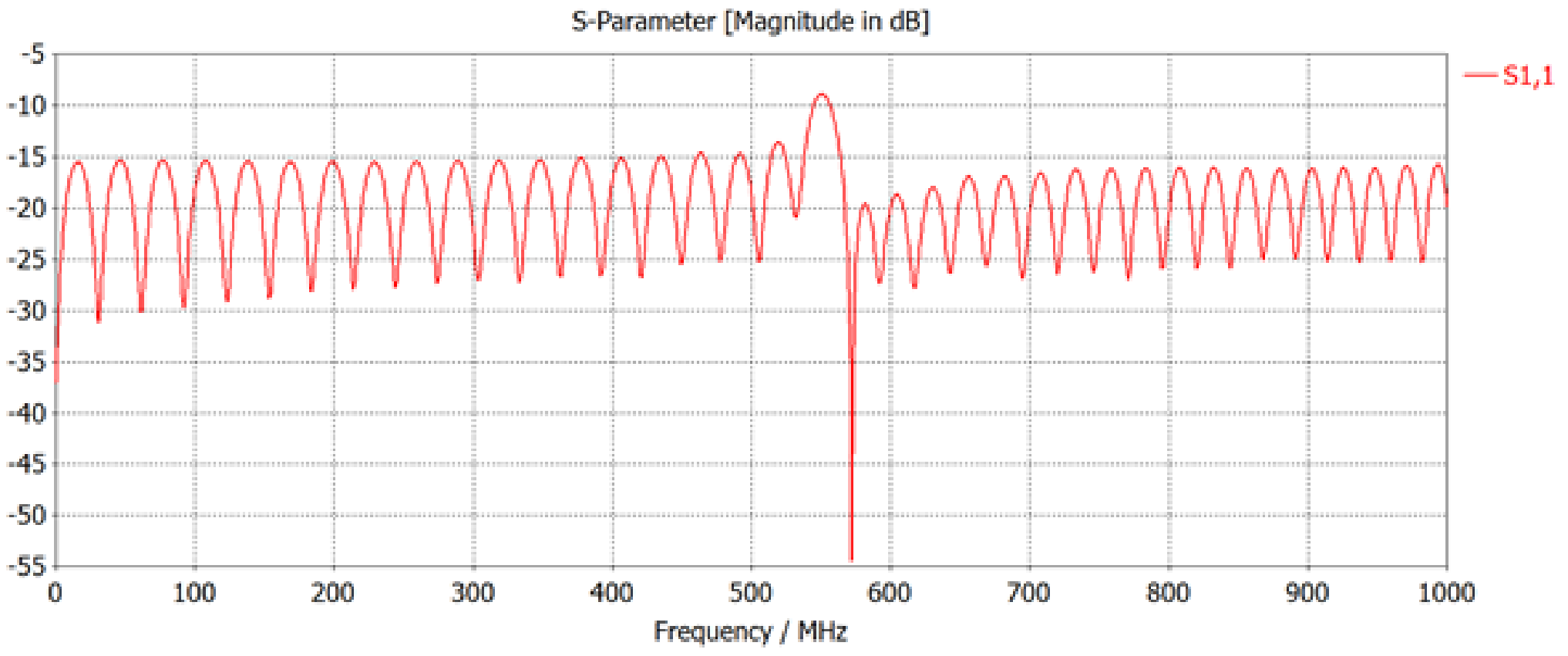

*Fig. 25: Magnitude of reflection coefficient ($S_{11}$) versus frequency.*

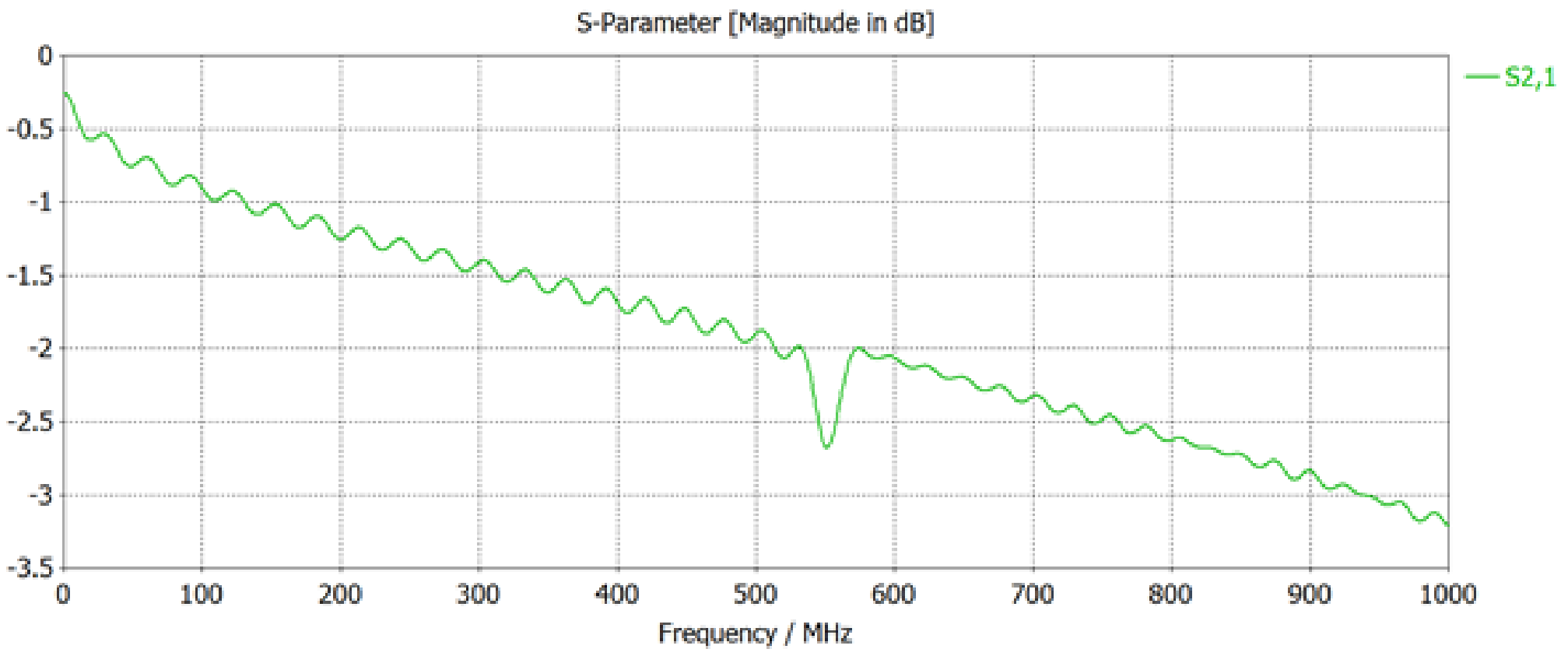

*Fig. 26: Magnitude of transmission coefficient ($S_{21}$) versus frequency.*

The time taken by 3 MeV (β=0.079) ion beam to traverse the length of the chopper is called the electrical length and given by $L_{chop}/\beta c$= 16.93 ns, where $L_{chop}$ is the length of chopper structure (40.6 cm) and $c$ is the speed of light. It is intended that the transverse electric field should also traverse the chopper length in the same time for efficient chopping. The electrical length has been estimated using the phase plot of transmission coefficient. The electrical length is a strong function of the transverse width of the metallic meander. For a width of 40.4 mm, the value of electrical length comes out to be around 16.44 ns, which is close to the analytical value.

Trajectory of the beam when the beam chopper is switched on is shown in Fig. 27.

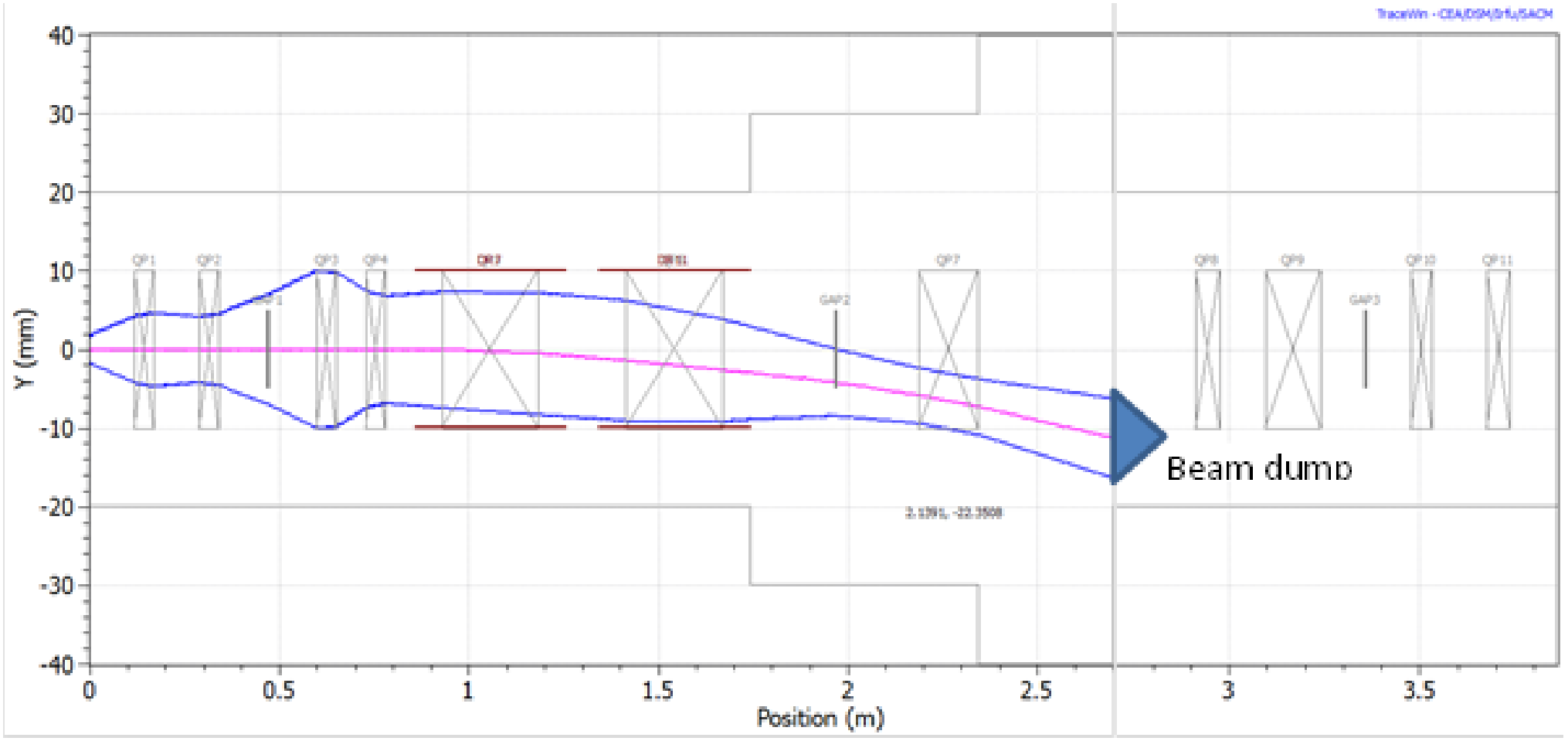

*Fig. 27: Trajectory of the beam when the chopper is switched on*

Electromagnetic design studies have been performed for the three buncher cavities for the MEBT. These cavities are designed for low power consumption i.e. high effective shunt impedance which in turn simplifies the cooling requirement. The peak surface electric field has been kept below 1.5 times the Kilpatrick limit, and the dimensions are chosen to ensure that the cavities fit inside the mechanical limitations of the MEBT. A coupled cavity linac (CCL) type of cavity is selected for the buncher cavity, as it offers higher shunt impedance and is easier to fabricate, as compared to the drift tube linac (DTL) type buncher cavity. Figure 28 shows the standard CCL cell. The various parameters of the buncher cavity are summarized in Table 14. Table 15 lists the figures of merit of the cavity.

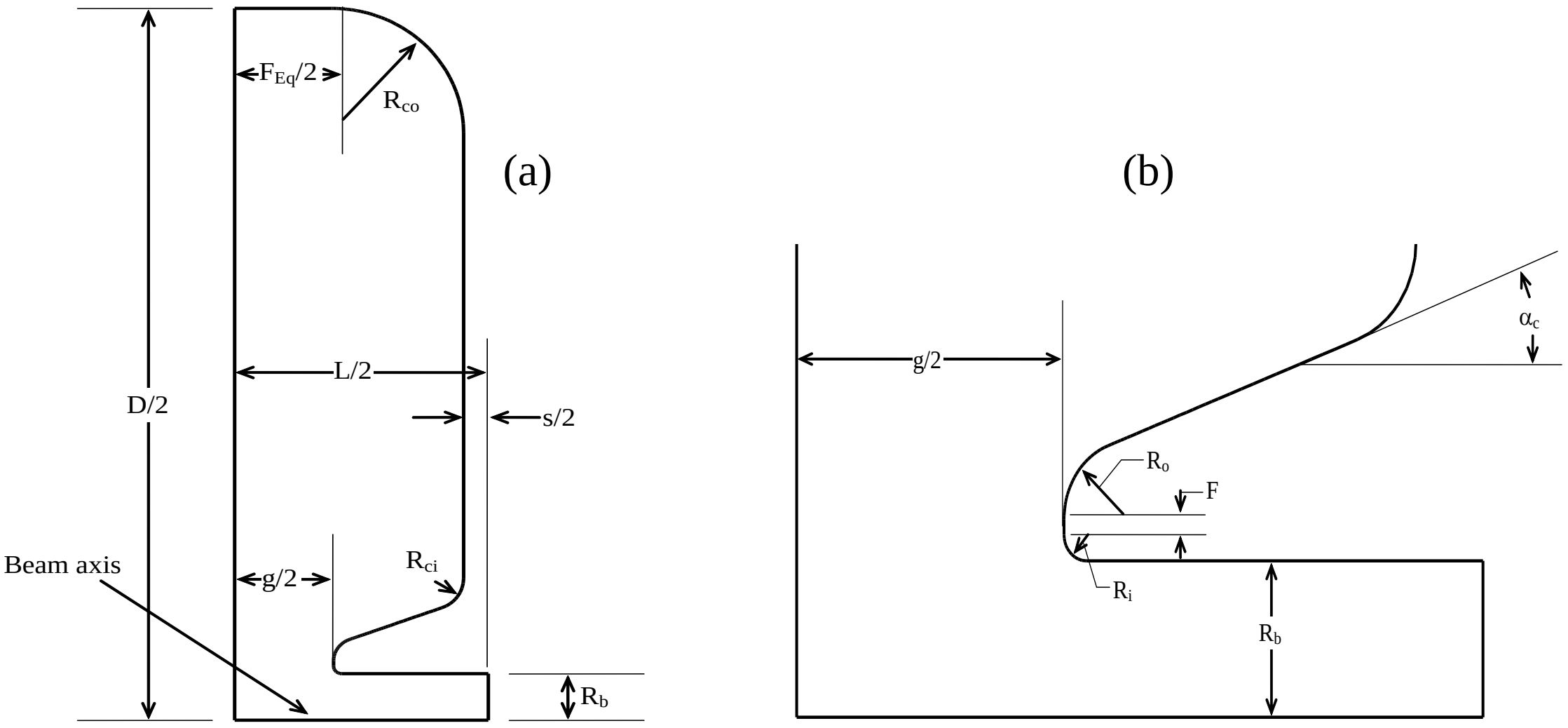

*Fig. 28 (a) The CCL half cell (b) Detail near the nose in a CCL cell.*

*Table 14: Parameters of the MEBT buncher cavity*

| | |
|---|---|
| Cavity diameter (*D*) | 55.77 cm |
| Bore Radius ($R_b$) | 1.5 cm |
| Cavity length (*L*) | 17.5 cm |
| Gap length (*g*) | 1.4 cm |
| Inner Corner radius ($R_{ci}$) | 2.0 cm |
| Outer Corner radius ($R_{co}$) | 7.0 cm |
| Inner nose radius ($R_i$) | 0.3 cm |
| Outer nose radius ($R_o$) | 0.3 cm |
| Flat length (*F*) | 0.1 cm |
| Cone angle ($\square_c$) | 25° |

*Table 15: Figures of merit of the MEBT buncher cavity*

| | Buncher 1 | Buncher 2 | Buncher 3 |
|---|---|---|---|
| Frequency (MHz) | 325 | 325 | 325 |
| Effective voltage (kV) | 94 | 75.19 | 101.57 |
| Quality factor | 27916 | 27916 | 27916 |
| Transit time factor | 0.6267 | 0.6267 | 0.6267 |
| Effective shunt impedance (MΩ/m) | 14.98 | 14.98 | 14.98 |
| r/Q (Ω) | 46.94 | 46.94 | 46.94 |
| Power dissipation (kW) | 18.35 | 11.76 | 21.46 |
| Peak electric field (MV/m) | 17.15 | 13.74 | 18.56 |
| Kilpatrick | 0.96 | 0.77 | 1.04 |

## 5. Superconducting Single Spoke Resonators

The 3 MeV beam from the RFQ, after undergoing the required phase space transformations and chopping in MEBT is accelerated in a series of independently phased superconducting single spoke resonators (SSRs) and elliptic cavities (ECs). Three classes of SSRs with $\beta_g$ = 0.11, 0.22 and 0.42, and two classes of ECs with $\beta_g$ = 0.61 and 0.9 will be used. Figure 29 shows the transit time factor as a function of $\beta$, which is the speed of charged particle in unit of speed of light. The range of energy for which a particular class of cavities will be used, is decided such that the transit time factor is more than half of the maximum value.

Electromagnetic design of 325 MHz SSRs has been done [10,11], where geometrical parameters are optimized such that the peak field ratios $E_{pk}/E_{acc}$ and $B_{pk}/E_{acc}$ are minimum, in order to achieve the maximum acceleration gradient, multipacting is not significant at the operating value of accelerating field, and higher order modes do not have significant strength. Figure 30 shows the different views of the SSR to explicitly show the different geometrical parameters, and also the topology of a typical SSR cavity. Optimized value of geometrical parameters and RF parameters are shown in Table 16 and 17 respectively for $\beta_g$=0.11, 0.22 and 0.42. The value of unloaded quality factor $Q_0$ for operation at 2K is $3.77 \times 10^9$, $6.85 \times 10^9$ and $8.34 \times 10^9$, for SSR0, SSR1 and SSR2, respectively. Surface resistance is taken as 11 nΩ for the superconducting material at the operating temperature and frequency.

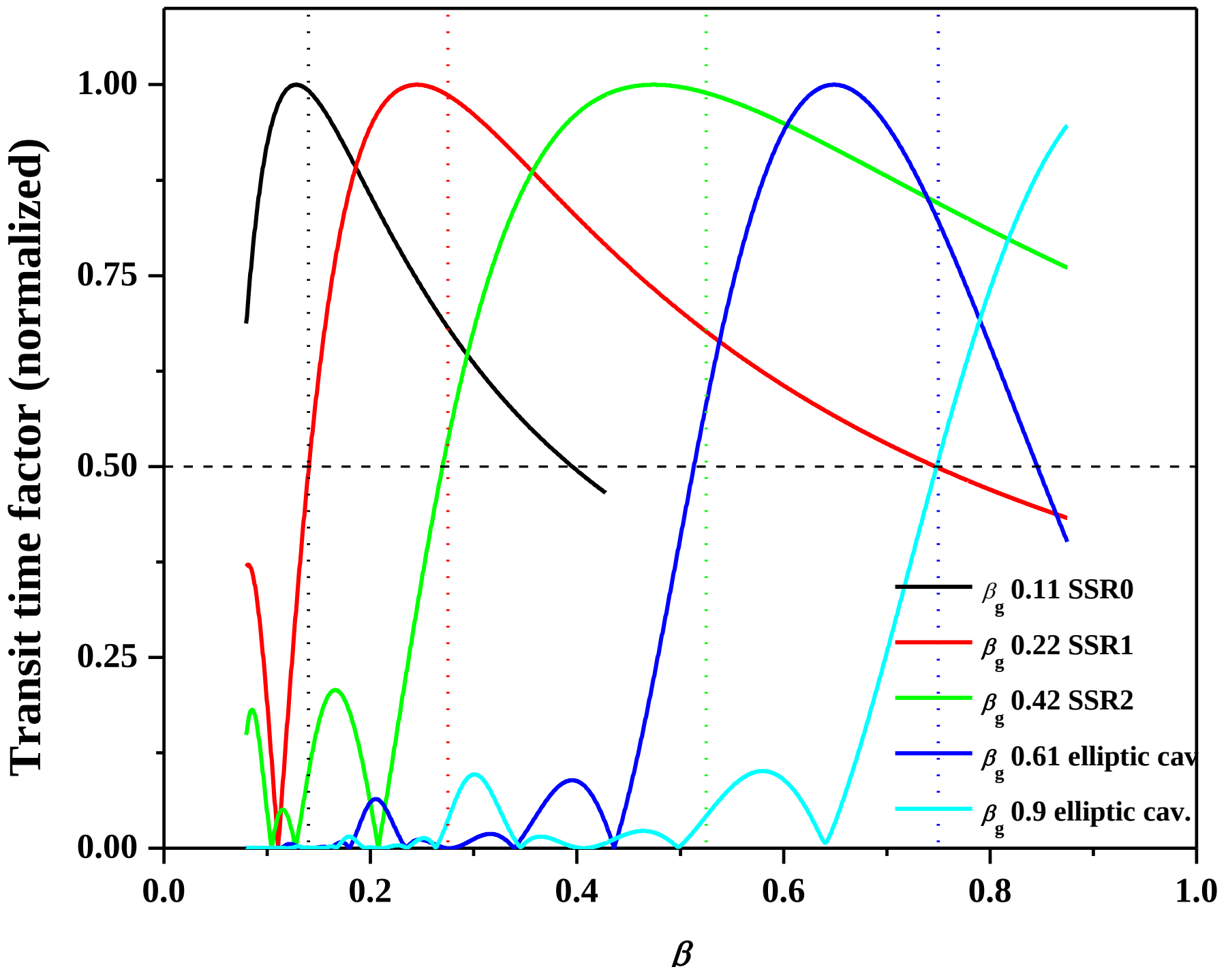

*Fig. 29: Transit time factor, normalized against its maximum value, as a function of β for 5 different sections used in the ISNS main linac. Dotted lines show the transition β from one section to another.*

For higher order dipole and quadrupole modes, the *R*/*Q* is less than 2 $\Omega/cm^2$ and 0.01 $\Omega/cm^2$ respectively, and for monopole modes except the operating mode, *R*/*Q* is typically less than 10 Ω. Calculations have been performed for the frequencies up to 2.5 GHz.

Multipacting studies have been performed and maximum growth rate in SSR2, SSR1 and SSR0 is calculated as 0.1 $ns^{-1}$, 0.14 $ns^{-1}$ and 0.15 $ns^{-1}$, respectively. This is similar to the growth rate achieved in the design of other projects such as CSNS. Figures 31 and 32 show the growth rate of multipacting for SSR2 and SSR1 respectively.

RF power will be coupled through a 3-inch co-axial transmission line and probe coupler, having an antenna terminated by a disk of diameter 60 mm [12]. A schematic of RF power coupler is shown in Fig. 33. The value of $Q_{ext}$ for critical coupling in the presence of beam is $9.3\times10^4$ to $7.5\times10^5$ for SSR0, 2.2 $\times10^5$ to 1.7 $\times$ $10^6$ for SSR1, and 4.0 $\times10^5$ to 2.2 $\times10^6$ for SSR2. This can be achieved by varying the antenna penetration depth from -40 mm to +100 mm. Multipacting studies of RF power coupler are in progress.

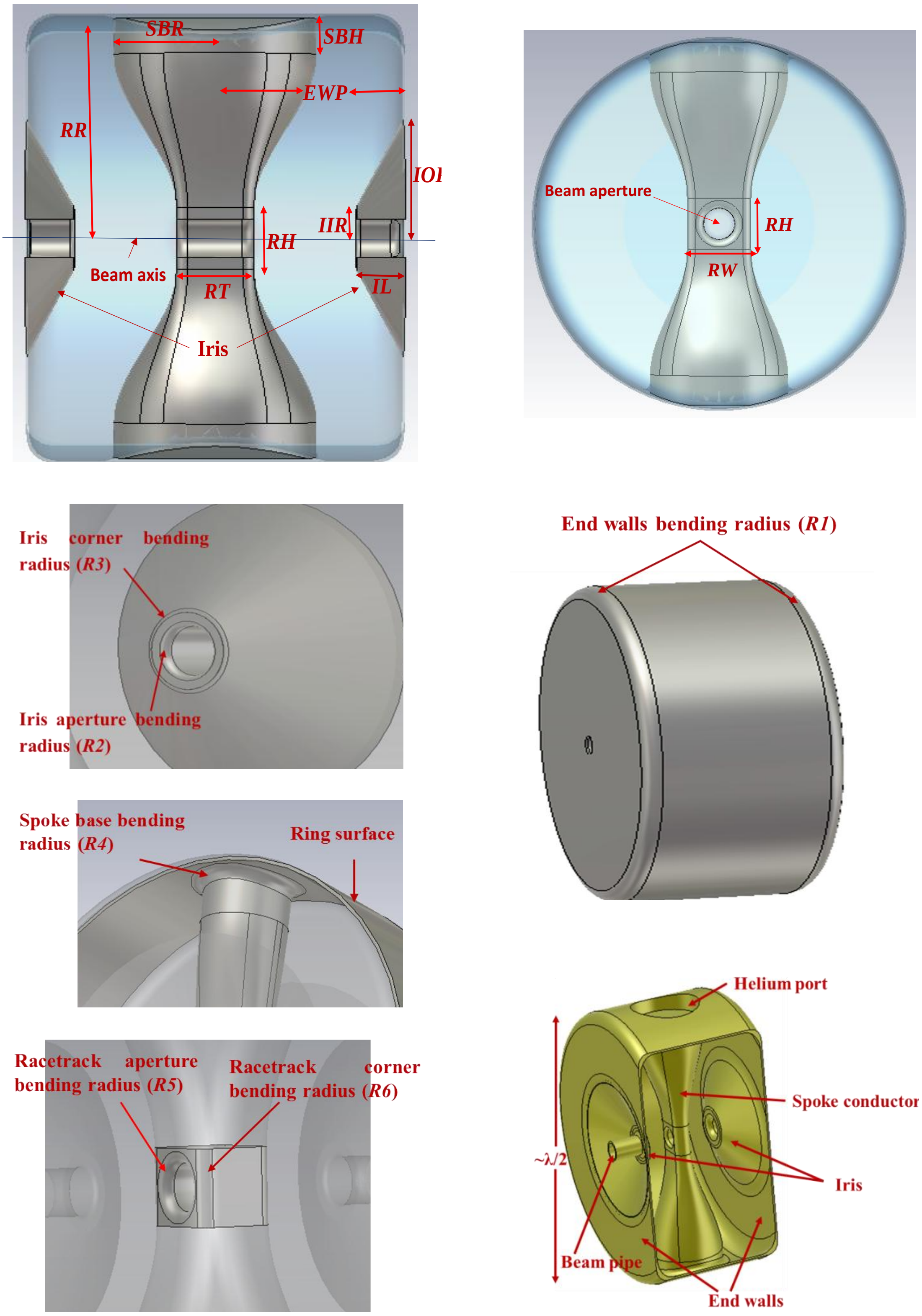

*Fig. 30: Different views of a SSR, explaining the important geometrical parameters and the topology.*

*Table 16: Optimized geometrical parameters of SSR0, SSR1 and SSR2*

| Geometrical Parameter | SSR2 | SSR1 | SSR0 |
|---|---|---|---|
| Racetrack height (*RH*) | 80.00 mm | 54.00 mm | 80.00 mm |
| Racetrack width (*RW*) | 100.00 mm | 70.00 mm | 66.00 mm |
| Racetrack thickness (*RT*) | 83.20 mm | 63.60 mm | 33.00 mm |
| Spoke base height (*SBH*) | 50.03 mm | 32.72 mm | 39.90 mm |
| Spoke base radius (*SBR*) | 109.60 mm | 65.00 mm | 38.00 mm |
| End Wall Position (*EWP*) | 220.00 mm | 117.00 mm | 65.85 mm |
| Ring Radius (*RR*) | 293.03 mm | 249.42 mm | 217.27 mm |
| Iris inner radius (*IIR*) | 41.00 mm | 36.00 mm | 33.00 mm |
| Iris outer radius (*IOR*) | 155.00 mm | 125.00 mm | 150.00 mm |
| Iris length (*IL*) | 67.80 mm | 47.30 mm | 34.25 mm |
| End wall corner bending radius (*R1*) | 90.00 mm | 20.00 mm | 10.00 mm |
| Iris aperture bending radius (*R2*) | 10.00 mm | 10.00 mm | 5.00 mm |
| Iris corner bending radius (*R3*) | 10.00 mm | 10.00 mm | 20.00 mm |
| Spoke base bending radius (*R4*) | 15.00 mm | 5.00 mm | 10.00 mm |
| Racetrack aperture bending radius (*R5*) | 12.00 mm | 5.00 mm | 5.00 mm |
| Racetrack corner bending radius (*R6*) | 12.00 mm | 15.00 mm | 13.00 mm |
| Beam aperture diameter | 50.00 mm | 30.00 mm | 30.00 mm |
| Loft tangency | 0.2 | 0.0 | 0.2 |
| Power coupler diameter | 76.71 mm | 76.71 mm | 76.71 mm |
| Vacuum port diameter | 76.71 mm | 76.71 mm | 76.71 mm |
| Helium port diameter | 219.20 mm | 130.20 mm | 76.00 mm |

*Table 17: Optimized RF parameters for SSR2, SSR1 and SSR0*

| Parameter | SSR2 | SSR1 | SSR0 |
|---|---|---|---|
| $E_p/E_{acc}$ | 3.4 | 3.3 | 4.1 |
| $B_p/E_{acc}$ | 5.9 mT/(MV/m) | 5.1 mT/(MV/m) | 6.1 mT/(MV/m) |
| *R/Q* | 270 Ω | 231.56 Ω | 128 Ω |
| $G$ (= $QR_s$) | 112 Ω | 80 Ω | 48 Ω |
| Maximum acceleration gradient, $E_{acc,max}$ | 10.2 MV/m | 11.82 MV/m | 9.8 MV/m |
| Geometric beta, $\beta_g$ | 0.42 | 0.22 | 0.11 |
| Optimum beta, $\beta_{opt}$ | 0.48 | 0.25 | 0.14 |
| Operating temperature | 2 K | 2 K | 2 K |
| Peak power per cavity (for 15 mA peak current) | 60 kW | 30 kW | 15 kW |
| Average power per cavity @ 10% DF | 6 kW | 3 kW | 1.5 kW |
| Cryogenic load per cavity | 4.3 W | 1.5 W | 1 W |
| Operating accelerating gradient, Eacc | 10 MV/m | 10 MV/m | 9 MV/m |

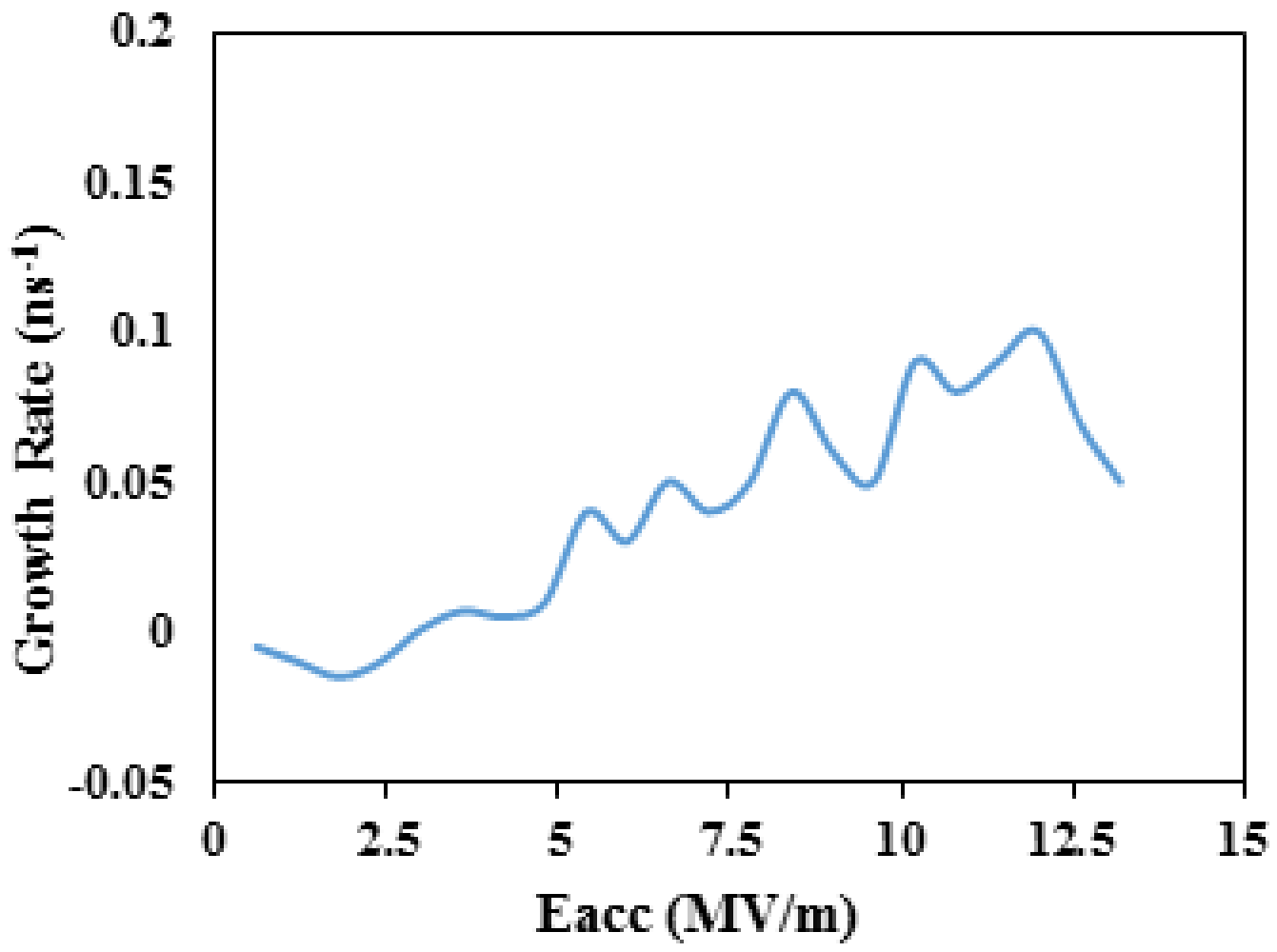

*Fig. 31: Multipacting growth rate as a function of accelerating electric field for SSR2.*

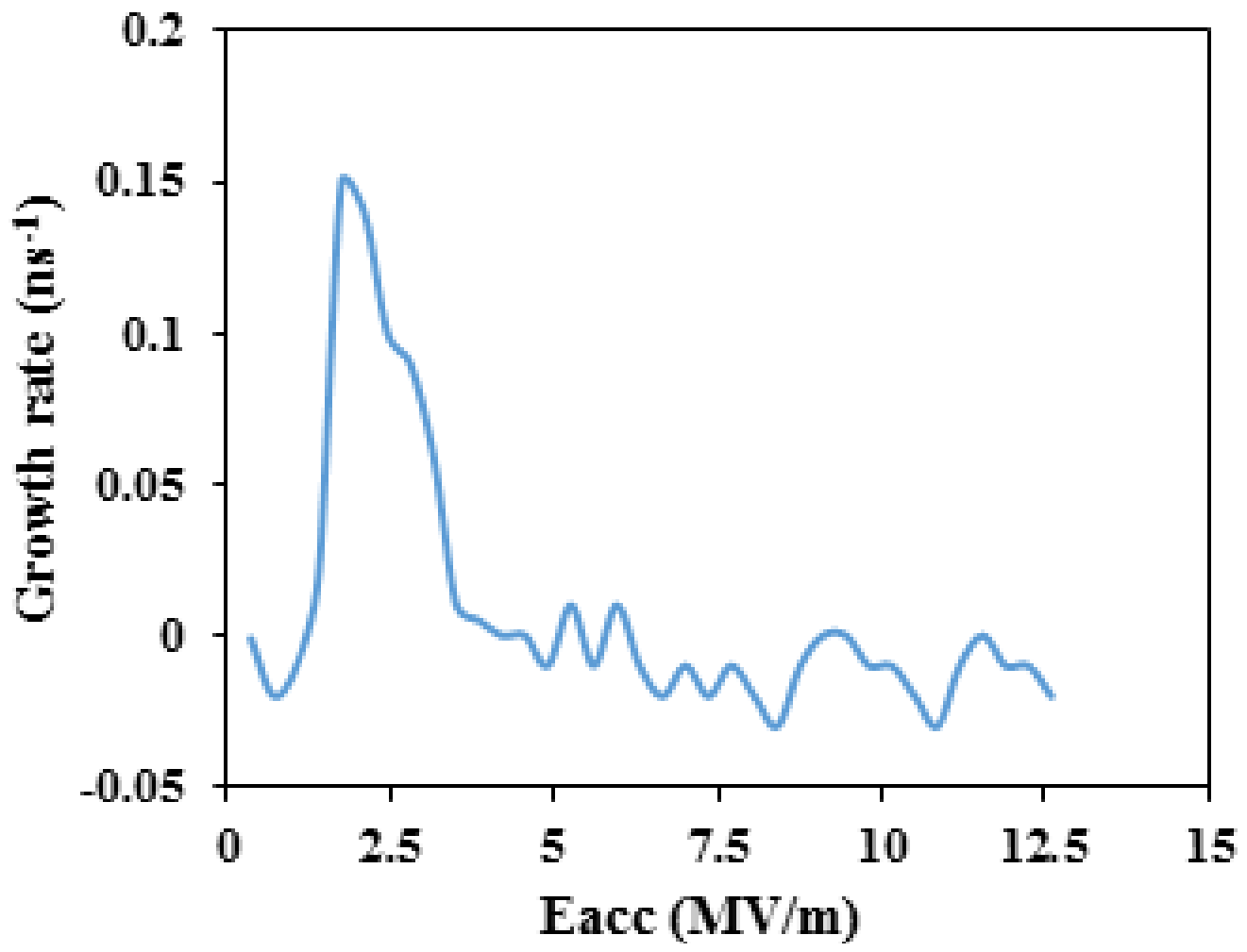

*Fig. 32: Multipacting growth rate as a function of accelerating electric field for SSR1.*

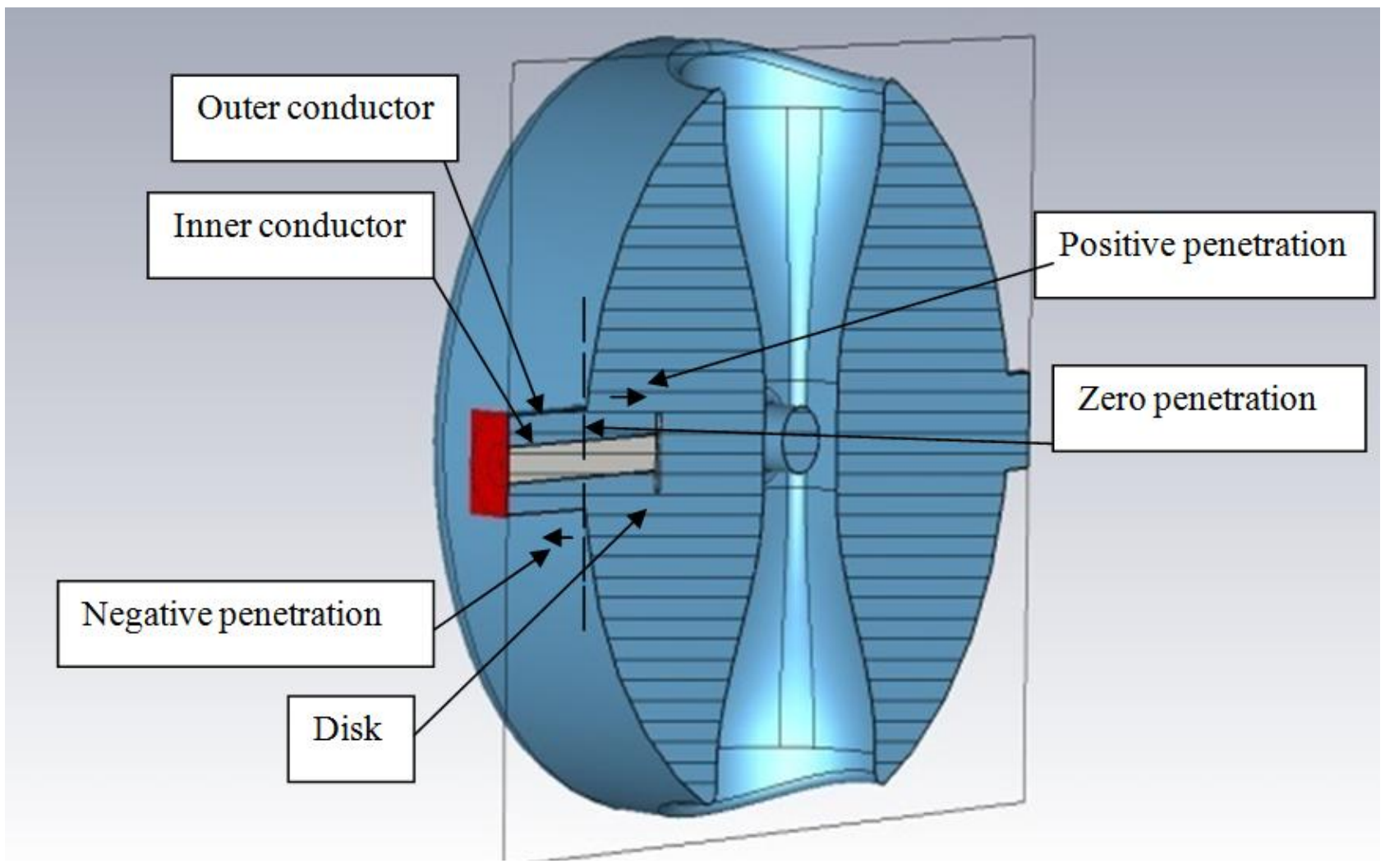

*Fig. 33: Schematic showing the fundamental RF power coupler of SSR.*

## 6. Superconducting Multi-cell Elliptic Cavities

The $H^-$ beam is accelerated in SSRs up to ~ 160 MeV, after which it is accelerated in superconducting multi-cell elliptic cavities. A typical model of a 5-cell, elliptic cavity, with cavity stiffener ring and helium vessel is shown in Fig. 34. The cross section of the half-cell, which is obtained by joining two elliptical arcs by a common tangent, is shown in Fig. 35.

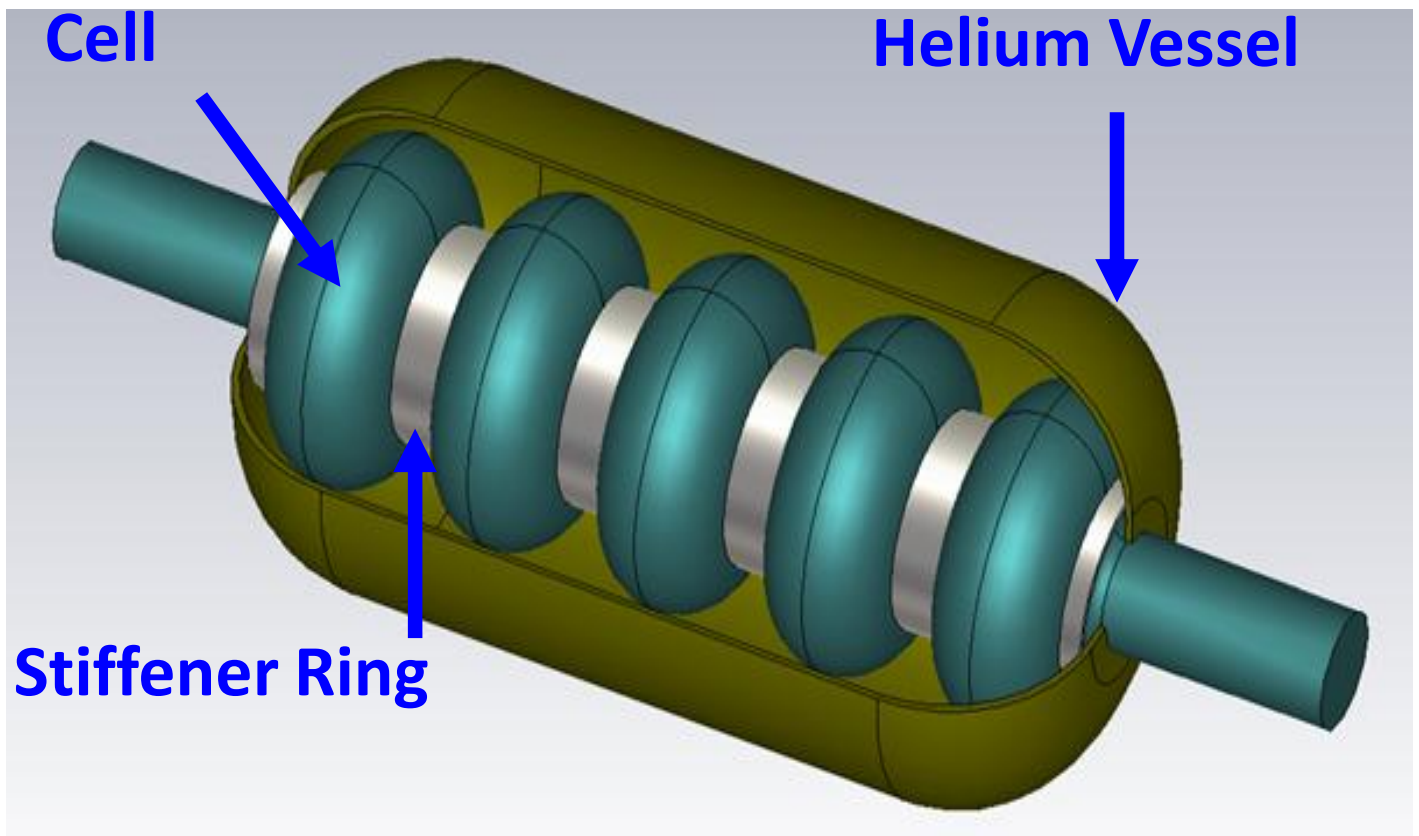

*Fig. 34: A 3D model of a 5-cell superconducting elliptic cavity.*

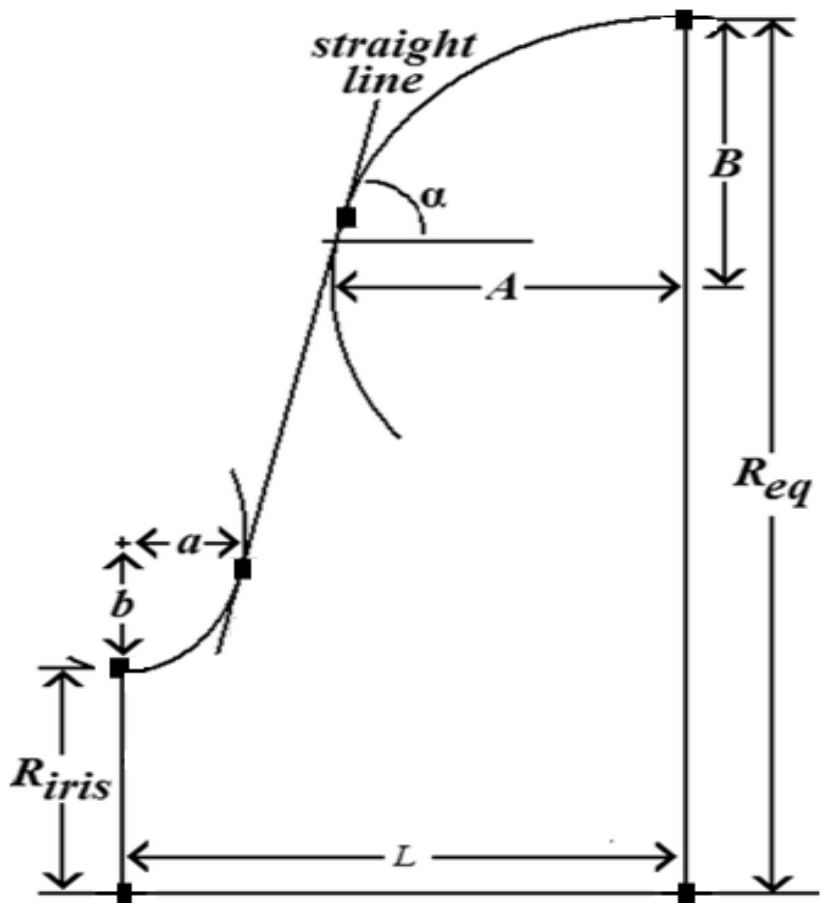

*Fig. 35: Schematic of the cross-section of half-cell.*

The geometrical parameters of $\beta_g$ = 0.61 and $\beta_g$ = 0.9, 5-cell 650 MHz cavity have been optimized such that the peak field ratios $E_{pk}/E_{acc}$ and $B_{pk}/E_{acc}$ are minimum to achieve the maximum acceleration gradient, higher order modes do not have significant strength, and there are no trapped higher order modes [13,14]. The geometry and shape have been further refined to ensure that there is no multipacting at the operating value of acceleration gradient. Tables 18 shows the optimized value of geometrical parameters for $\beta_g$ = 0.9 and 0.61 cavities. The RF parameters for the optimized design are summarized in Table 19.

*Table 18: Optimized geometrical parameters of 650 MHz, 5-cell superconducting elliptic cavity*

| **Parameter** | **$\beta_g$ = 0.61 cavity** | | | **$\beta_g$ = 0.9 cavity** | |
|---|---|---|---|---|---|
| | Mid-cell | End-cell (entry) | End-cell (exit) | Mid-cell | End-cell |
| $R_{iris}$ (mm) | 44.00 | 44.00 | 44.00 | 50.00 | 50.00 |
| $R_{eq}$ (mm) | 195.591 | 195.591 | 195.591 | 199.93 | 199.93 |
| $L$ (mm) | 70.336 | 71.55 | 71.24 | 103.77 | 105.80 |
| $A$ (mm) | 52.64 | 52.64 | 52.25 | 83.26 | 83.26 |
| $B$ (mm) | 55.55 | 55.55 | 55.55 | 84.00 | 84.00 |
| $a$ (mm) | 15.28 | 15.28 | 15.28 | 16.79 | 16.79 |
| $b$ (mm) | 28.83 | 28.83 | 28.83 | 29.45 | 29.45 |

*Table 19: RF parameters of elliptic cavities for the optimized geometry*

| **Parameter** | **$\beta_g$ = 0.61** | **$\beta_g$ = 0.9** |
|---|---|---|
| $E_{acc}$ (MV/m) | 15.4 | 18.6 |
| $E_{pk}/E_{acc}$ | 2.36 | 2.0 |
| $B_{pk}/E_{acc}$ (mT/(MV/m)) | 4.56 | 3.78 |
| $k_c$ | 0.8% | 0.8% |
| $R/Q$ (Ω) | 328 | 609 |
| Cryogenic load | 16 W | 20 W |

Lorentz force detuning (LFD) studies have been performed, and the thickness of cavity wall and helium vessel have been optimized to 4 mm and 5 mm respectively [15]. The optimized value of radial location of stiffener ring, taking dynamic LFD into account has been found to be 124 mm for the $\beta_g$ = 0.61 cavity and 119.5 mm for the $\beta_g$ = 0.9 cavity. Helium vessel will be made of titanium, and it will be joined with cavity using a transition piece made of 55Ti-45Nb alloy. Diameter of helium vessel is chosen as 504 mm, with end closure in “tori-spherical” shape with torus radius of 120 mm.

Calculations have been performed for heat load due to resonant excitation of HOMs due to the beam pulse structure. The value of $Q_{ext}$ is required to be kept below $10^7$ to ensure that heat load due to HOMs is not significant.

RF power will be coupled to these cavities through a 3 inch co-axial transmission line, and a probe coupler. For $\beta_g$ = 0.9 cavity, the axis of the co-axial transmission line will be at 48.4 mm from the end-cell exit, and required coupling coefficient can be achieved by varying the penetration depth in the range - 40 mm to 0 mm, for a disk type end tip coupler with optimum disk diameter of 55 mm [16]. A schematic showing the fundamental RF power coupler is shown in Fig. 36. For the $\beta_g$ = 0.61 cavity, axis of the co-axial transmission line will be at 41.5 mm from the end-cell exit.

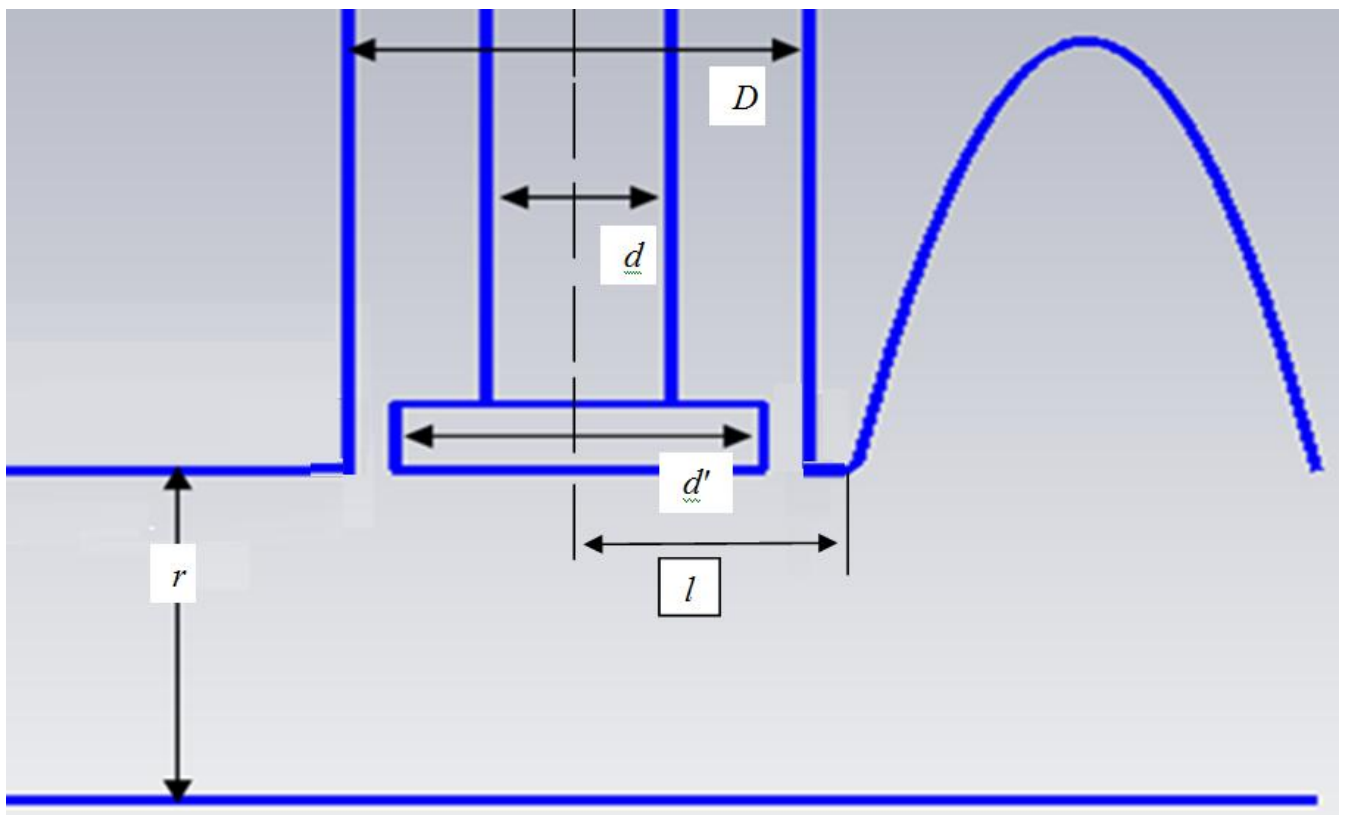

*Fig. 36: Schematic showing the fundamental power coupler for the superconducting cavity. Here D = 76.2 mm and d = 32.15 m.*

Electromagnetic design of HOM coupler is currently in progress.

Multipacting studies have been performed for the $\beta_g$ = 0.61 and 0.9 cavities, and by making suitable refinement in the geometry (i.e., by giving a convex radius of deformation of 10 mm at the equator), it has been ensured that there will be no multipacting in the cavity in the operating range of acceleration gradient, as seen in Fig. 37. Multipacting analysis of power couplers is in progress.

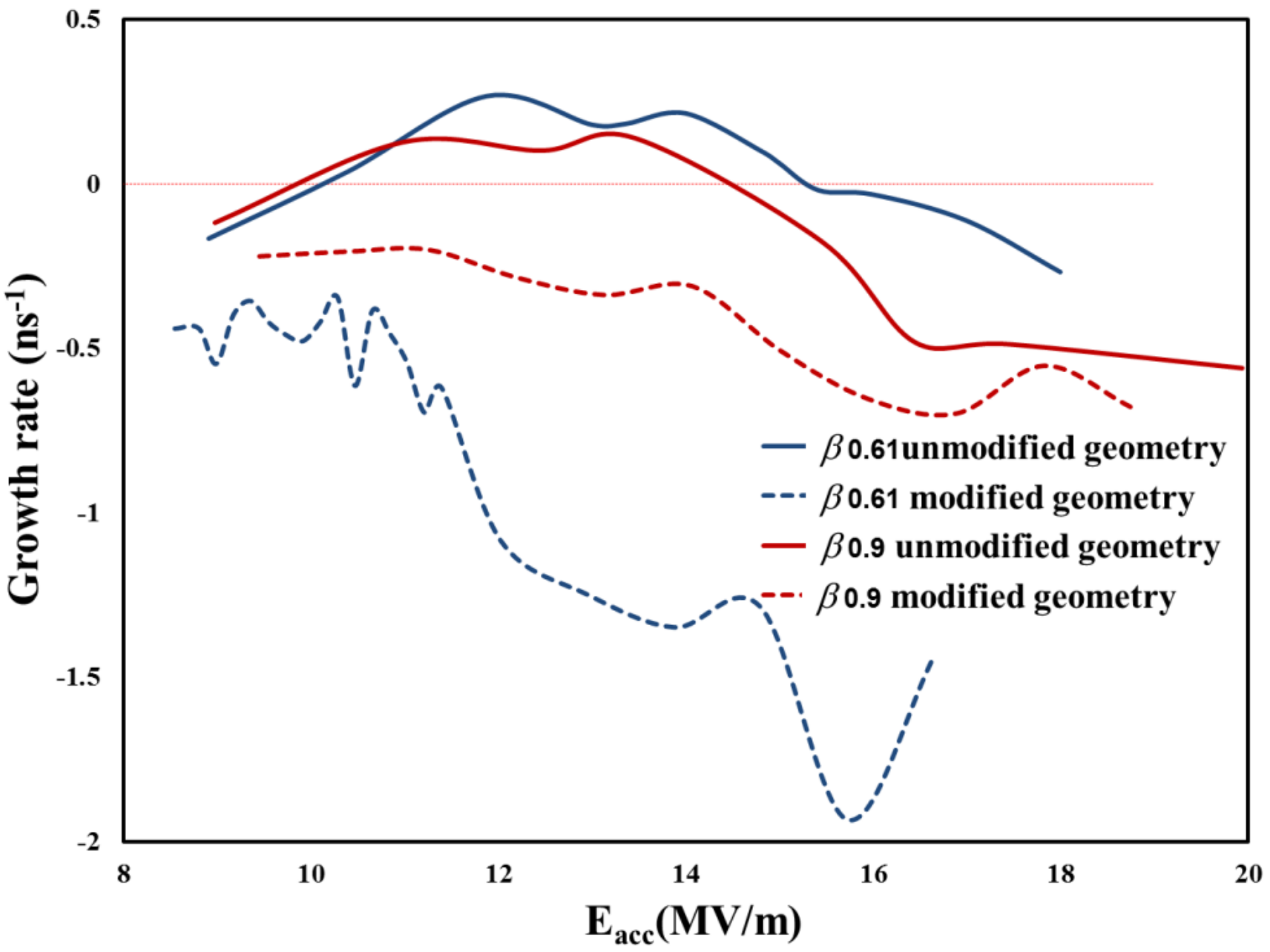

*Fig. 37: Multipacting growth rates in $\beta_g$=0.61 and 0.9 cavity for unmodified, as well as modified geometry.*

For the high beta section, an alternate option is to use $\beta_g$ = 0.81 cavity, which is more efficient for the energy range 500 MeV – 1 GeV. Electromagnetic design for 5-cell, 650 MHz, $\beta_g$ = 0.81 cavity has been completed and geometrical parameters have been optimized.

## 7. Lattice design of 1 GeV injector linac

In order to accelerate the 3 MeV $H^-$ beam up to 1 GeV in the low beta section having SSRs, followed by medium beta ($\beta_g$ = 0.61) and high beta ($\beta_g$=0.9) section, and keeping it adequately focused in the longitudinal as well as transverse directions, a suitable lattice of cavity and focusing magnets has been designed [17]. In the low beta section, superconducting solenoids will be used for transverse focusing. In the medium and high beta section, quadrupole doublets will be used for transverse focusing. The lattice configurations in different sections are shown in Fig. 38. The lattice has been optimized such that all the stringent beam dynamics criteria to minimize the beam loss are satisfied. Lengths of corresponding elements in the optimized lattice are also shown in Fig. 38.

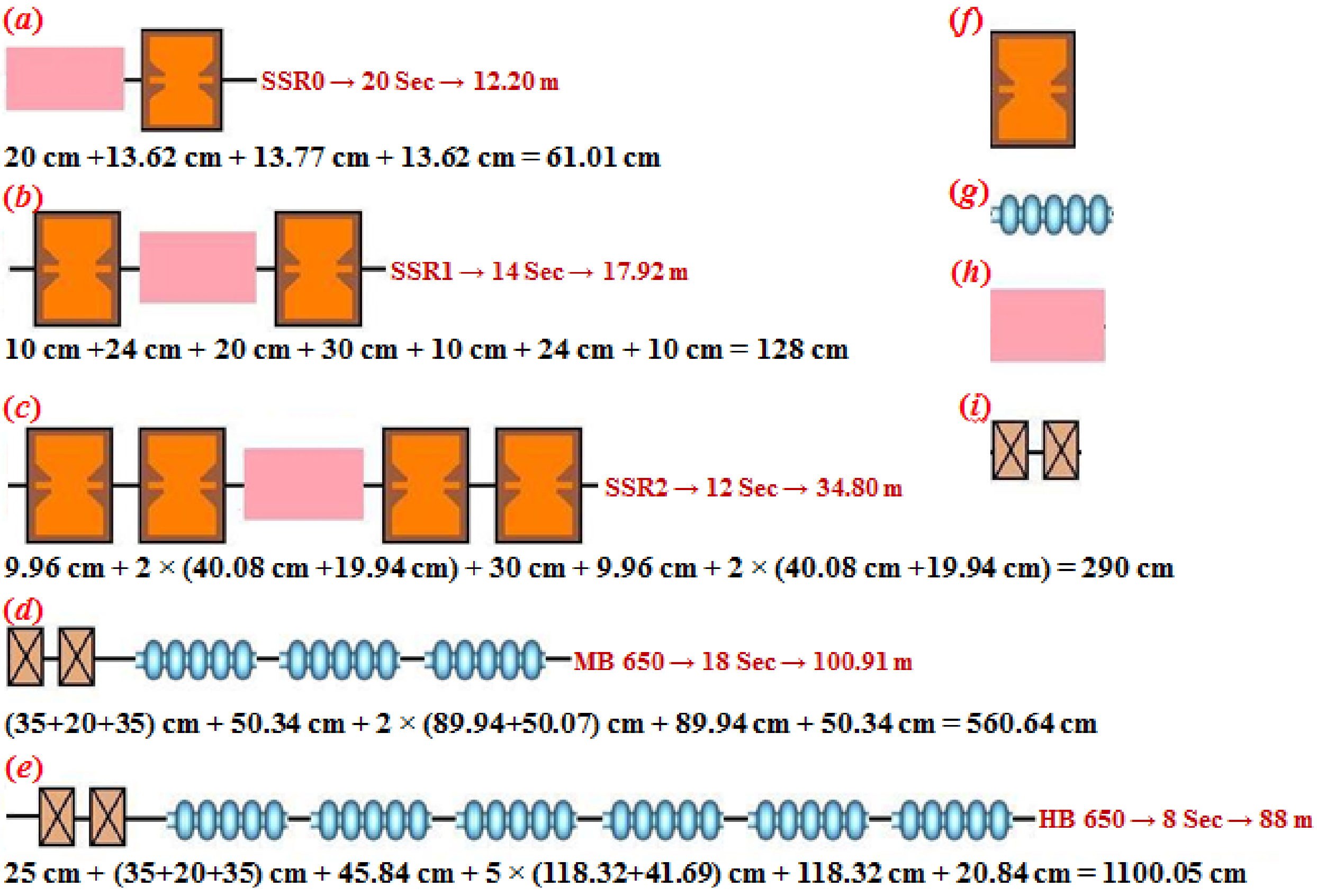

*Fig. 38: Schematic of the lattice structures used in the different sections of the main linac. Here the sections are (a) $\beta_g$ = 0.11 SSR0, (b) $\beta_g$ = 0.22 SSR1, (c) $\beta_g$ = 0.42 SSR2, (d) $\beta_g$ = 0.61 and (e) $\beta_g$ = 0.9 elliptical cavity sections. Here, (f) and (g) are the spoke and elliptical cavity resonators and (h) and (i) are the solenoid and the quadrupole doublet respectively.*

The SSR0 section has twenty periods, which will be housed in a 12.2 m long cryomodule. This will be followed by a matching section, and then the SSR1 section, having fourteen periods, which will be housed in two cryomodules, each having a length of 9.0 m. Lattice of the matching section between SSR0 and SSR1 sections is described in Fig. 39. Note that in the SSR1 section, after the seventh and the fourteenth period, there will be an additional 0.25 m long drift space, in order to accommodate the cryomodules. Next, there will be the SSR2 section, having twelve periods, which will be housed in four cryomodules, each having a length of 9.0 m. Note that there will an additional drift space of 0.25 m, after the third, sixth and ninth period, in order to accommodate the cryomodule. After the SSR2 section, there will be a matching section having one quadrupole and drift spaces as shown in Fig. 40, and then the medium beta section having eighteen periods. In the medium beta section, the quadrupole doublets will be warm, and will therefore not be inside the cryomodule. The cryomodules in this section will be 4.2 m long, and each cryomodule will house three cavities. After the medium beta section, there will be a drift space of 0.4 m, and then the high beta section, having eight periods. In this section also, the quadrupole doublets will be warm, and will be outside the cryomodule. There will be eight cryomodules in this section, each having a length of 9.6 m, and will house six cavities.

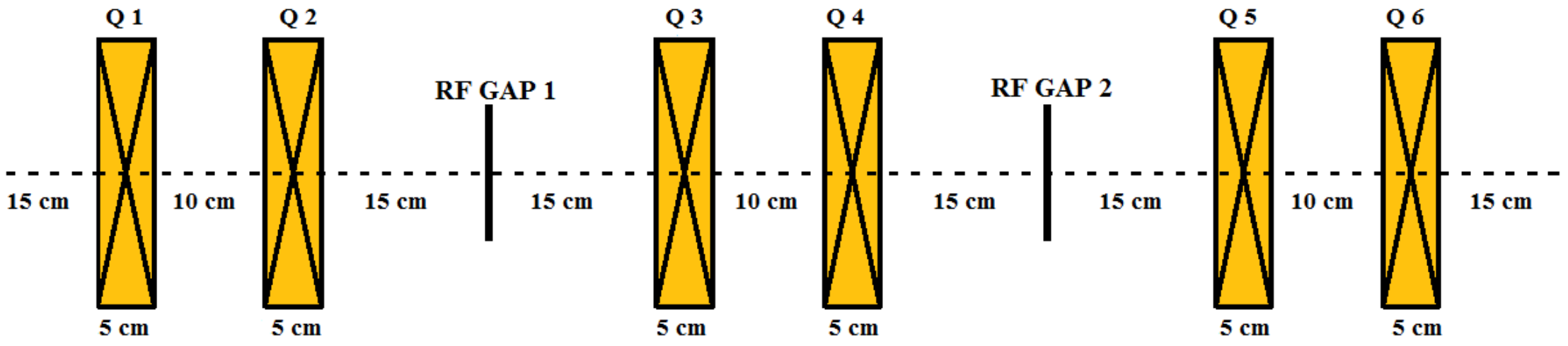

*Fig. 39: Schematic of the matching section between SSR0 and SSR1.*

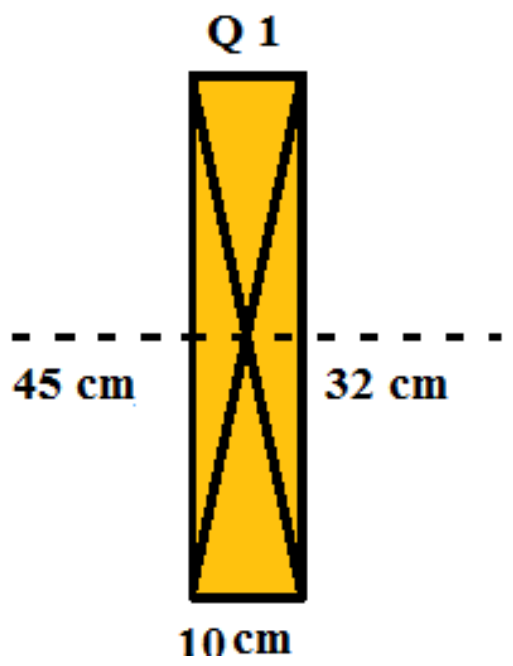

*Fig. 40: Schematic of the matching section between SSR2 and Medium beta section.*

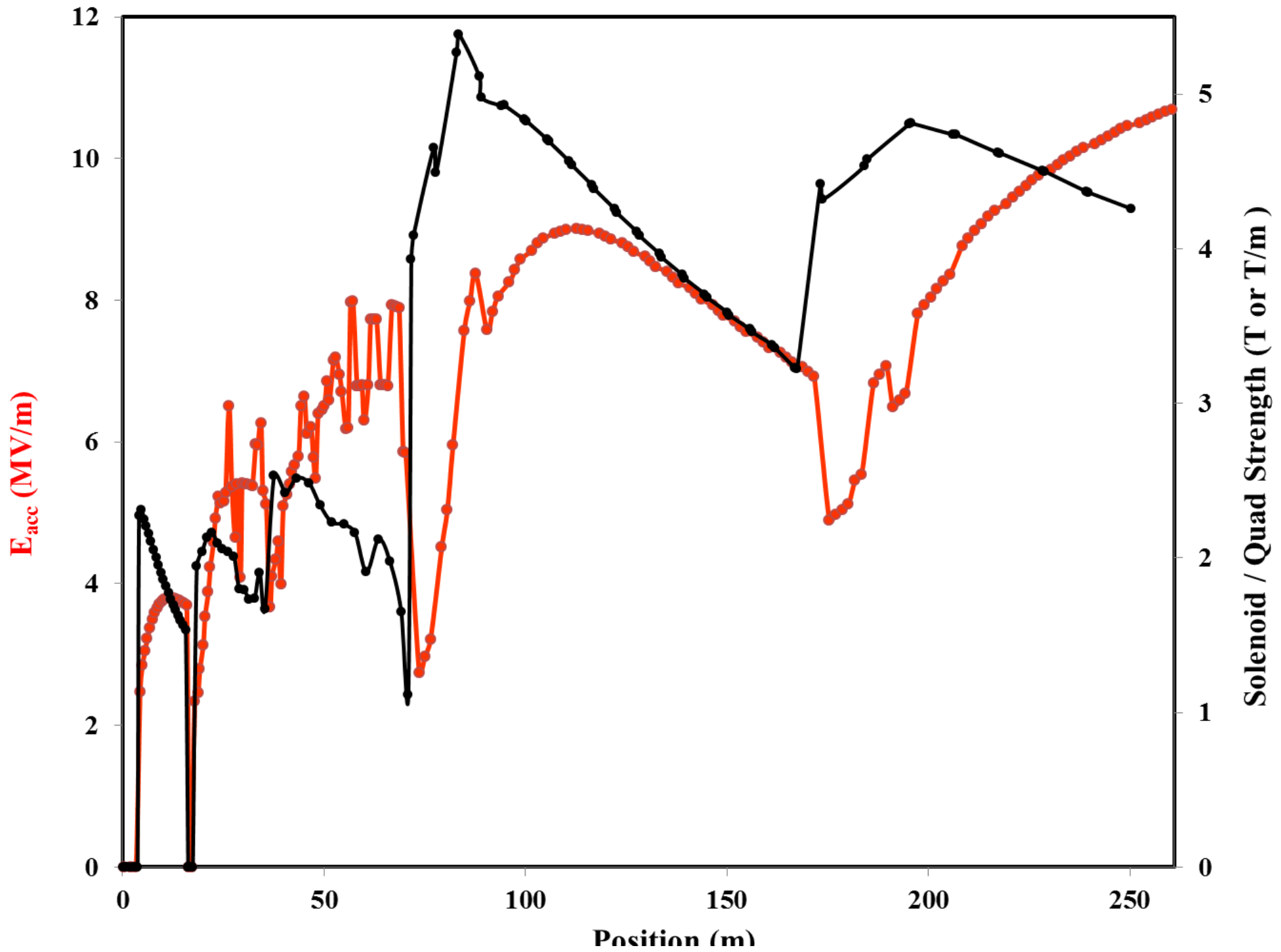

*Fig. 41: Effective accelerating field ($E_0T$), and magnet strength along the linac.*

Beam dynamics simulations have been performed for the optimized lattice, and the focusing strengths of magnets, as well as acceleration gradients and synchronous phases in cavities have been calculated, satisfying all the stringent beam dynamics criteria and constraints. Figure 41 shows the effective acceleration gradient in the cavities, and the magnet strengths along the length of the linac. Energy gain per cavity along the length of the linac is shown in Fig. 42.

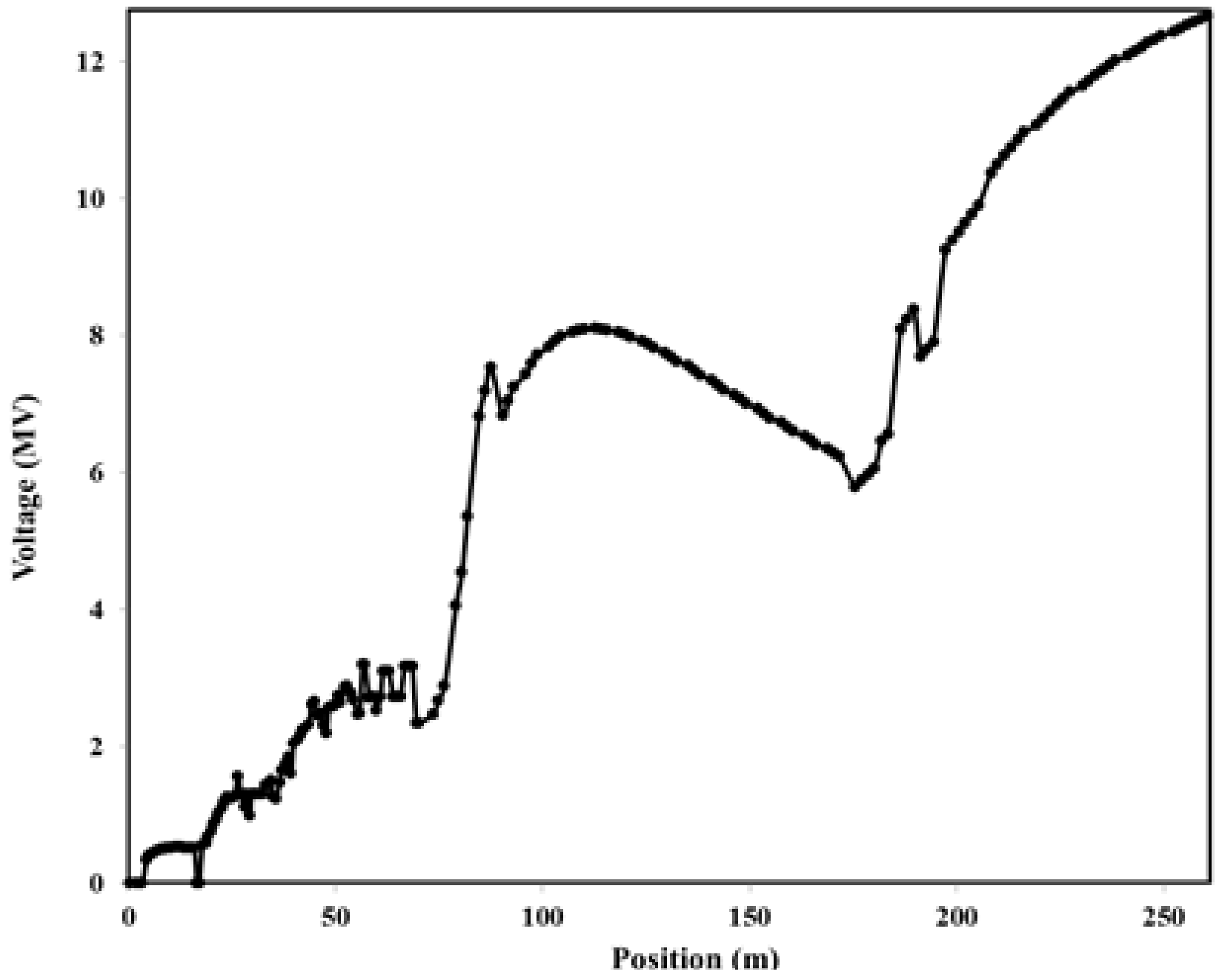

*Fig. 42: Energy gain per cavity along the linac.*

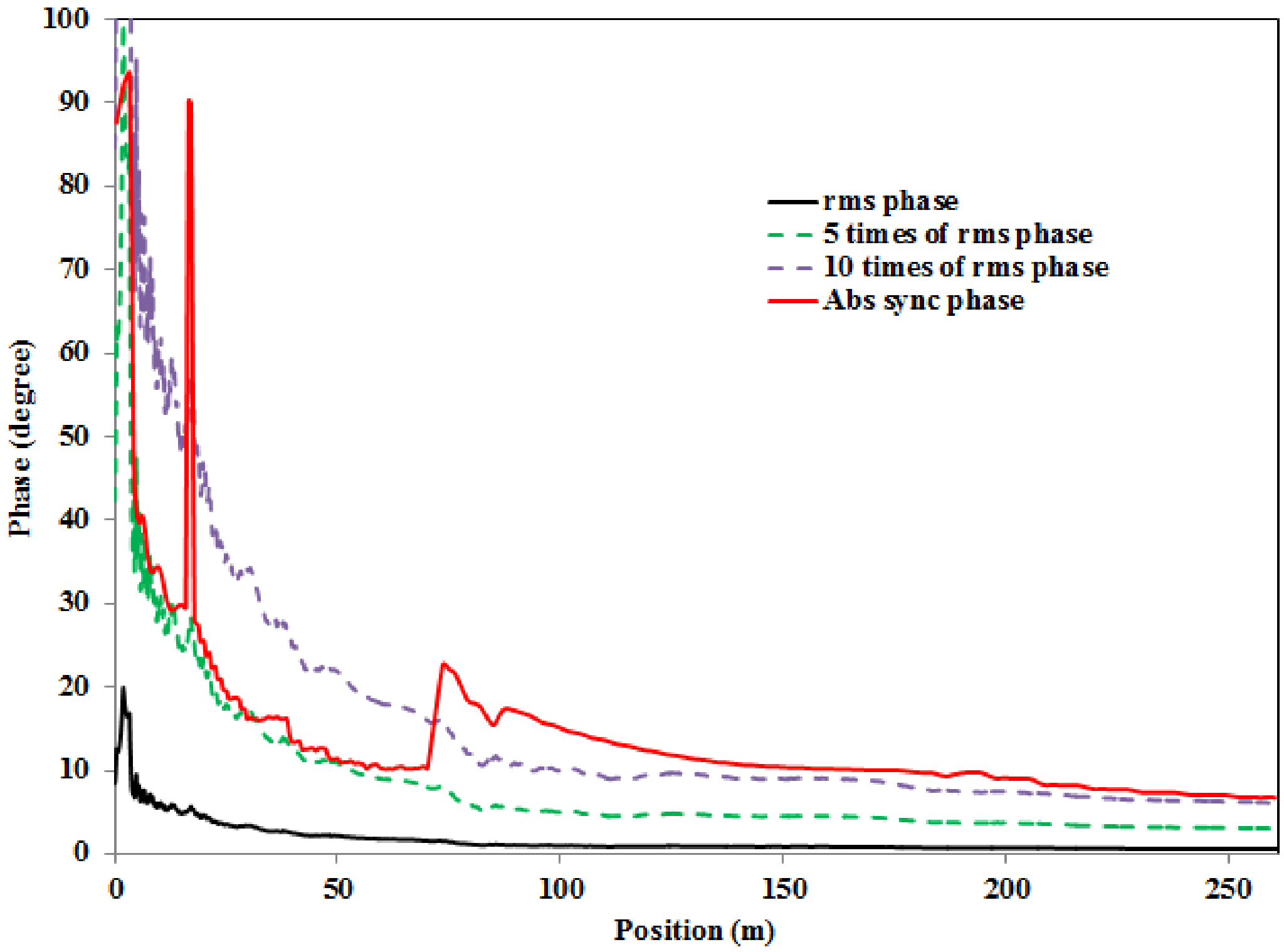

*Fig. 43: Absolute synchronous phase and rms phase width along the linac.*

.

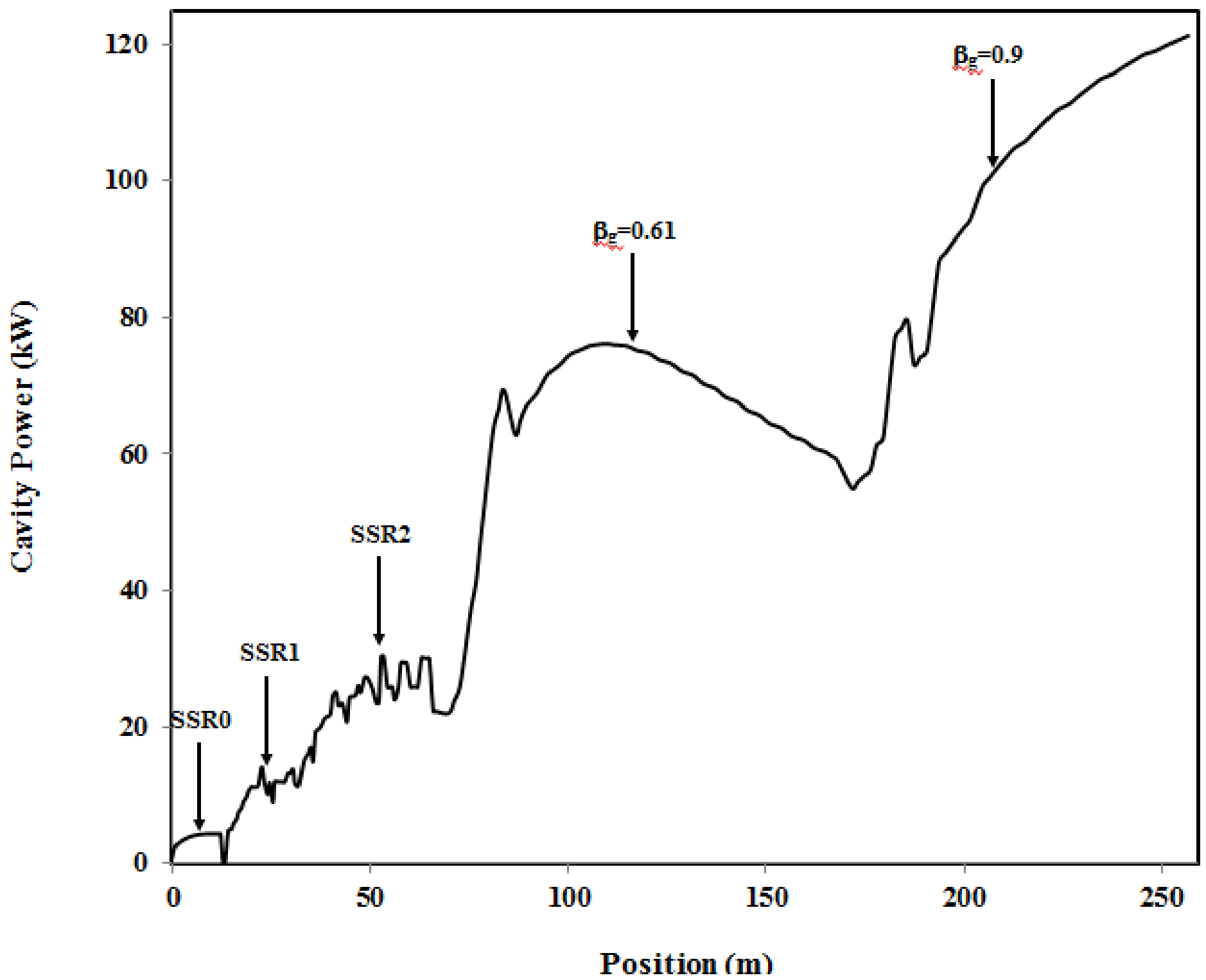

*Fig. 44: Required RF power per cavity along the linac.*

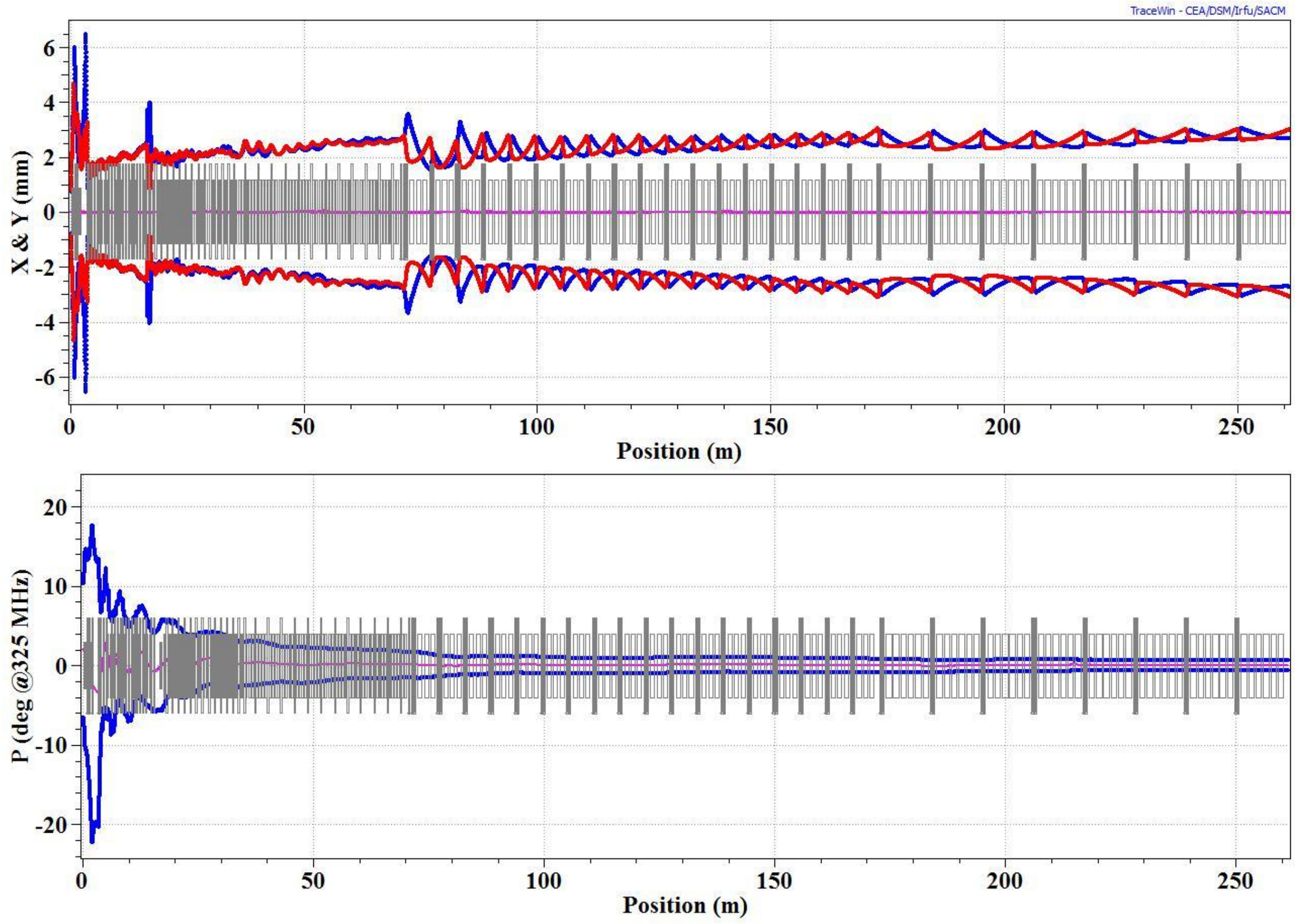

*Fig. 45: Evolution of vertical (top, red), horizontal (top,blue) and longitudinal (bottom) rms beam size.*

Variation of synchronous phase along the length of the linac is shown in Fig. 43. It can be seen that we are able to keep the rms phase width less than one sixth of the absolute synchronous phase along the linac. This ensures that all the particles are in the stable region of the longitudinal phase space. The required RF power per cavity along the linac is shown in Fig. 44. Note that the calculation of cavity power has been done assuming average macro-pulse current of 10 mA, whereas beam dynamics calculations are done assuming a midi-pulse current of 15 mA. Contribution due to microphonics has not been included here. Evolution of beam envelope in transverse as well as longitudinal directions is shown in Fig. 45.

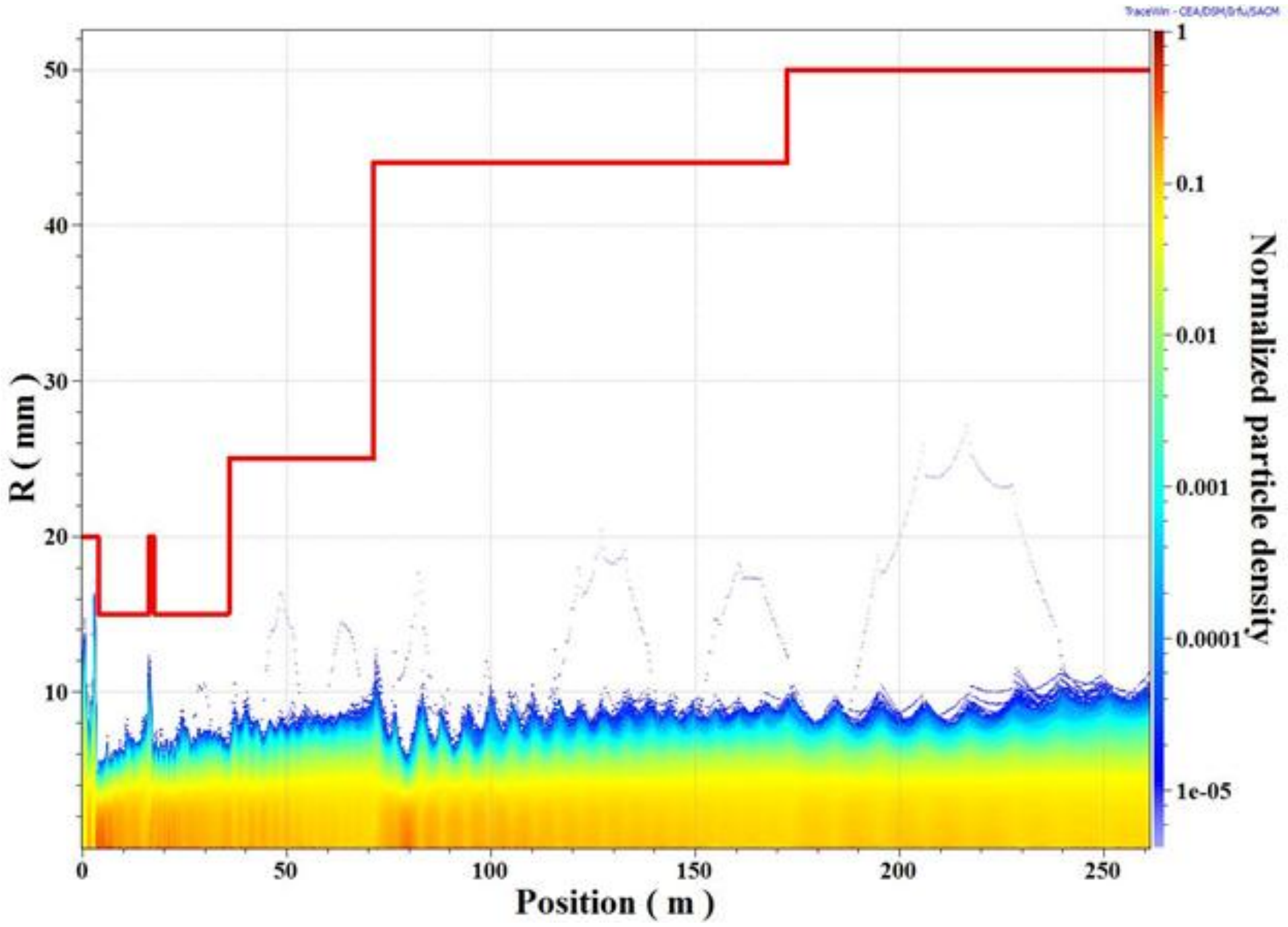

*Fig. 46: Variation of radial density of particles along the length of the linac.*

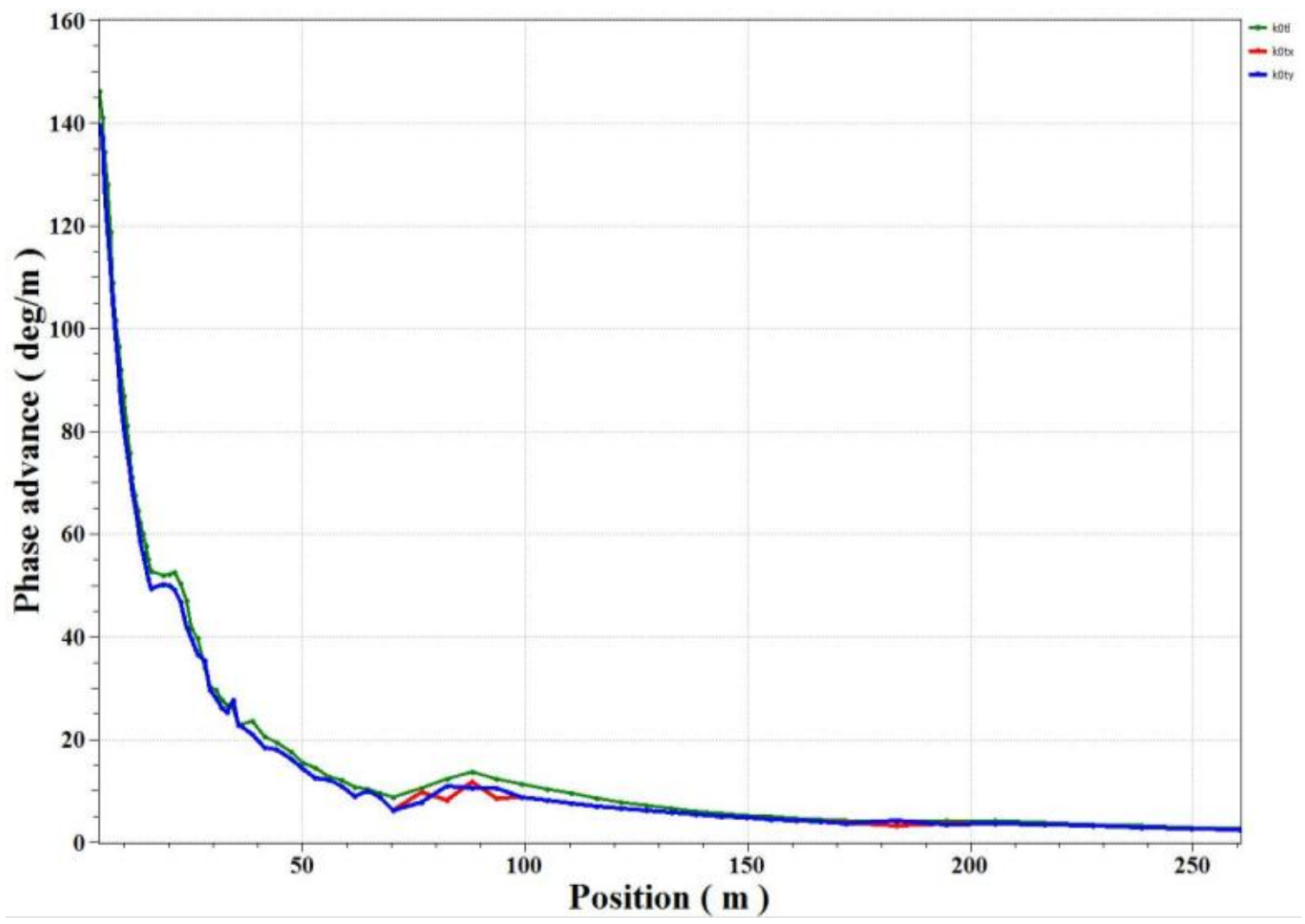

*Fig. 47: Phase advance per unit length along the length of the linac.*

Transverse particle distribution is shown in Fig. 46. Here the dark line represents the apertures at respective linac sections. It is clearly seen in the figure that except MEBT and the matching section used in between SSR0 and SSR1, beam radius remains several times less than that of the available aperture. Phase advance per unit length along the linac has a smooth variation along the length of the linac, as seen in Fig. 47, which is an important beam dynamics criterion. In addition, the beam dynamics criteria require that the phase advance per period should be between $20^0$ and $90^0$, which is also satisfied. The growth of emittance is plotted in Fig. 48.

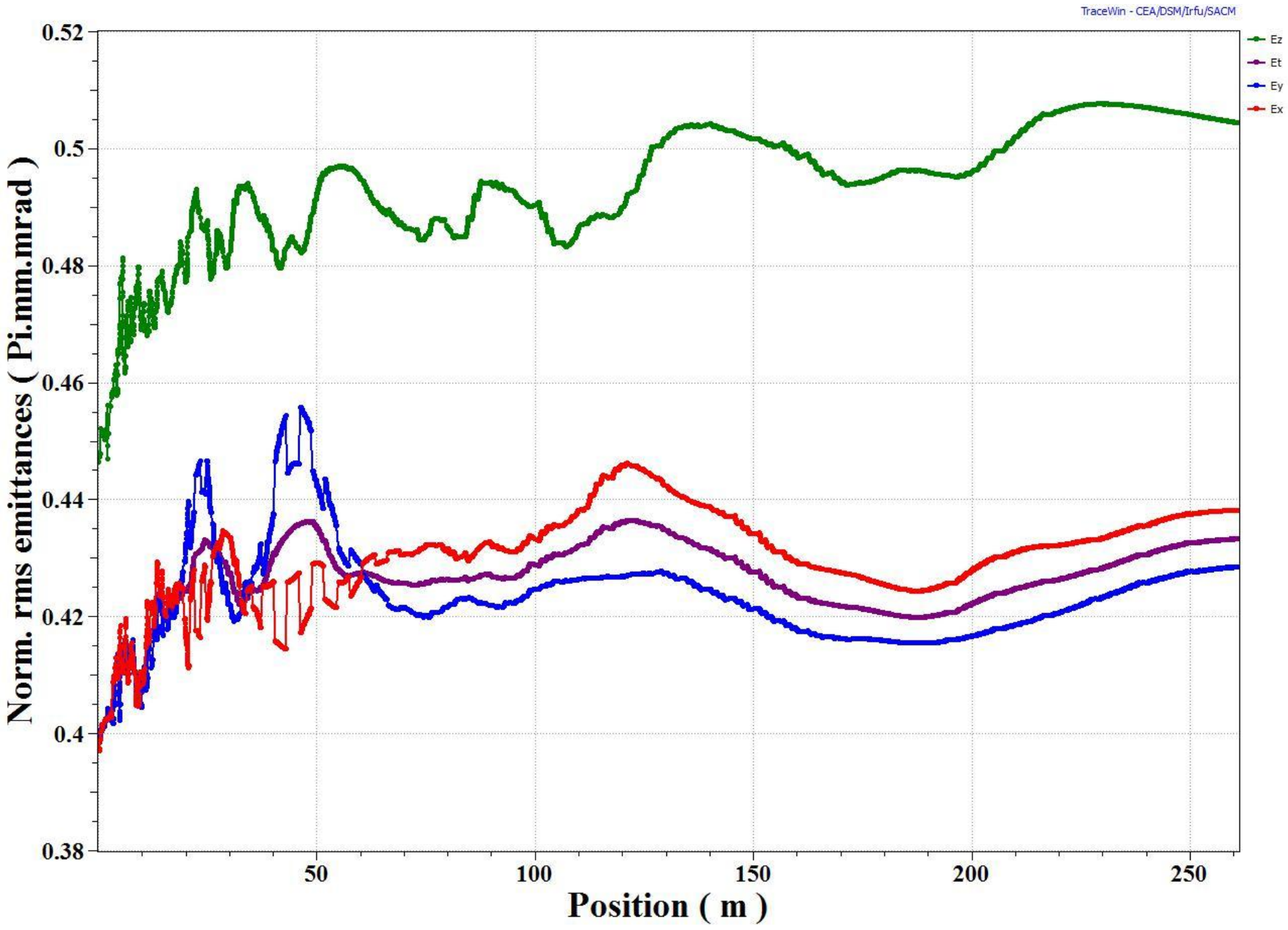

*Fig. 48: Growth of beam emittance in longitudinal (green), vertical (blue) and horizontal (red) directions along the linac.*

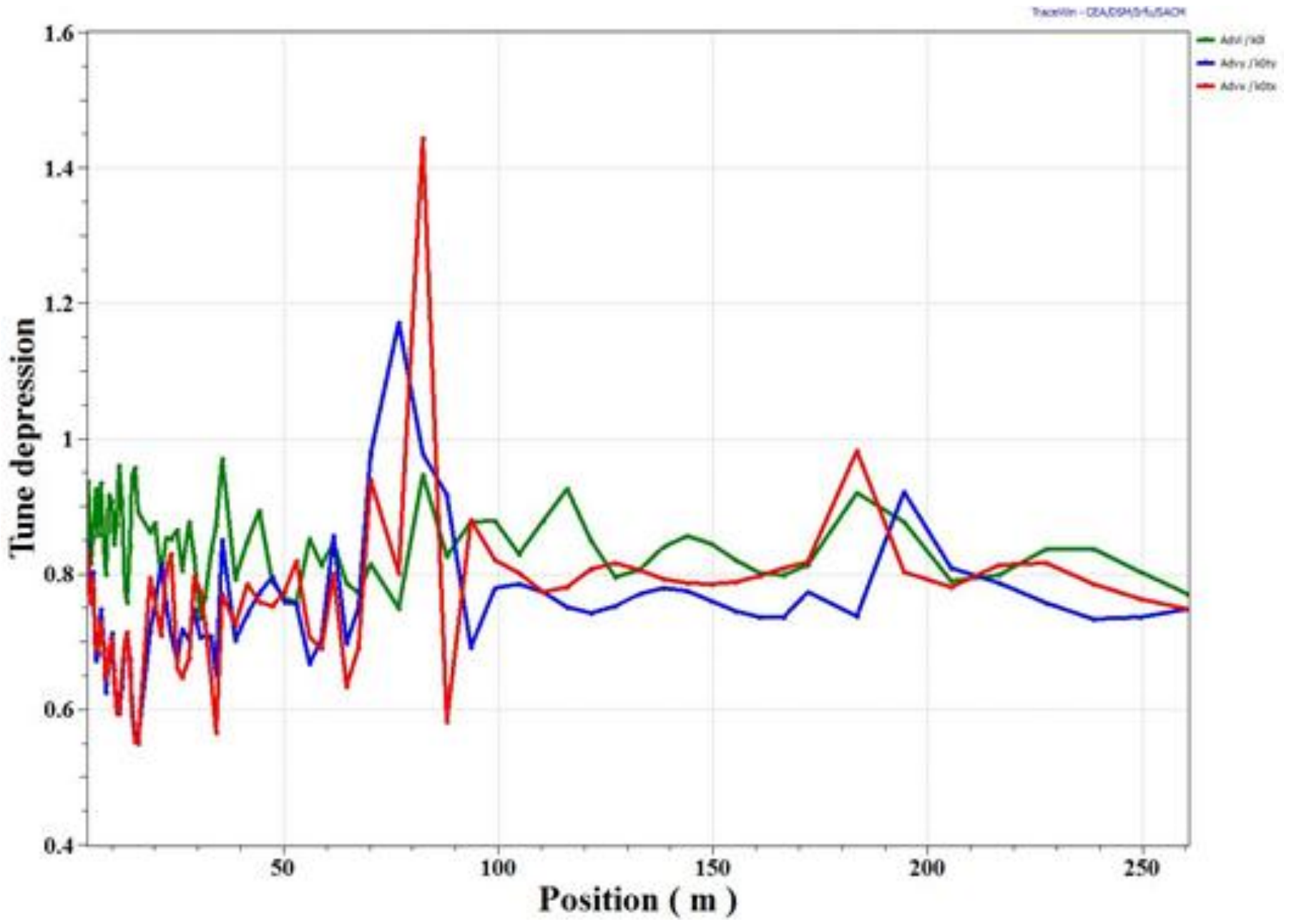

*Fig. 49: Tune depression in longitudinal (green), vertical (blue) and horizontal (red) directions, along the linac.*

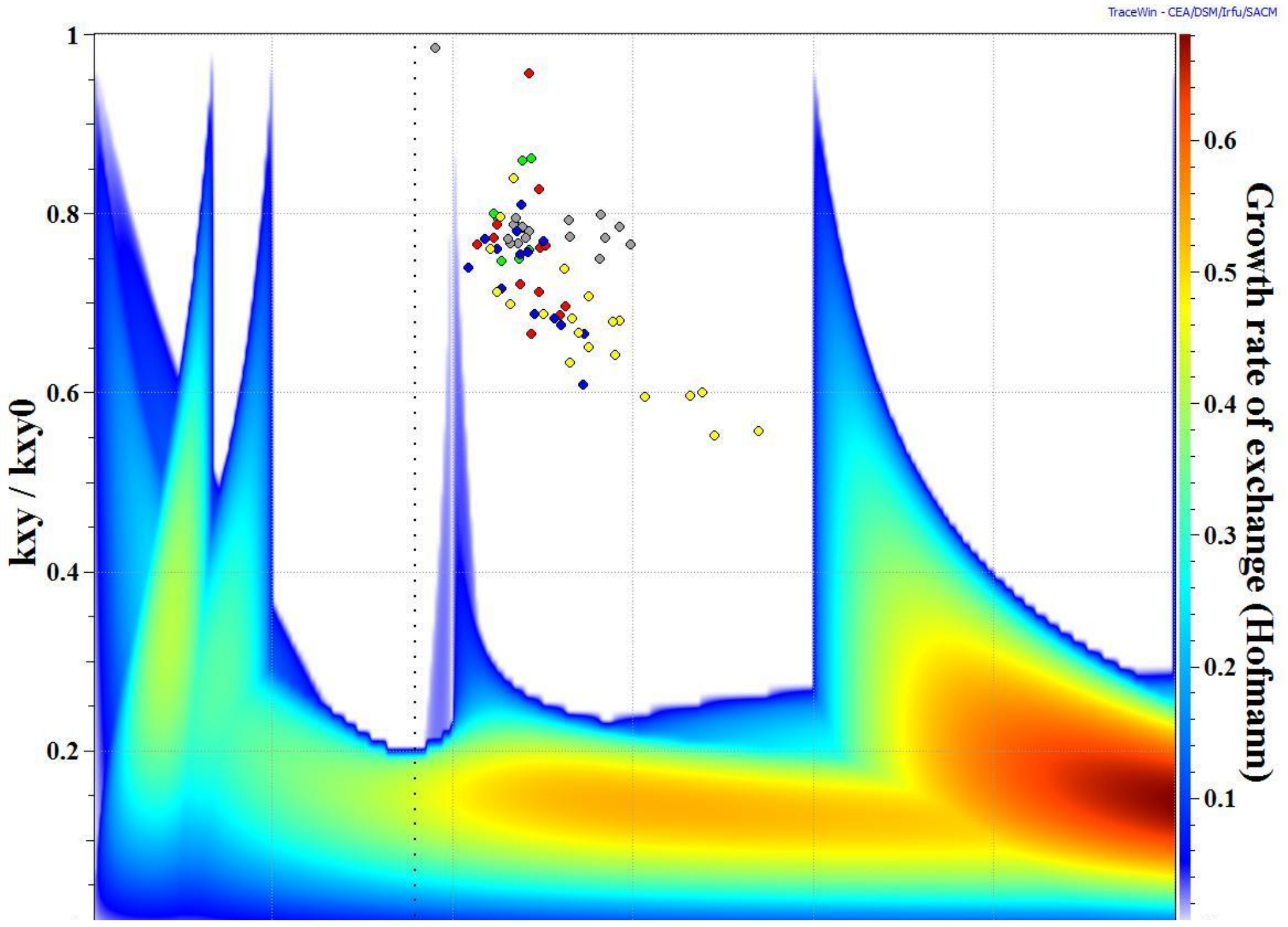

*Fig. 50: Tune footprint of the injector linac.*

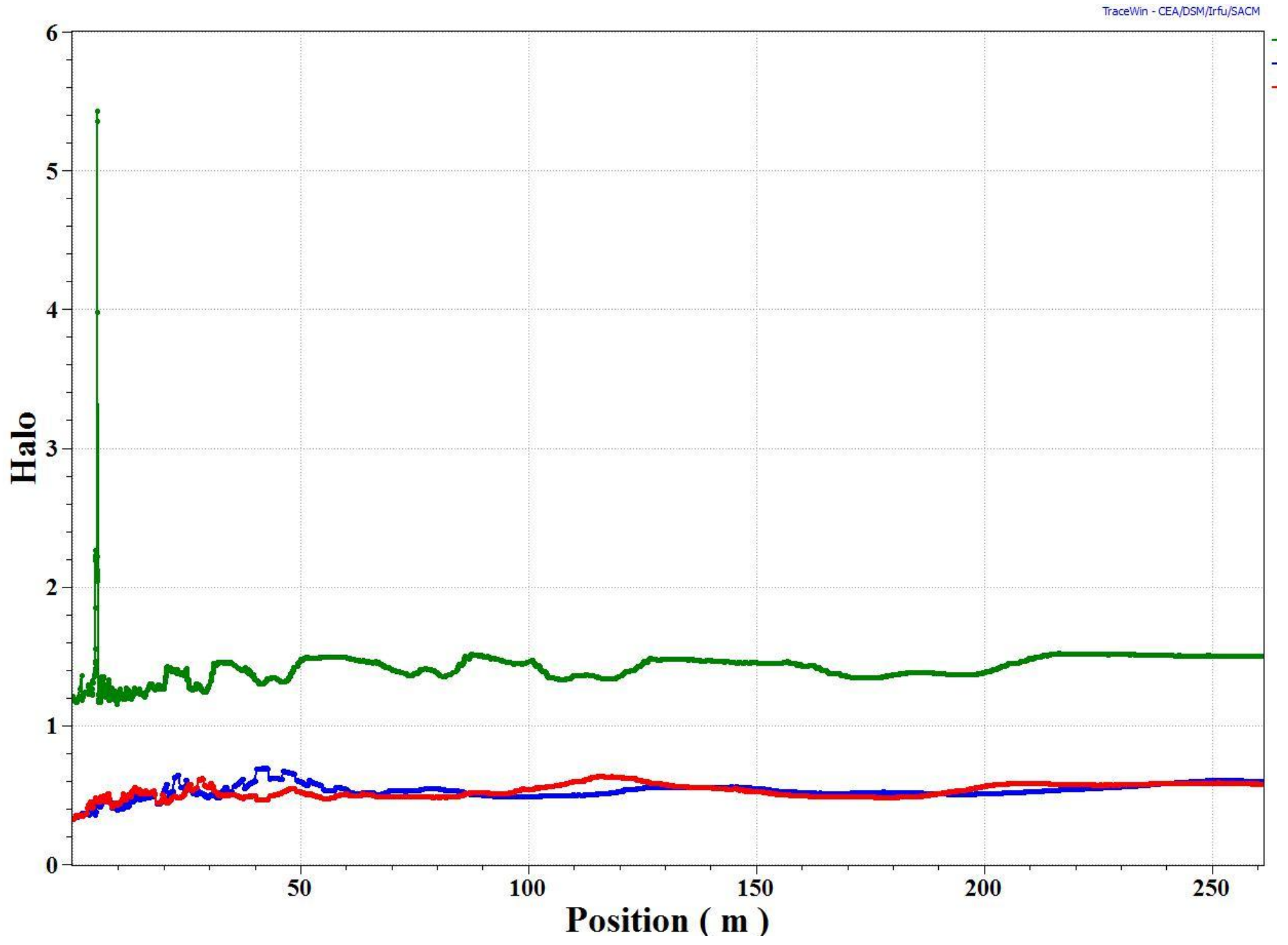

*Fig. 51: Beam halo in longitudinal (green), vertical (blue) and horizontal (red) directions, along the injector linac.*

Tune depression along the length of the linac is shown in Fig. 49. Except at the transition from the SSR2 to the medium beta section, the tune depression is maintained between 0.5 and 1. Operating tune points are in the stable zone of the Hoffman diagram, as seen in the tune footprint of the injector linac in Fig. 50. Beam halo studies have also been performed, and a plot showing the evolution of halo parameter is shown in Fig. 51. Fig. 52 and 53 shows the beam distribution at the entrance of the MEBT and at the linac exit respectively.

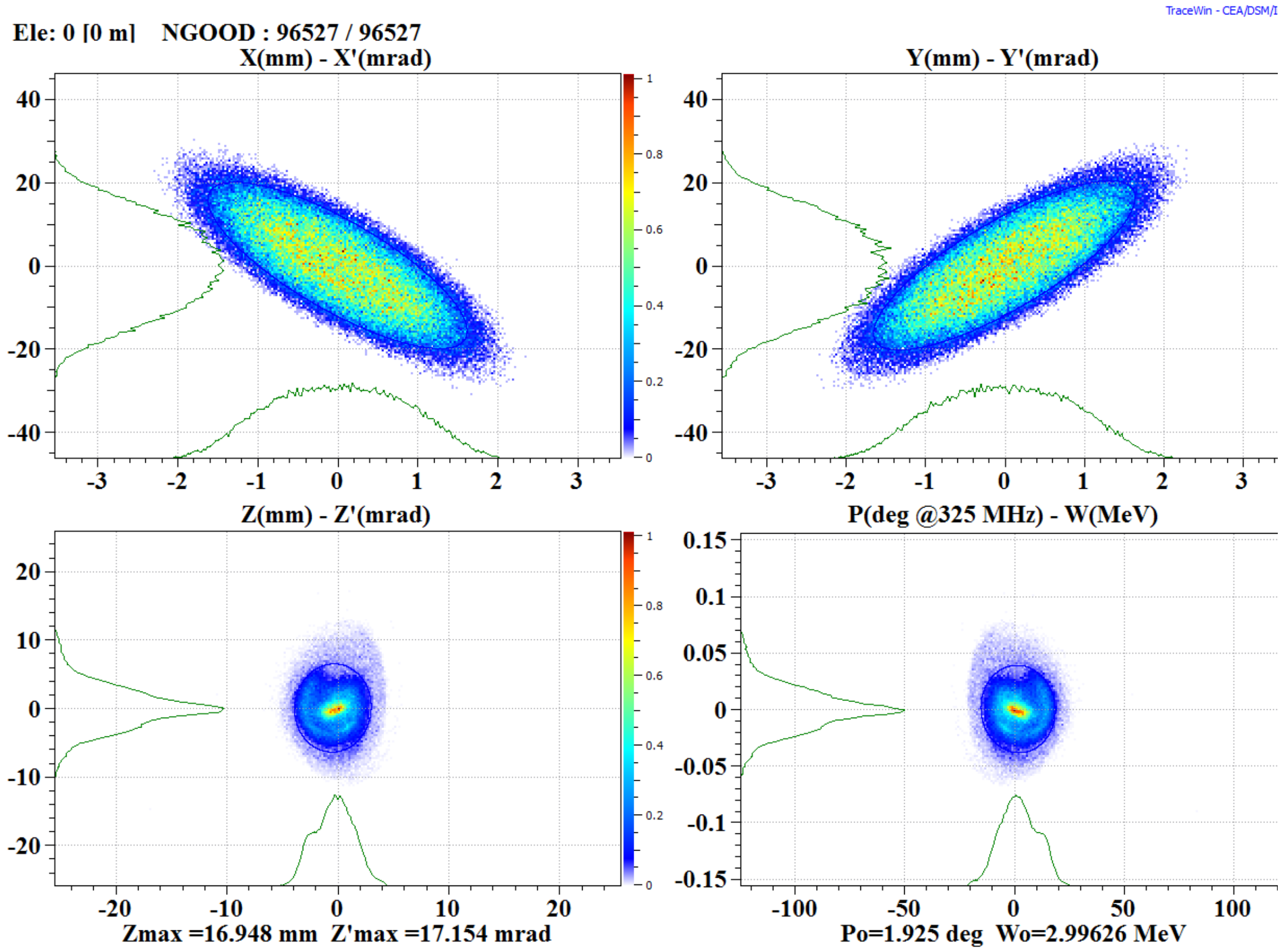

*Fig. 52: Beam distribution at the MEBT entrance.*

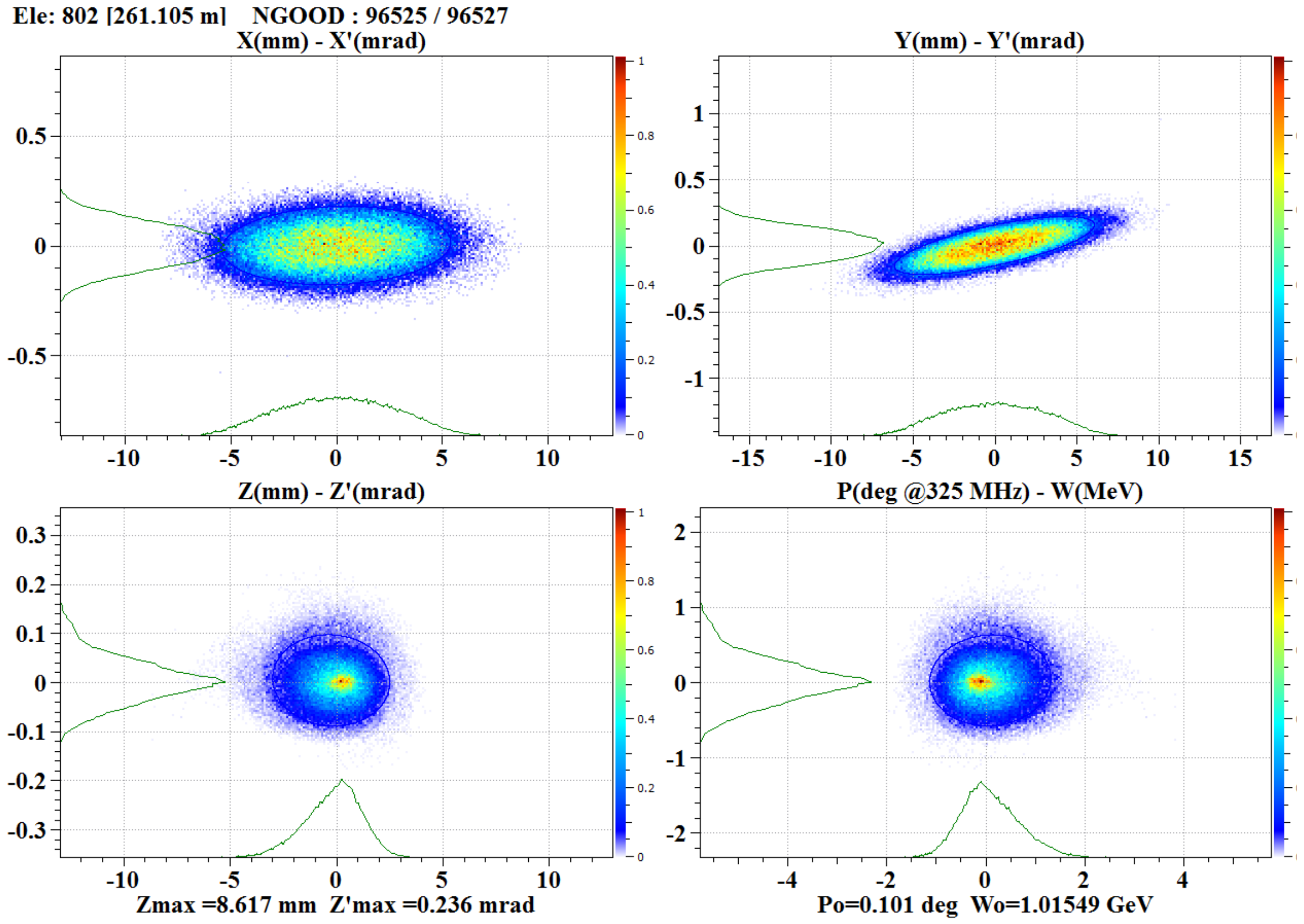

*Fig. 53: Beam distribution at the linac exit.*

Finally, matched beam parameters at the entrance of MEBT section, and the linac exit are shown in Table 20. The phase advances per period in the transverse and longitudinal directions are 26.3° and 29° respectively. It is also important to mention that the phase advances per meter in the transverse and longitudinal directions are 2.4° and 2.6°, respectively.

*Table 20: Matched beam parameters at the entrance and exit*

| Parameter | MEBT entrance | LINAC exit |
|---|---|---|
| Pulse beam current [mA] | 15 | 15 |
| $\alpha_x$ | 1.2887 | -0.0945 |
| $\beta_x$ [m/rad] | 0.1321 | 30.4471 |
| $\varepsilon_{x, n, rms}$ [mm-mrad] | 0.3973 | 0.4382 |
| $\alpha_y$ | -1.3502 | -1.0076 |
| $\beta_y$ [m/rad] | 0.1360 | 39.6226 |
| $\varepsilon_{y, n, rms}$ [mm-mrad] | 0.3996 | 0.4285 |
| $\alpha_z$ | -0.0266 | 0.0432 |
| $\beta_z$ [deg/MeV] | 437.5215 | 2.0234 |
| $\beta_z$ [m/rad] | 0.5379 | 29.6082 |
| $\varepsilon_{z, rms}$ [deg-MeV] | 0.1637 | 0.1849 |
| $\varepsilon_{z, n, rms}$ [mm-mrad] | 0.4467 | 0.5045 |

## 8. High Energy Beam Transport (HEBT) line

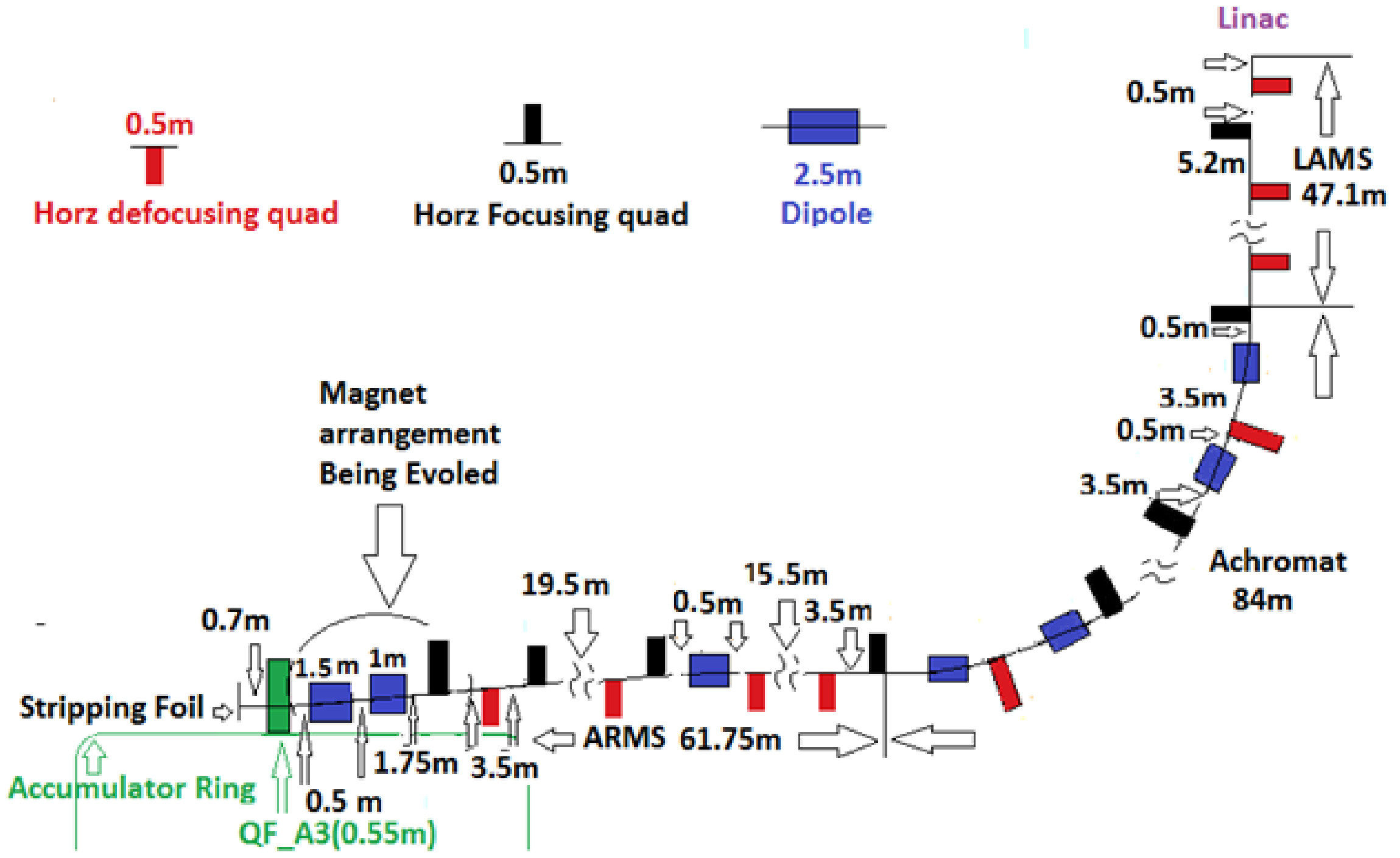

Fig. 54: Schematic layout of HEBT, based on elementary considerations.

A beam transport line from Linac to Ring, which will transfer the beam at high energy, is known as High Energy Beam Transfer Line (HEBT). The major task of this transfer line is to send the beam with minimum losses to the ring with matching. Besides matching, the other requirement of HEBT is to accommodate betatron and momentum collimators, energy jitter correction cavity, energy spreader cavity, beam characterization system for incoming Linac beam, and to protect the AR in case of faulty condition. Optics of the line should incorporate magnets with the magnetic field below the threshold value to avoid the magnetic stripping of $H^-$ ion. Due to these requirements, the HEBT may typically be very long, as in the case of SNS at Oak Ridge, where its length is ~180 m [18]. Design studies for the HEBT for the ISNS project are currently ongoing. Figure 54 shows the schematic of HEBT, based on the elementary considerations. The HEBT consists of three sections – (i) Linac to Achromat Matching Section (LAMS), (ii) Achromat Section, and (iii) Achromat to Ring Matching Section (ARMS). In the current scheme shown in Fig. 54, length of the LAMS, Achromat and ARMS is ~ 47 m, ~ 84 m and ~ 61 m respectively.

Initial beam optics optimization studies have been carried out to keep the beta function below 40 m, as seen in Fig. 55, and the peak dispersion up to 4 m to make an efficient momentum collimation. Currently, at the end of the line, i.e. near the injection point of AR, beta functions are higher in order to match the optical parameters of the ring. In future, near the injection point of AR, the geometrical layout, as well as the beam optics, has to be optimized further.

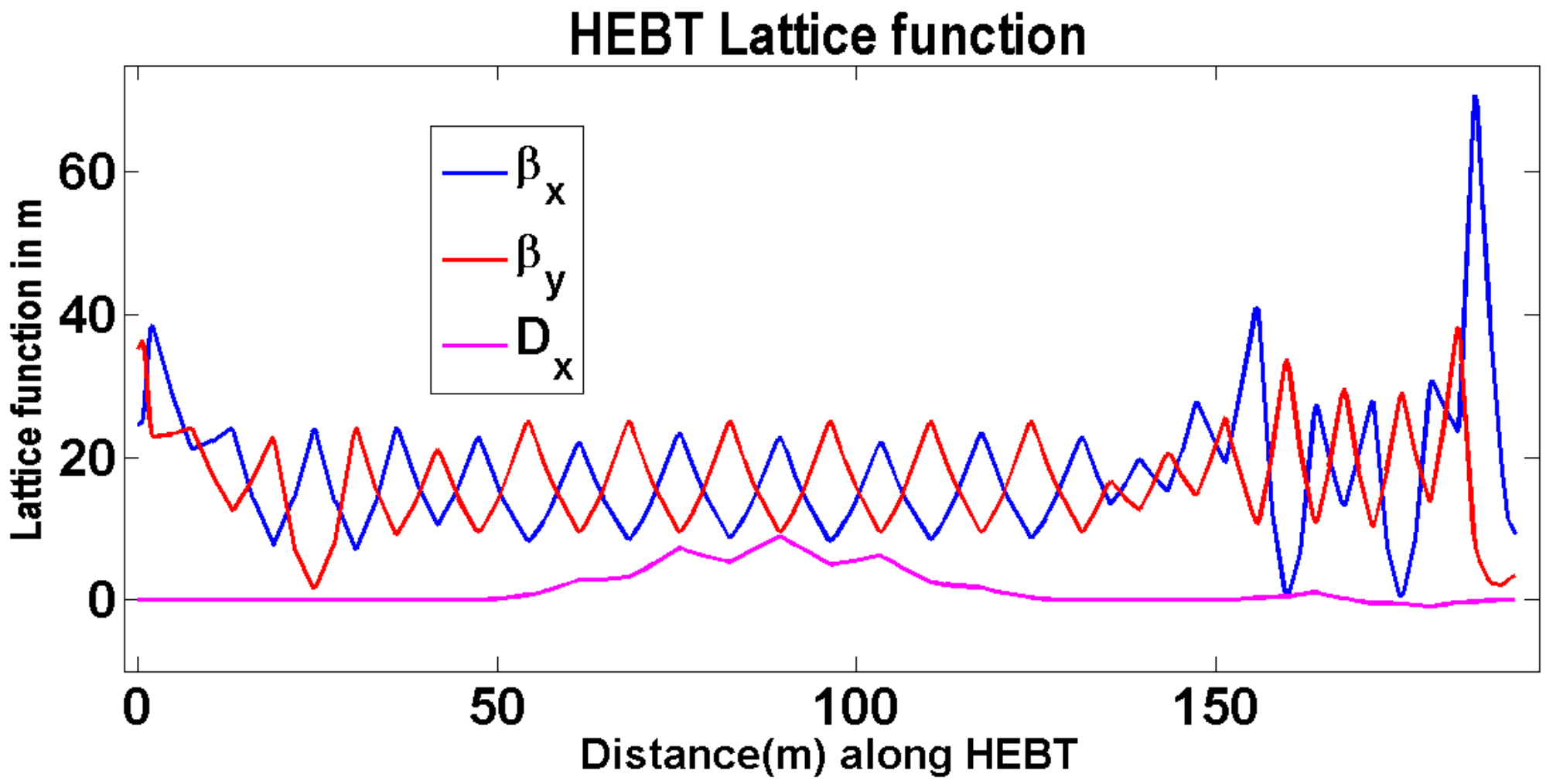

*Fig. 55: Beta functions and dispersion in HEBT.*

## 9. Lattice Design of Accumulator Ring

### 9.1 Transverse Beam Optics

The 2 ms long, $H^-$ beam pulses with a pulse structure described earlier will be injected into an accumulator ring, using multi-turn, charge exchange injection scheme, and will be compressed to 1 μs. In order to keep the beam focused in the accumulator ring in the transverse direction, a suitable magnetic lattice is designed. Focusing in the longitudinal direction is achieved with the help of RF cavity. Two types of lattices are currently being considered – FODO lattice and hybrid lattice. In the case of FODO, the arcs, as well as dispersion-free straight sections, use FODO cells. In the case of hybrid, the arc uses FODO cells, whereas the dispersion-free straight section uses quadrupole doublets/triplets. In both the schemes, dispersion is matched using missing dipole scheme.

A schematic of the layout of magnetic lattice of the accumulator ring for the FODO case is shown in Fig. 56, and the basic parameters of the accumulator ring are specified in Table 21 [19]. There are four dispersion free sections to accommodate four major subsystems - (1) beam injection system, (2) RF system (3) beam collimators, and (4) beam extraction system. Circumference of the accumulator ring is chosen as 262.1 m, such that the revolution frequency in the ring is an integral sub-multiple of linac frequency. In each superperiod, the arc length is chosen as 40 m, and the length of the straight section is chosen as 25.6 m. The space should be sufficient to accommodate all the required sub-systems. As seen in Fig. 56, the straight sections S1, S2, S3 and S4 will be used for RF cavities, beam extraction system, collimators and scrapers, and injection system respectively.

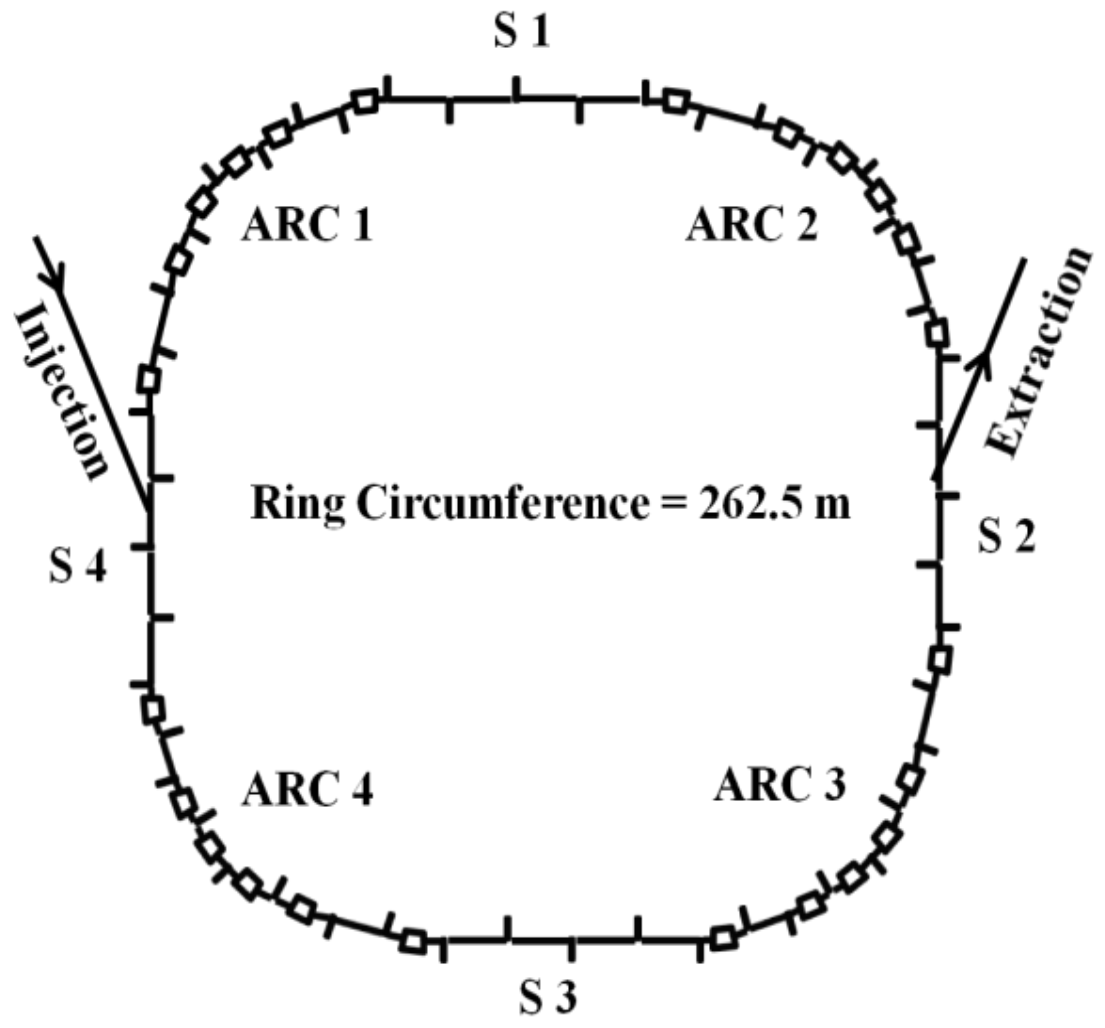

*Fig. 56: Layout of the FODO lattice of accumulator ring. Rectangular boxes represent bending magnets, and outward and inward bars represent focusing and defocusing quadrupoles, respectively.*

*Table 21: Basic parameters of the accumulator ring*

| **Parameter** | **FODO lattice** | **Triplet (hybrid) lattice** |
|---|---|---|
| Circumference | 262.1 m | |
| Periodicity | 4 | |
| Dipole magnet length | 2.25 m (sector type) | |
| Bending angle | 15° | |
| No. of dipole magnets | 24 | |
| No. of quadrupole magnets | 48 | 60 |
| No. of quadrupole families | 5 | 7 |
| Revolution time | 1 µs | |
| RF frequency (h = 1) | 1 MHz | |
| RF frequency (h = 2) | 2 MHz | |

The schematic of one superperiod for the FODO lattice is shown in Fig. 57. Table 22 summarizes the main parameters of FODO lattice at different working points. Twiss parameters of FODO lattice at one tune point is shown in Fig. 58.

The advantage of hybrid lattice scheme is that it has long uninterrupted straight sections with zero dispersion. This makes the incorporation of injection chicane relatively easier in hybrid lattice, as compared to FODO lattice. However, matching of optics and obtaining large tuning range will be difficult for the hybrid lattice. A better matching at different tune points with lowering the maximum beta functions is carried out and a revised lattice for hybrid configuration is obtained. The schematic of one superperiod of the revised Hybrid lattice is shown in Fig. 59. Table 23 shows the major parameters of hybrid lattice at different working points. Figure 60 shows lattice parameters of Hybrid lattice at one representative tune point.

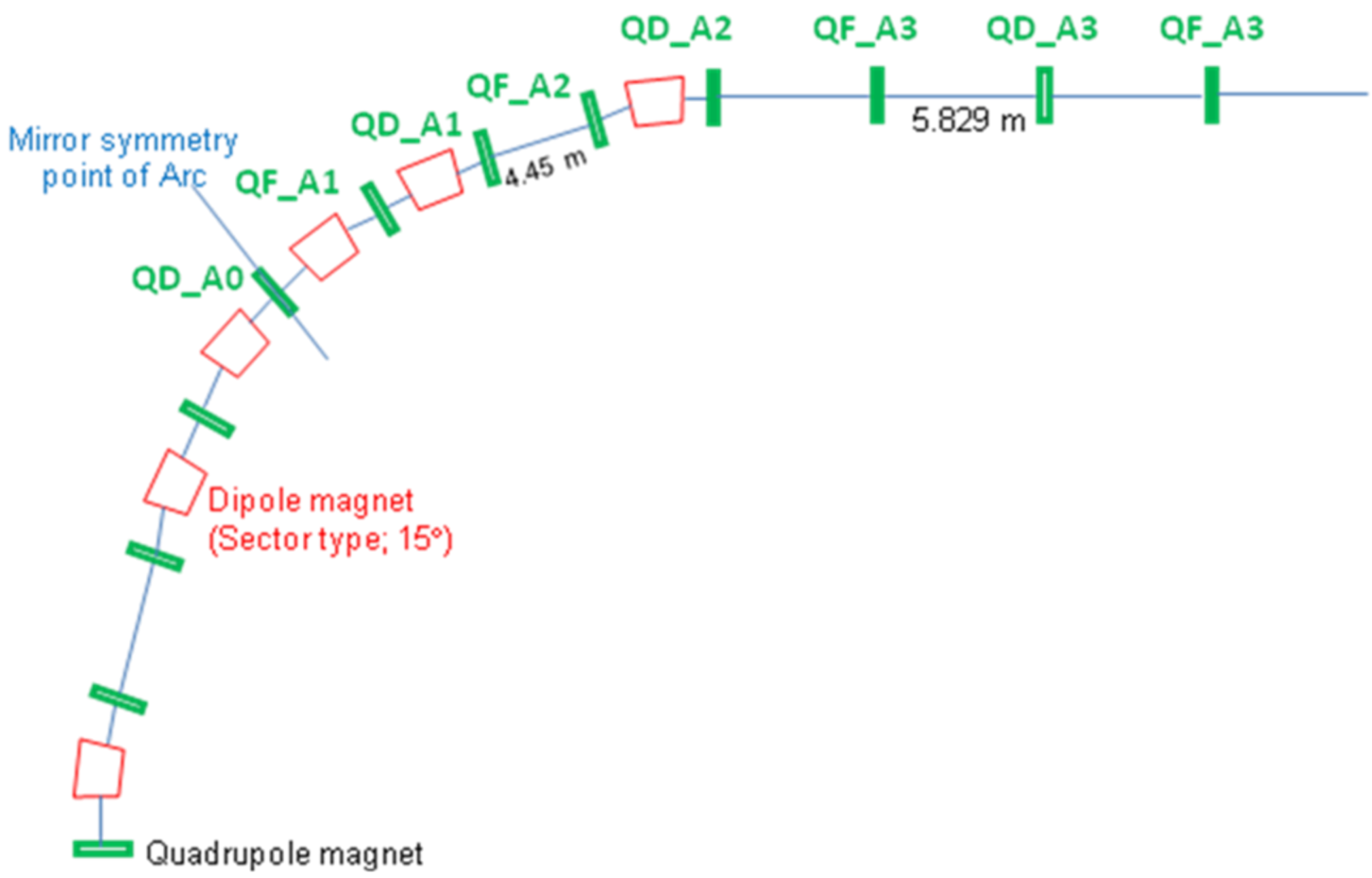

*Fig. 57: Schematic optics configuration of a superperiod of FODO Lattice. Red and green blocks show dipole magnets and quadrupole magnets, respectively.*

*Table 22: Quadrupole magnet strengths and lattice parameters for FODO lattice. Normalized strength (g/B$\rho$ i.e. in $m^{-2}$) of the quadrupole is used here.*

| **Parameter** | **Tune-1** | **Tune-2** | **Tune-3** | **Tune-4** | **Tune-5** |
|---|---|---|---|---|---|
| QD_A0 | -0.6642 | -0.6459 | -0.6009 | -0.5730 | -0.5886 |
| QF_A1 | 0.5903 | 0.5837 | 0.5749 | 0.5855 | 0.5551 |
| QD_A1 | -0.6781 | -0.6820 | -0.6172 | -0.6051 | -0.6432 |
| QF_A2 | 0.5966 | 0.5857 | 0.5670 | 0.6053 | 0.5136 |
| QD_A2 | -0.5153 | -0.4818 | -0.4566 | -0.4808 | -0.4530 |
| QF_A3 | 0.4382 | 0.3958 | 0.4112 | 0.4216 | 0.3824 |
| QD_A3 | -0.3560 | -0.3197 | -0.3509 | -0.3603 | -0.3128 |
| $\beta_{x,max}$ (m) | 21.95 | 24.96 | 22.25 | 24.81 | 24.11 |
| $\beta_{y,max}$ (m) | 24.95 | 24.97 | 24.84 | 23.21 | 24.97 |
| $D_{x,max}$ (m) | 3.36 | 3.43 | 3.38 | 3.12 | 3.77 |
| Tune X | 7.20 | 6.82 | 6.82 | 7.20 | 6.20 |
| Tune Y | 7.20 | 6.82 | 6.20 | 6.20 | 6.20 |
| $\gamma_t$ (Gamma Transition) | 5.33 | 5.26 | 5.28 | 5.50 | 4.96 |

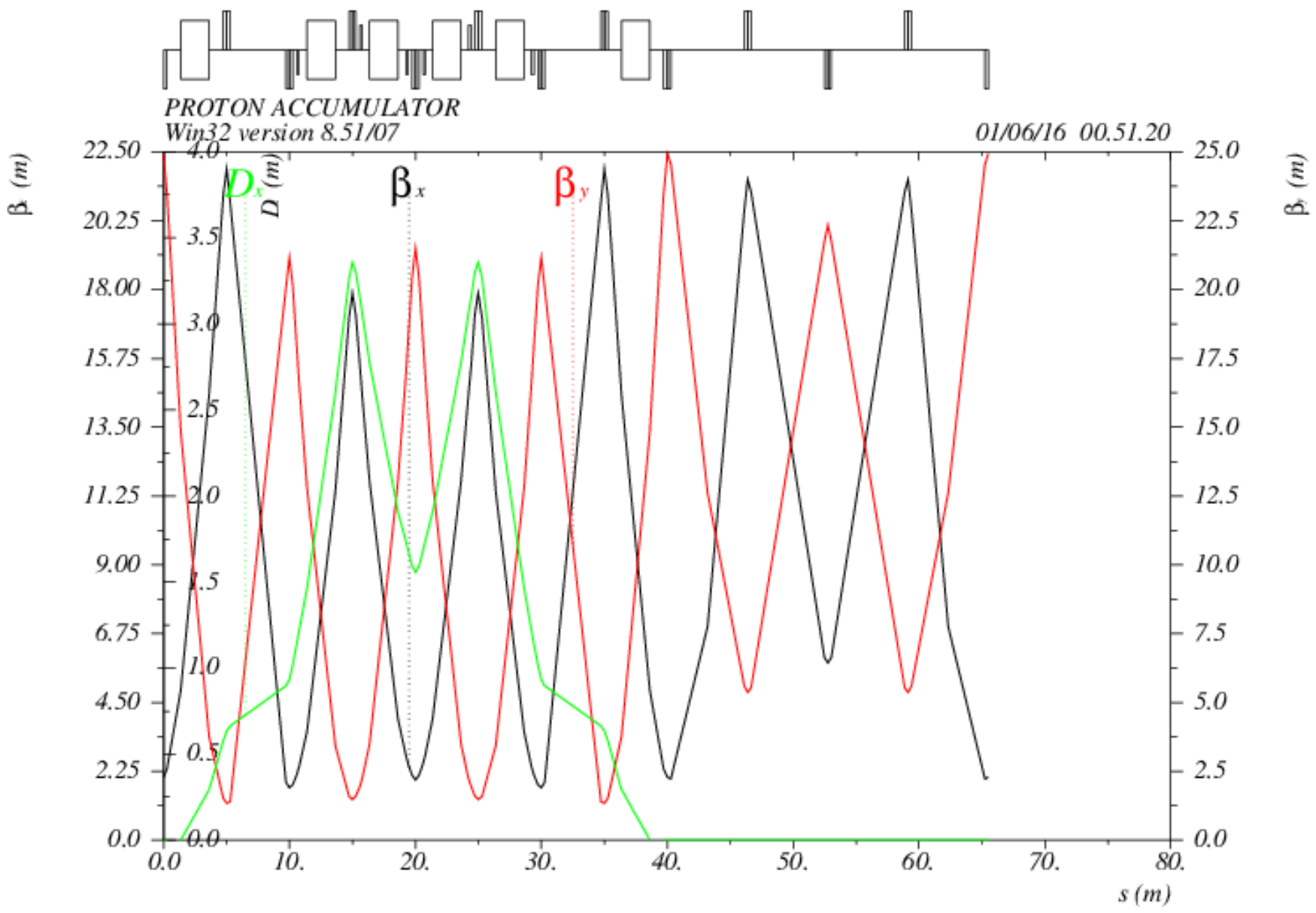

*Fig. 58: Lattice parameters of FODO lattice for $\nu_x$ = 7.20, $\nu_y$ = 7.20.*

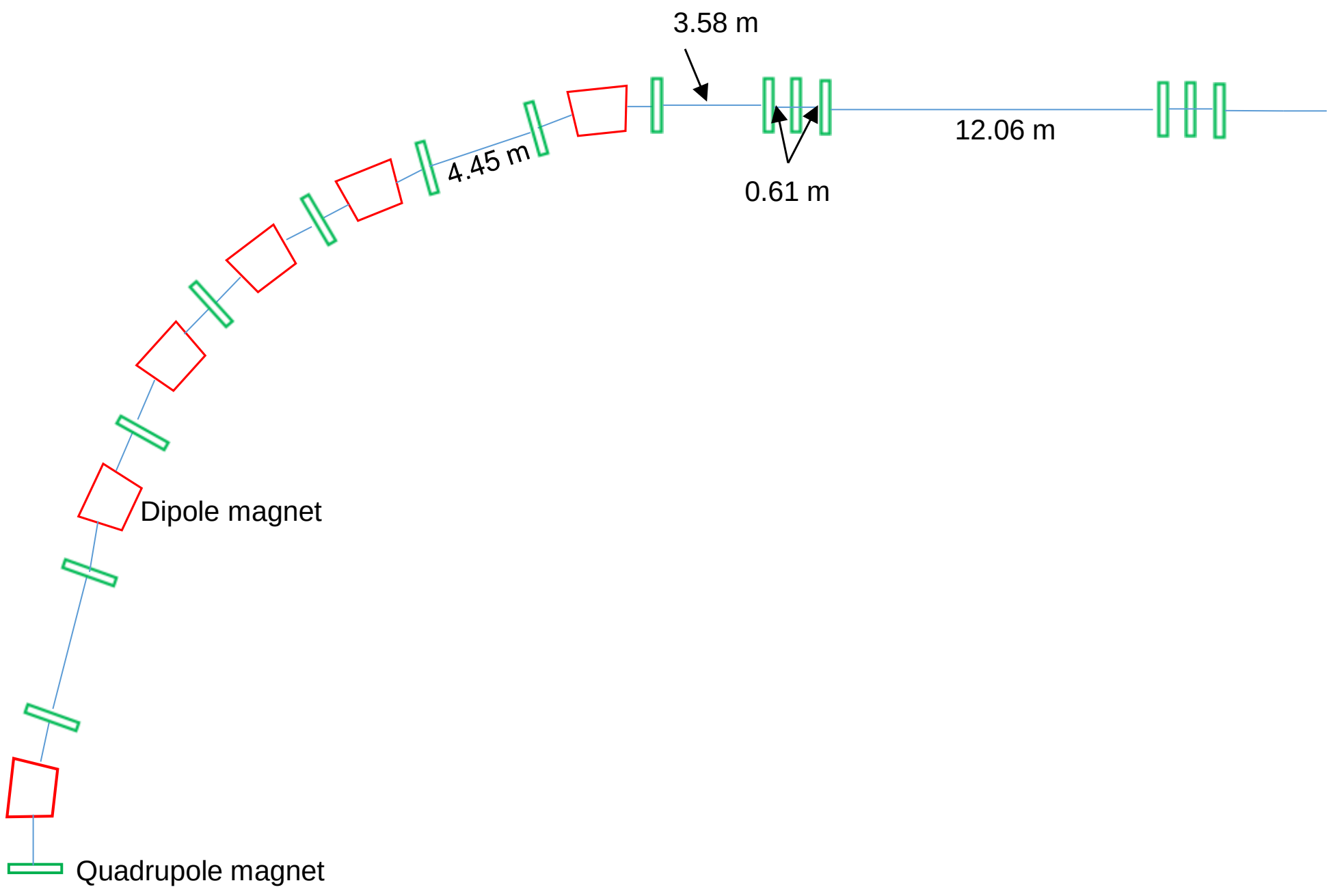

*Fig. 59: Schematic optics configuration of a superperiod of Hybrid Lattice (revised). Red and green blocks show dipole magnets and quadrupole magnets, respectively.*

*Table 23: Quadrupole magnet strengths and lattice parameters for triplet lattice. Normalized strength (g/Bρ i.e. in $m^{-2}$) of the quadrupole is used here.*

| **Parameters** | **Tune-1** | **Tune-2** | **Tune-3** | **Tune-4** |
|---|---|---|---|---|
| QDA | -0.6660553 | | | -0.6266441 |
| QFAM | 0.5762587 | | | 0.6065027 |
| QDAM | -0.6586081 | | | -0.6453917 |
| QFA | 0.5840824 | | | 0.5911141 |
| QDAS | -0.6619022 | | | -0.6104985 |
| QDAST | -0.2265688 | | | -0.2209521 |
| QFD | 0.7078894 | 0.7078894 | 0.8337533 | 0.7005811 |
| QDD | -0.9426116 | -1.002572 | -1.076533 | -0.9123858 |
| QFD1 | 0.4930294 | 0.5988049 | 0.5296152 | 0.4939277 |
| $\beta_{x\,max}$ (m) | 22.01 | 19.01 | 22.97 | 20.13 |
| $\beta_{y\,max}$ (m) | 23.46 | 25.80 | 26.86 | 24.27 |
| $D_{x\,max}$ (m) | 3.43 | 3.43 | 3.43 | 3.21 |
| Tune X | 6.88 | 7.20 | 7.22 | 7.20 |
| Tune Y | 6.88 | 6.88 | 7.22 | 6.20 |
| $\gamma_{tr}$ | 5.26 | 5.26 | 5.26 | 5.47 |

The good field region of the magnets has been evaluated considering the lattice function obtained, and assuming a painted emittance of 210 π mm-mrad, momentum spread of 2% and a margin larger than 20 mm for COD. Taking the acceptance as twice the emittance, the required good field region in the dipole is obtained as ±103 mm and ±92 mm in the horizontal and the vertical plane, respectively. Similarly, the required good field region in the quadrupole magnets is obtained as ±103 mm and ±119 mm in the zero dispersion and non-zero dispersion region, respectively. Taking the acceptance as 2.5 times the painted emittance, the required good field region for the dipole magnet is obtained as ±119 mm and ±103 mm, in horizontal and vertical plane, respectively. In this case, the required good field region of a quadrupole is ±119 mm and ±140 mm for zero and non-zero dispersion, respectively.

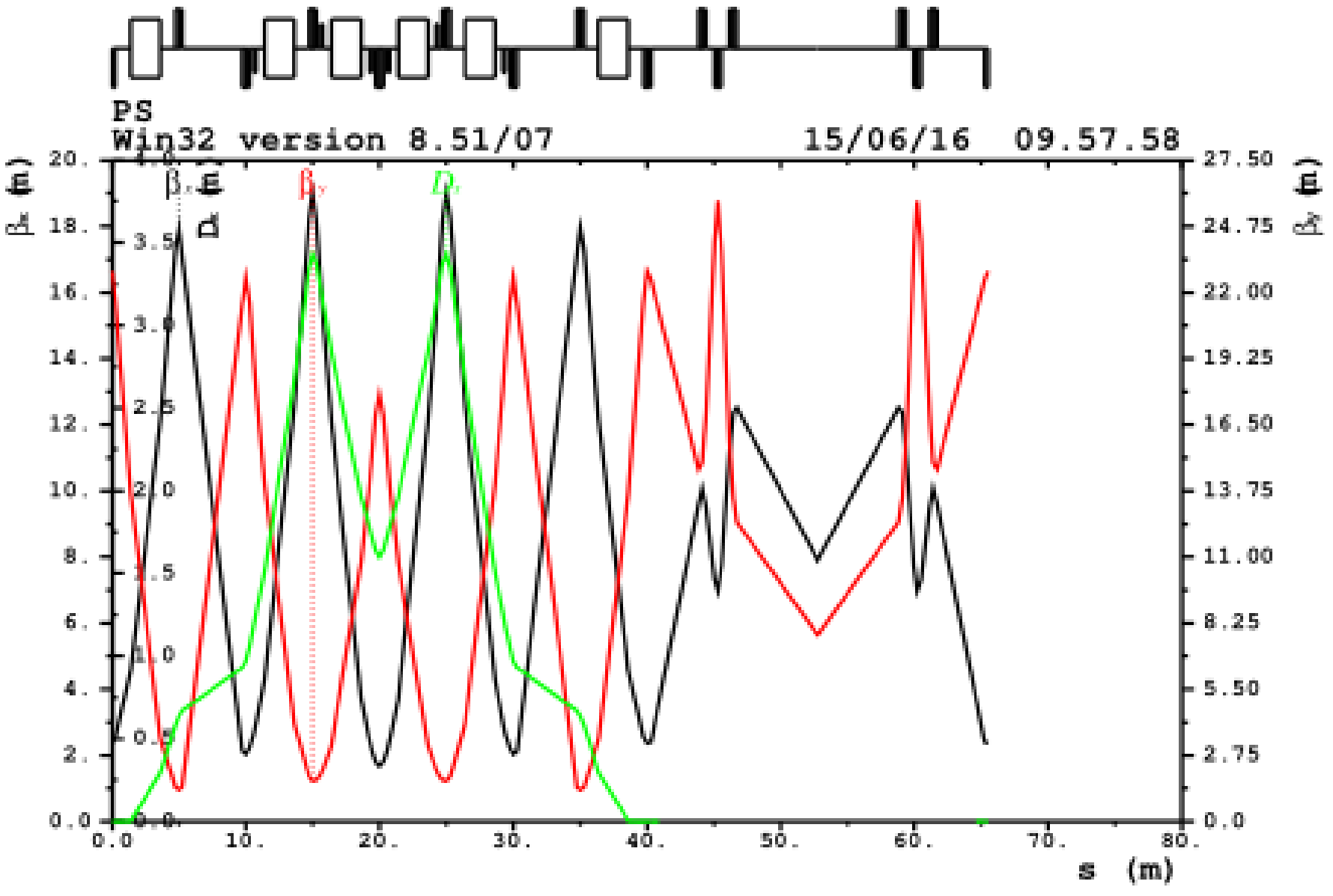

*Fig. 60: Lattice parameters of the revised Hybrid lattice for $\nu_x = 7.20$, $\nu_y = 6.88$.*

*Table 24: Tuning studies of different working points of FODO lattice*

| Horizontal Tune | Vertical Tune | Remarks |
|---|---|---|
| 7.20 | 7.20 | Good tuning range is available |
| 6.82 | 6.82 | Narrow tuning range is available |
| 6.82 | 6.20 | Adequate tuning range is available |
| 7.20 | 6.20 | Narrow tuning range is available |
| 6.20 | 6.20 | Adequate tuning range is available |

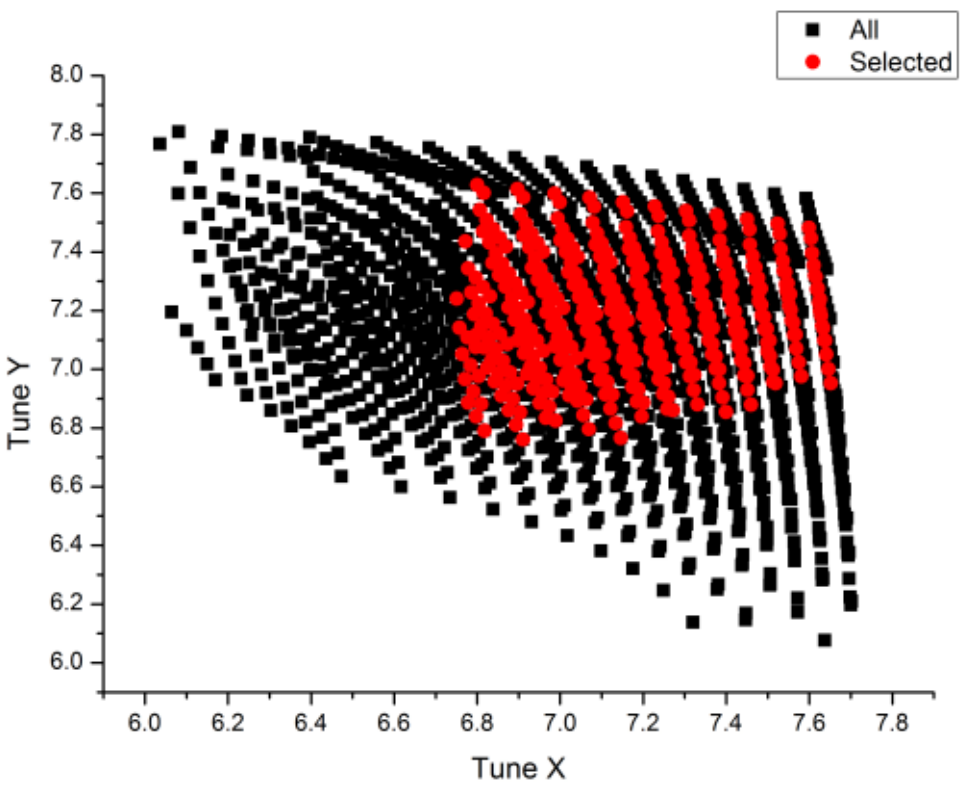

Tune (7.20, 7.20)

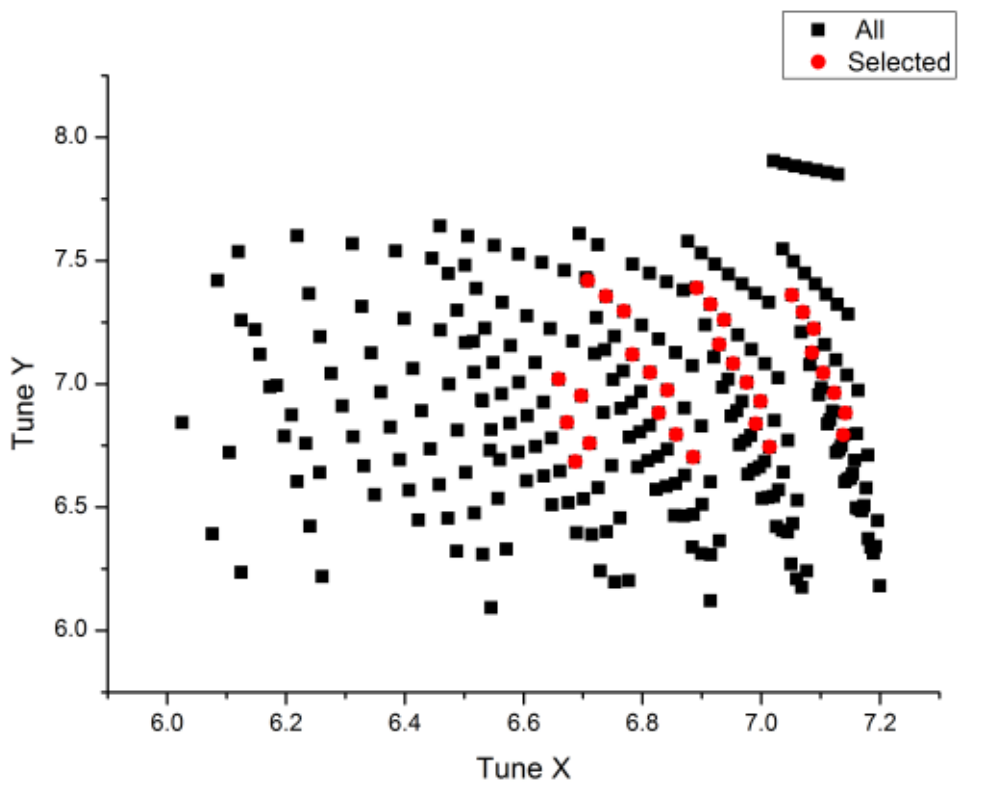

Tune (6.82, 6.82)

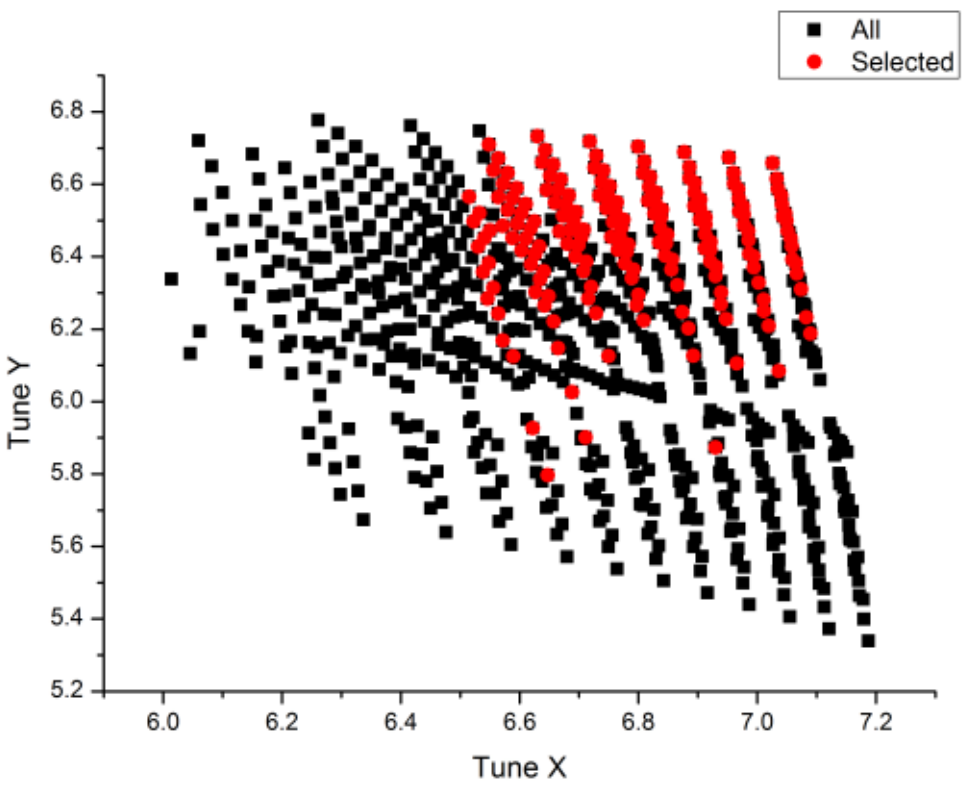

Tune (6.82, 6.20)

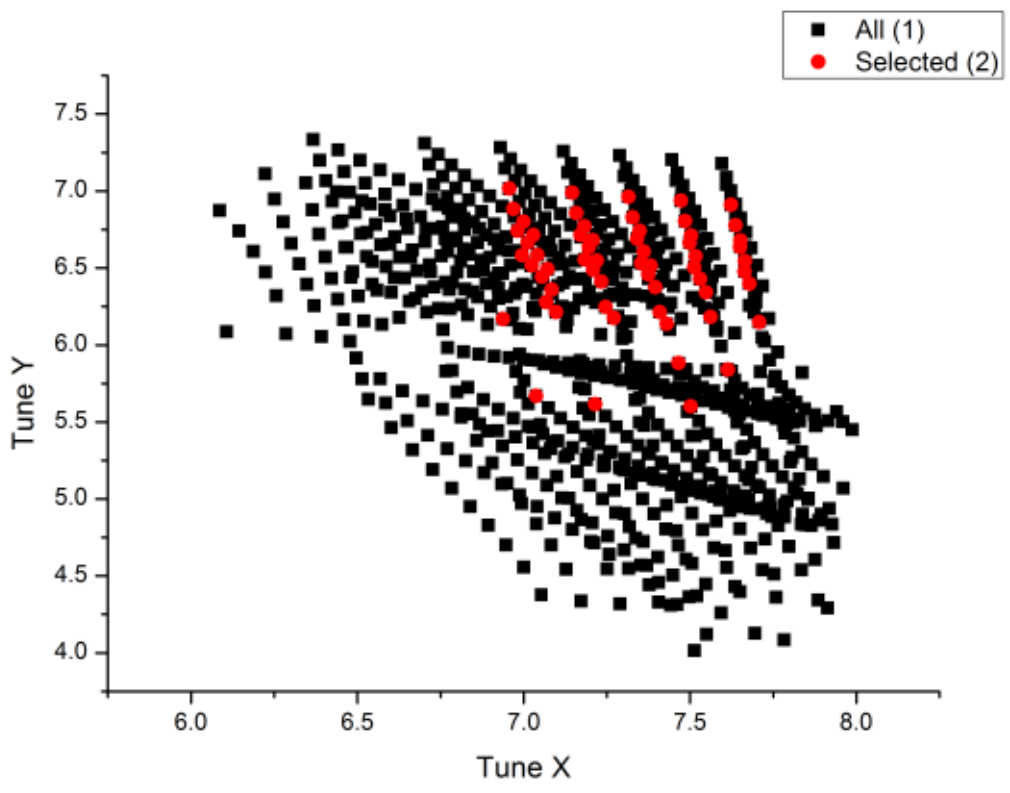

Tune (7.20, 6.20)

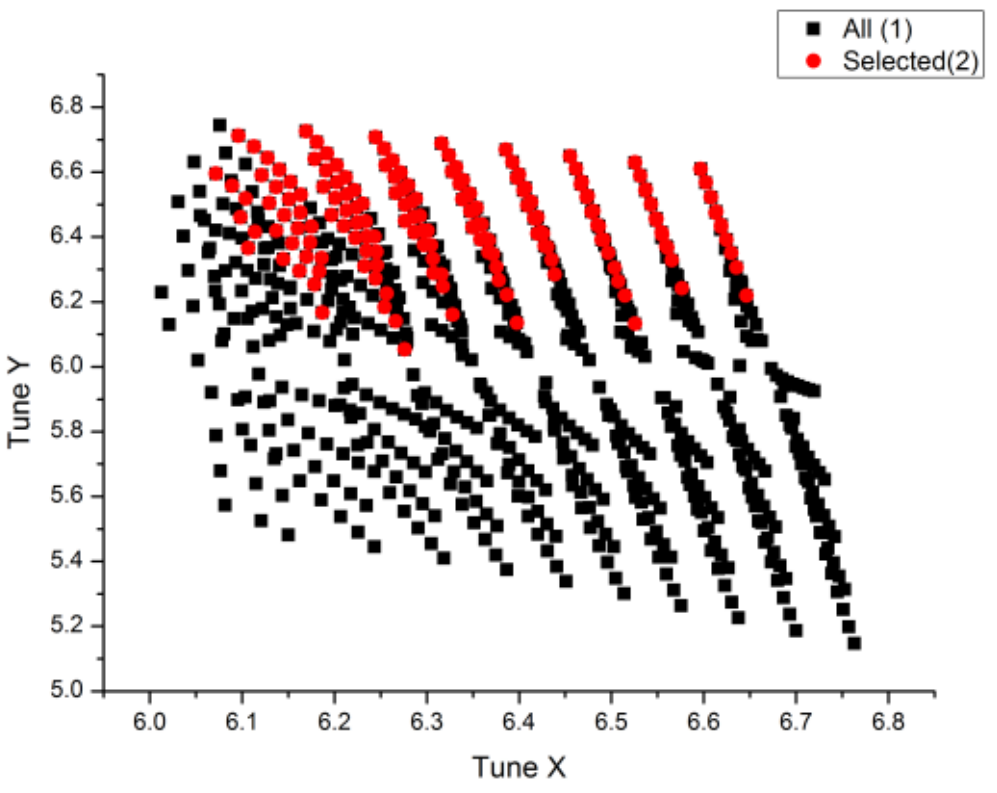

Tune (6.20, 6.20)

*Fig.61: Tuning range of FODO lattice around different working points. Red dots show the acceptable solution on the basis of maximum beta function and structure resonances up to fourth.*

Tuning range of FODO lattice has been explored. Results of the study around five tune points for the FODO lattice are summarized in Table 24, and shown in Fig. 61. These studies for revised Hybrid lattice are presently being done.

## 9.2 Longitudinal Beam Dynamics

Studies have also been performed on the longitudinal beam dynamics in the accumulator ring to evolve the RF parameters. The basic purpose of the RF system is to produce a longitudinal focusing of the proton bunch, and compresses it further down to ~700 ns from ~ 1µs to provide a time gap for extraction kicker rising, and improve the bunching factor to mitigate the detrimental effects of space charge.

The RF system at the start of injection should provide sufficient bucket height to accept the momentum spread of Linac beam without significant losses. To reduce the probability of beam loss due to space charge by accumulating more and more turns of injected beam, a sudden increase in bucket height is required and then after accumulation a proper voltage is necessary to compress the bunch. A detailed study of RF program is under progress. Table 25 shows the RF voltage requirement in AR for two different compression of bunches, for harmonic number *h*=1. The required compression depends on the extraction kicker rise time. The requirement increases by 30% due to increase of rise time of kickers from ~250 ns to ~300 ns. The results in the table are without considering second harmonic system.

*Table 25: Initial estimation of required RF voltage. Here, $t_{gap}$ is the available time gap for rising the extraction kicker pulse.*

| Relative momentum spread δ (%) | Voltage (kV) required to compress from 1µs to 0.75µs **($t_{gap}$=0.25µs)** | | Voltage (kV) required to compress from 1µs to 0.70µs **($t_{gap}$=0.3µs)** | |
|---|---|---|---|---|
| | Initial Voltage | Final Voltage | Initial Voltage | Final Voltage |
| ±1 | 42 | 133 | 42 | 175 |
| ±0.5 | 21 | 66 | 21 | 87 |

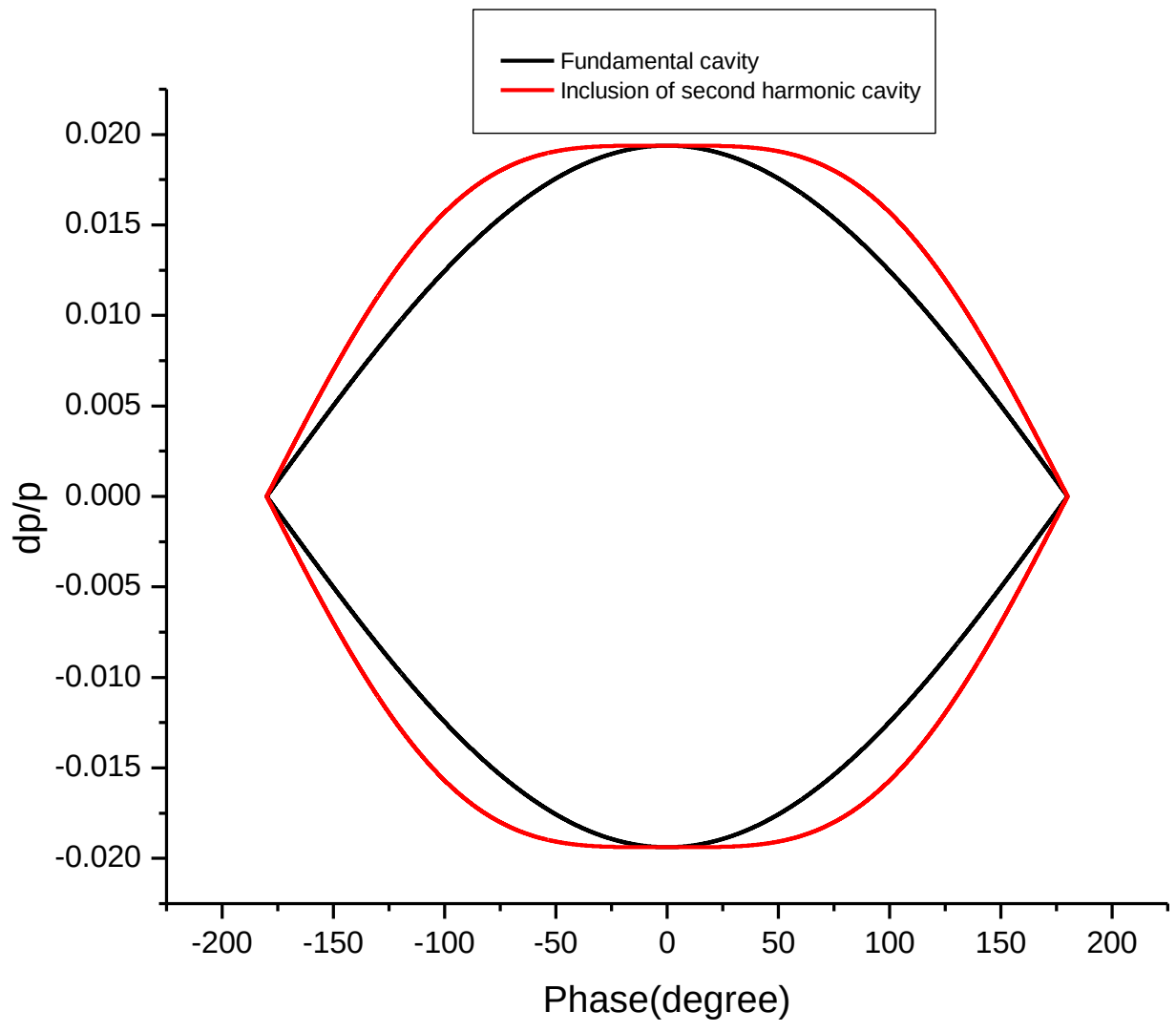

*Fig. 62: RF bucket with and without second harmonic RF system.*

Including second harmonic system, bunching factor can be improved, so that the transverse space charge tune shift becomes lower. Therefore, a second harmonic RF system will also be required. In general, the required peak voltage of second harmonic system will be nearly half of the fundamental RF voltage. How the second harmonic RF system changes the bucket is shown in Fig. 62, where phase difference between two RF waves is 180° and second harmonic RF voltage is half of the fundamental RF voltage of 42 kV.

Further studies including the effect of beam loading are currently underway. To carry out these studies, a computer code for simulating longitudinal beam dynamics under beam loading and space charge is under development. Results of single particle dynamics from this code have been benchmarked with the standard code ESME [20]. It was required to develop an indigenous code for longitudinal dynamics studies since the updated version of ESME for beam loading studies was not accessible to us. Beam loading module development has been completed and beam tracking under the beam loading with and without higher harmonic cavity is possible. Space charge module is also nearly complete. For computing the bunching factor from the tracked beam, smoothening of beam distribution is carried out using rectangular or un-weighted sliding-average smoothing algorithms up to first ten tracking injection turns, and triangular smoothing algorithms for remaining tracking turns. The smoothened beam distribution is also used for calculating the rate of change of the line charge density, which is used to determine the voltage generated due to space charge. For calculating the space charge effects in our code, we calculated the longitudinal space charge impedance defined by $Z_{sc} = -j\frac{Z_0 g}{2\beta\gamma^2}$, where $Z_0$ = 377 Ω is space impedance, *β, γ* are relativistic factors and geometric factor is defined as $g = 1 + \ln\left(\frac{Chamber\ Radius}{beam\ radius}\right)$ [21]. For our simulation, we have taken the average vacuum chamber radius for the machine as 0.1 m, and the beam radius as 0.0263 m, which give the longitudinal space charge impedance equal to –*j* 126 Ω. For the given beam current, the space charge impedance induced voltage seen by the particle in each turn is determined by knowing the rate of change of the current in the longitudinal direction as following [21]:

$$V_{sc} = \frac{dI}{d\phi}|Z_{sc}|$$

This space charge voltage provides the longitudinal defocusing force, which reduces the effective RF voltage within the bunch duration. Space charge effect is included in the code by determining the slope of the beam current distribution. Currently, the testing of beam loading and the space charge module is in progress.

Figure 63 shows the comparison of results from the developed code and results from the code ESME for the tracking of multi turn injected beam with and without second harmonic cavity.

<table>
<tr><td>Results of our developed code</td><td>Results of ESME code</td></tr>
<tr><td colspan="2">Fundamental cavity</td></tr>
<tr><td colspan="2">Phase space</td></tr>
<tr><td>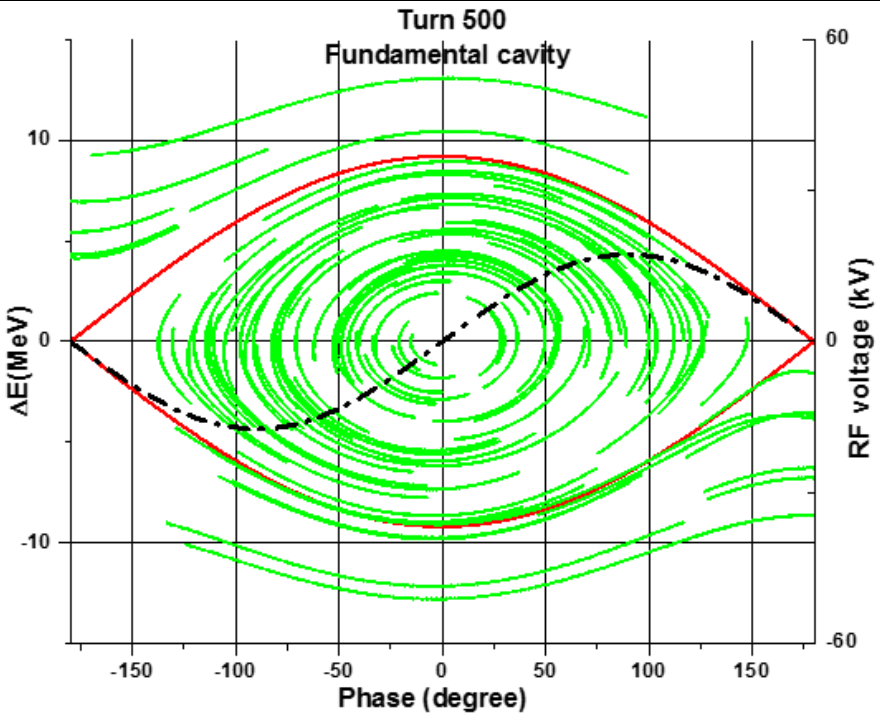
</td><td>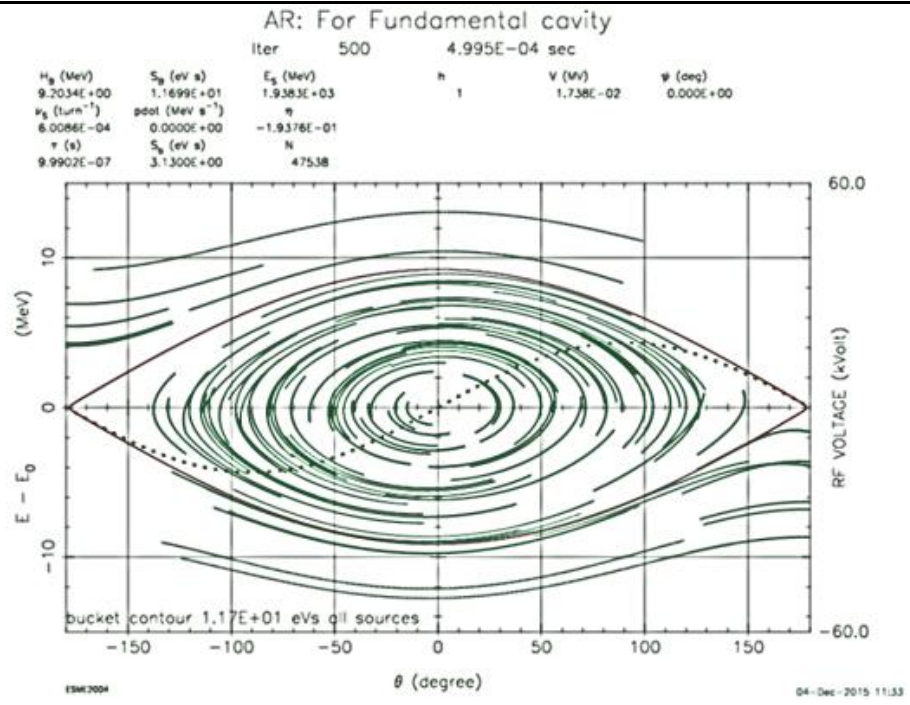
</td></tr>
<tr><td>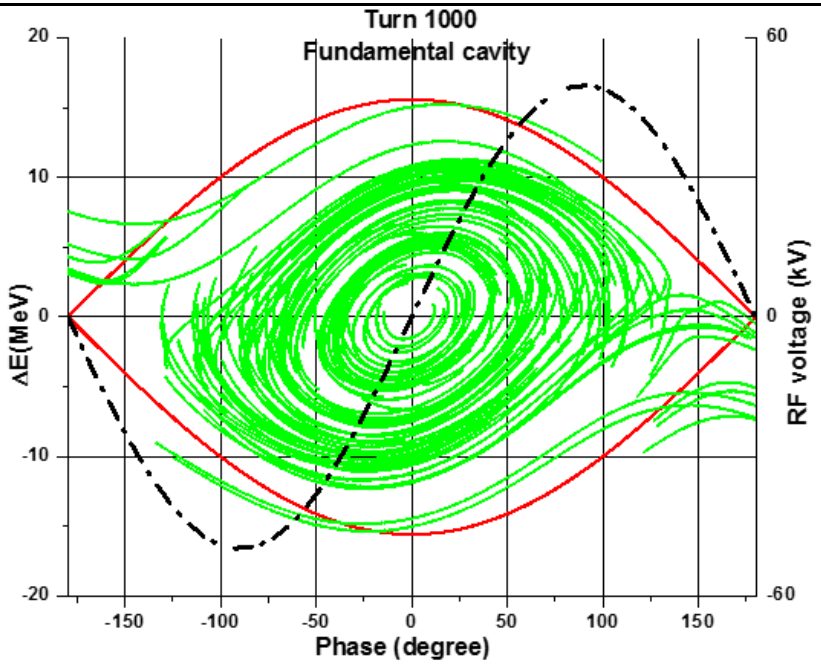
</td><td>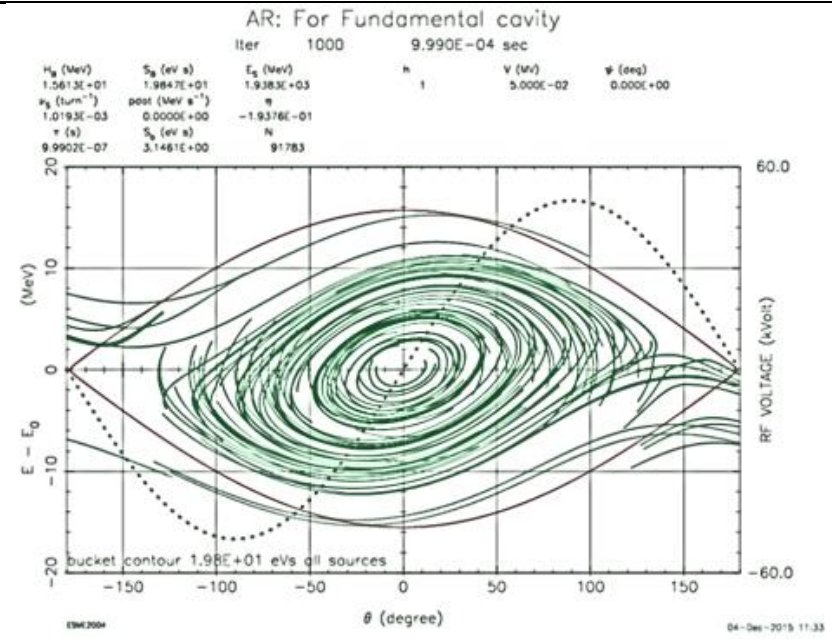
</td></tr>
<tr><td>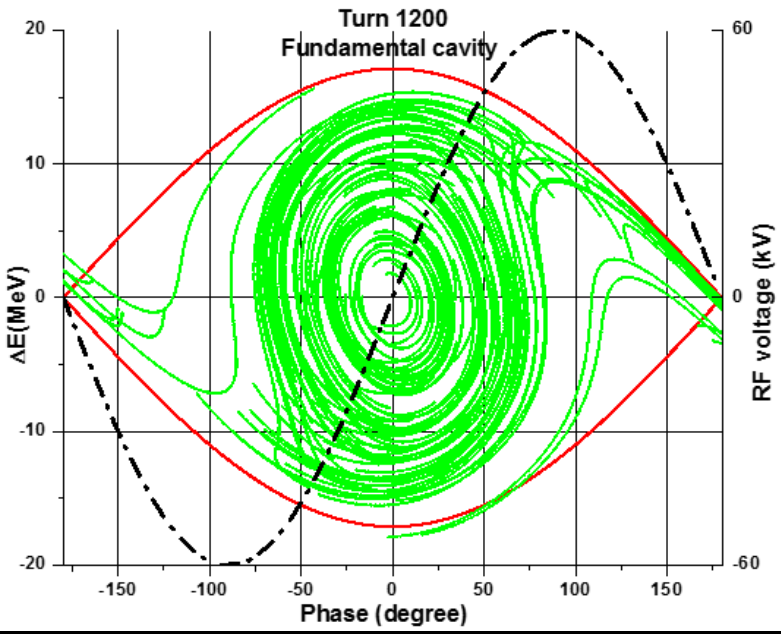
</td><td>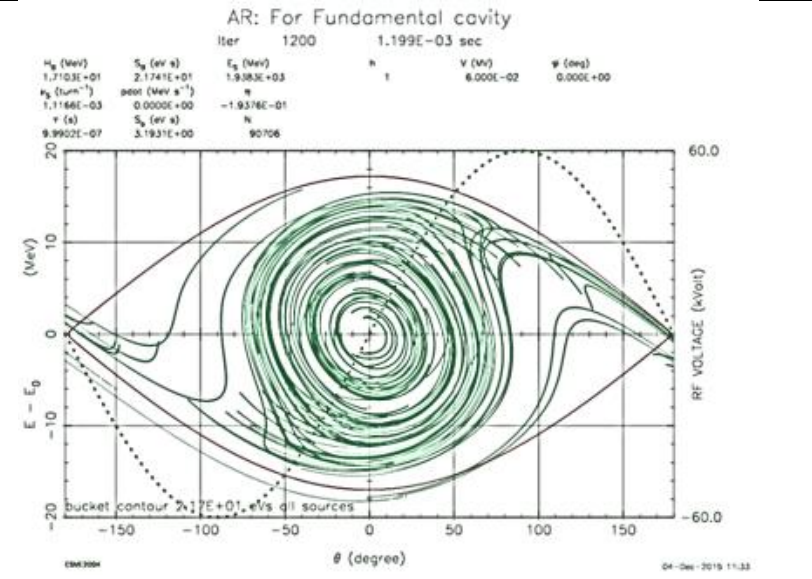
</td></tr>
<tr><td colspan="2">Synchrotron tune</td></tr>
<tr><td>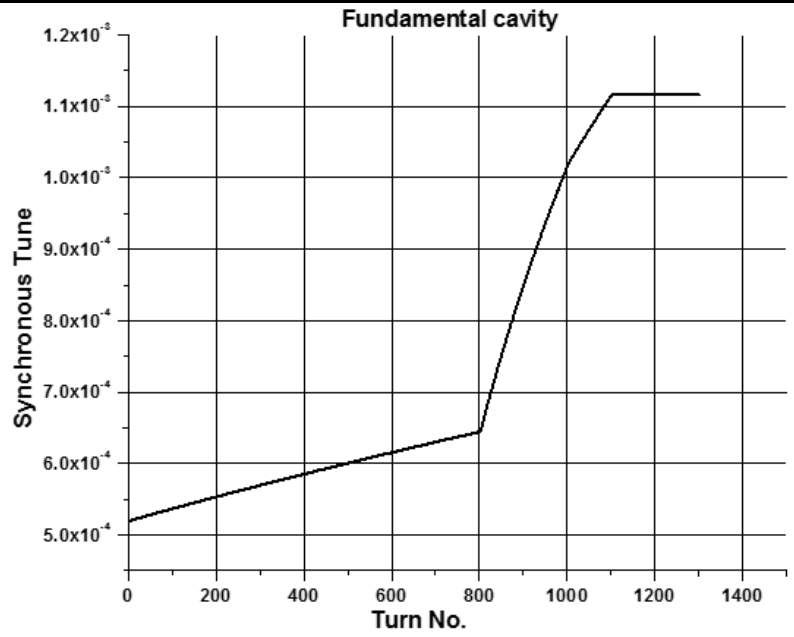
</td><td>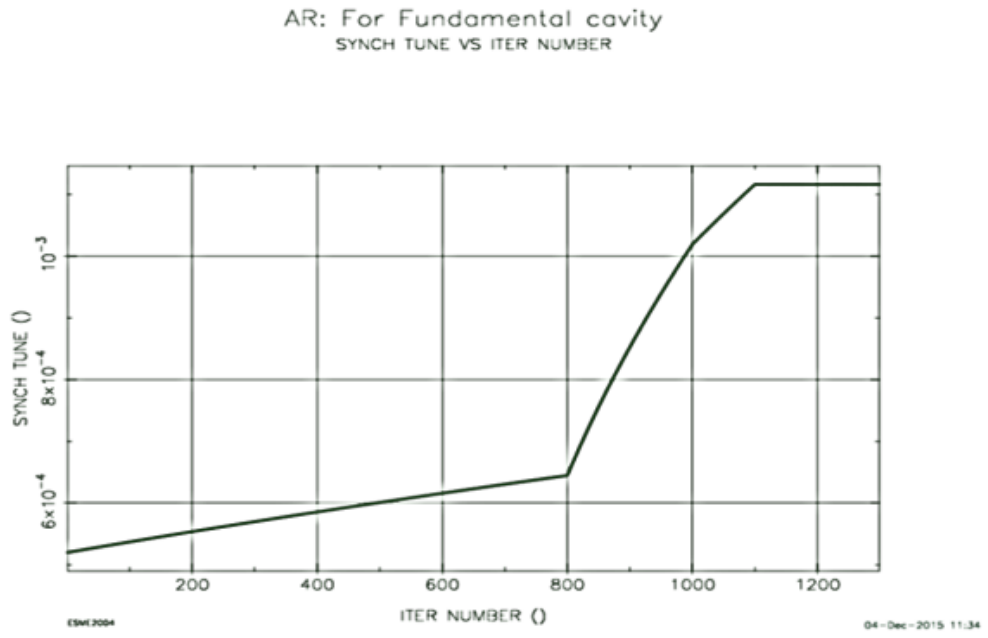
</td></tr>
<tr><td colspan="2">Fundamental cavity + Second harmonic cavity</td></tr>
<tr><td colspan="2">Phase space</td></tr>
</table>

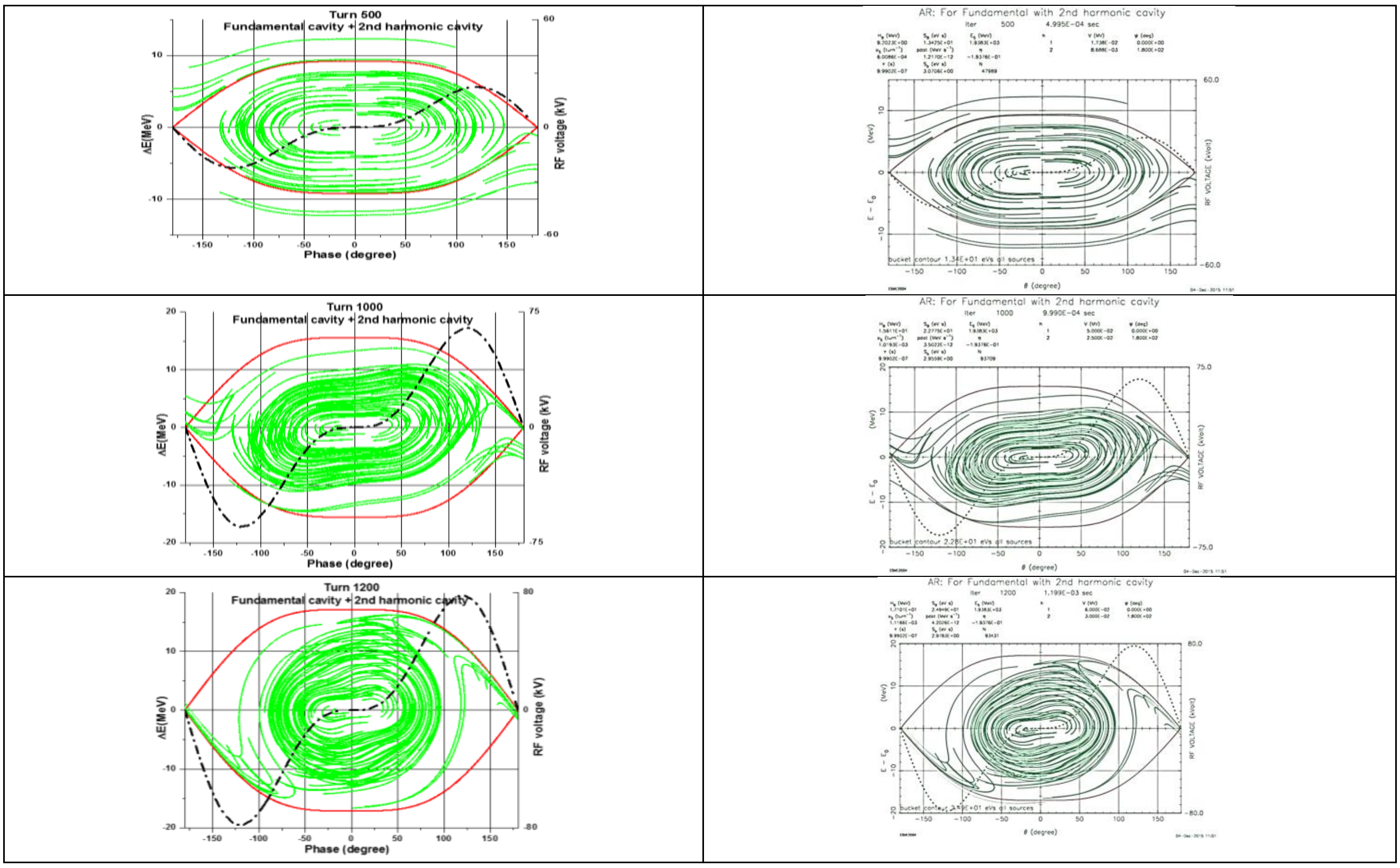

Fig. 63: Comparison of results from the developed code and from ESME.

Parameters taken in above simulations are shown in following Table 26.

*Table 26: Parameters used in simulation for comparison with the results from ESME*

| | | | |
|---|---|---|---|
| Total tracking periods | | 1300 turns | |
| Beam Energy | | 1 GeV | |
| Linear RF voltage profile | up to 800 turns | 13- 20kV | 6.5- 10kV |
| | up to 1000 turns | 20-50kV | 10-25kV |
| | up to 1100 turns | 50-60kV | 25-30kV |

# 10. Linear optics correction schemes in Accumulator Ring

## 10.1 Closed Orbit Correction

Studies have been performed on distortion in the closed orbit of the accumulator ring due to various errors in the magnetic elements. Table 27 shows the errors, assumed for the study. The maximum COD is evaluated as 25 mm in both the planes, assuming random distribution of the errors.

*Table 27: Different errors taken for studying the COD and its correction*

| **Type of error** | **Magnitude (1σ)** |
|---|---|
| Dipole to dipole mismatch (ΔB/B) | $5\times10^{-4}$ |
| Quadrupole misalignment | 100 μm |
| Dipole horizontal misalignment | 100 μm |
| Dipole rotation error around s-axis | 500 μrad |

For correcting the COD, scheme of correctors and beam position indicators (BPI) for FODO lattice is shown in Fig. 64. In the proposed scheme for FODO lattice, there are 24 steering magnets in each plane and 48 BPIs for monitoring the beam position. Different algorithms, MICADO, Harmonic correction, SVD and sliding bump technique were used to test the orbit correction scheme. Fig. 65 shows the results of SVD on different working points for FODO lattice [22]. After correction, the maximum COD is around 0.5 mm in both the planes. The maximum corrector kick required, is nearly 0.5 mrad. Study is extended to obtain the tolerances of steering magnets as well as resolution of BPIs. Similar studies for Hybrid lattice is also carried out [23]. The locations of steering magnets and BPIs are shown in Fig. 66. Fig. 67 shows the results of SVD on different working points for hybrid lattice. Table 28 shows the details of the scheme for orbit correction for both the lattices.

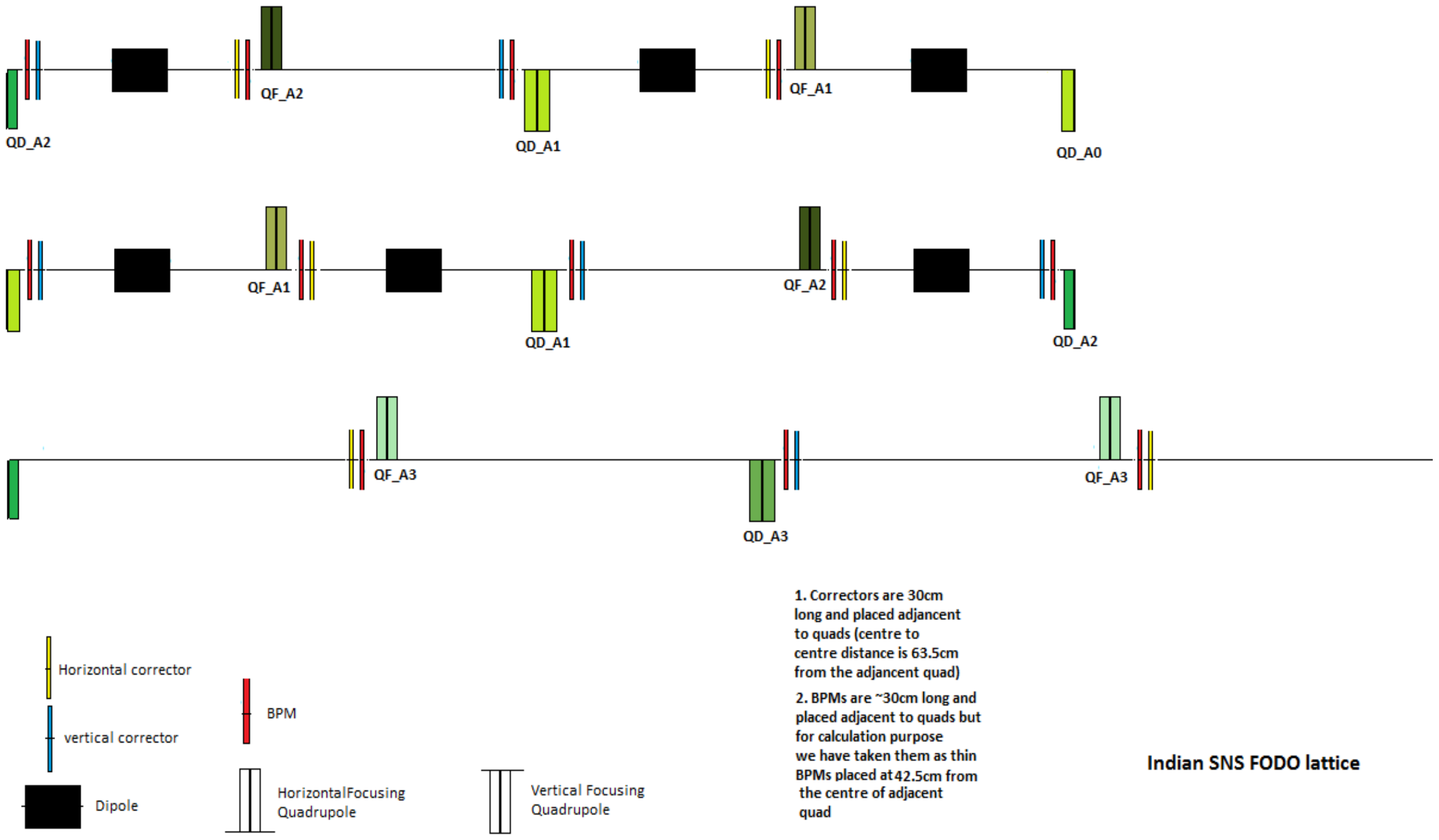

*Fig. 64: Schematic of locations for steering magnets and beam position indicators in a superperiod of a FODO lattice.*

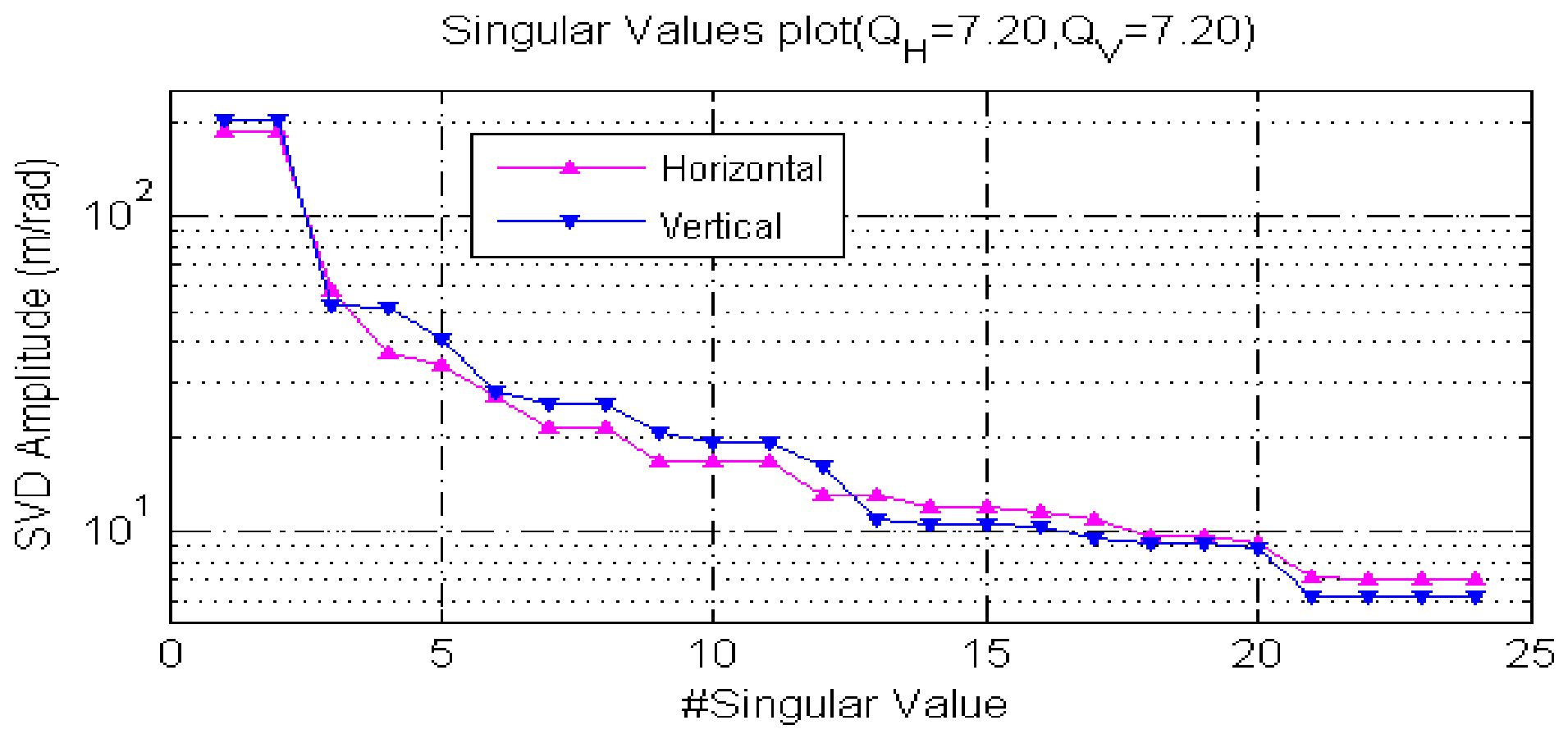

*Fig. 65: Singular values of orbit correctors at a working point in FODO lattice*

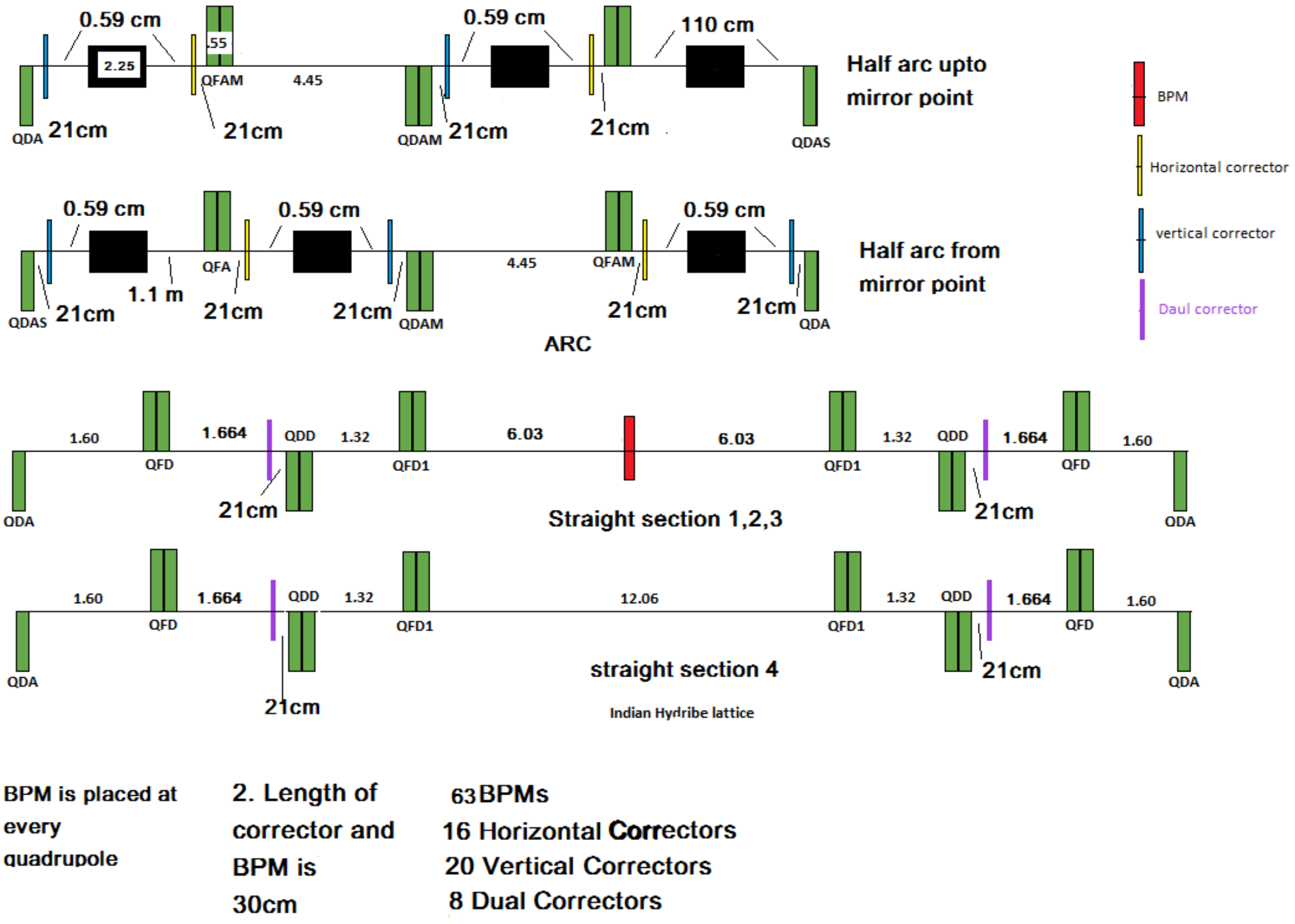

*Fig. 66: Schematic of locations for steering magnets and beam position indicators in a superperiod of a Hybrid lattice.*

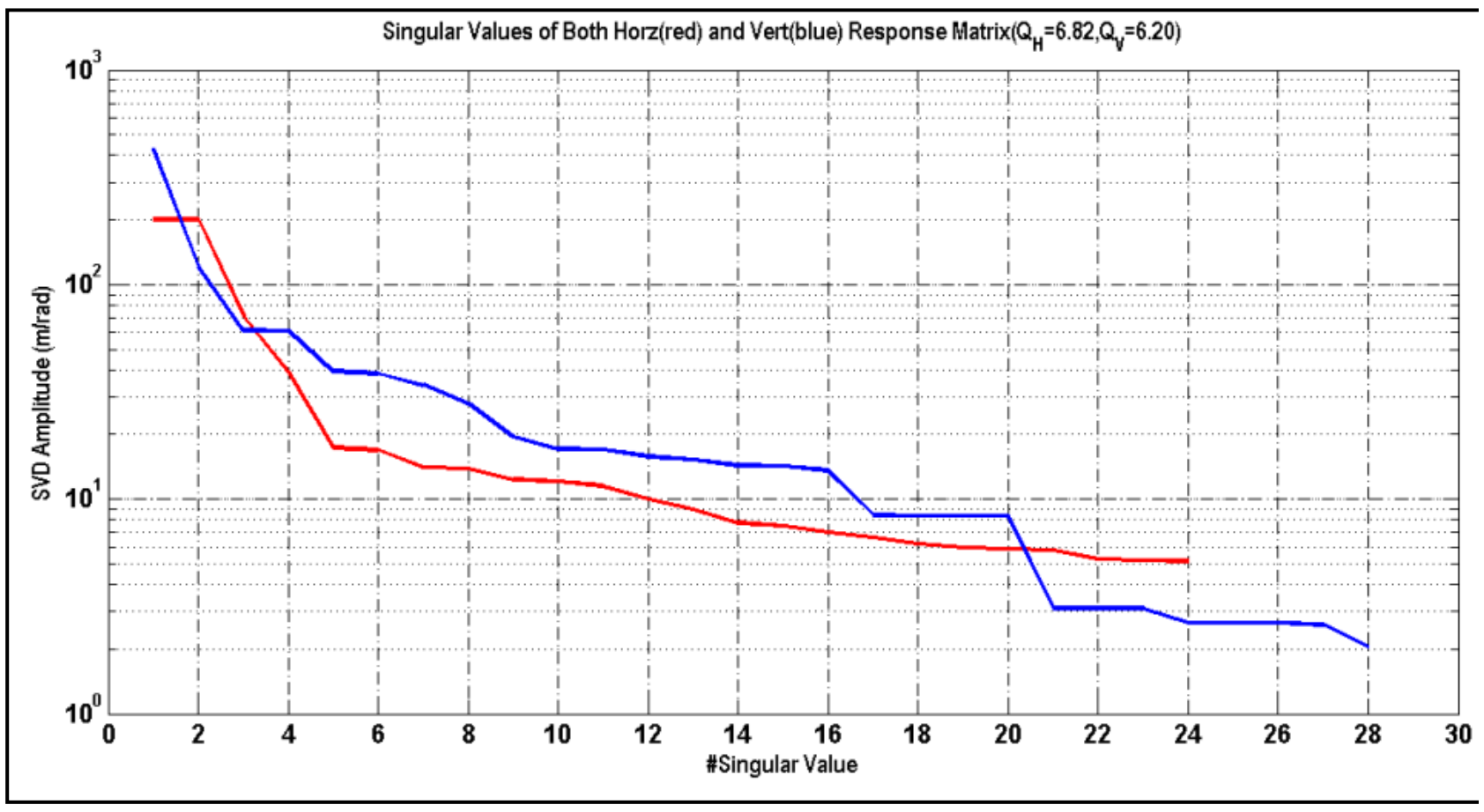

*Fig. 67: Singular values of orbit correctors at a working point in Hybrid lattice (red curve and blue curve are for horizontal and vertical plane, respectively).*

*Table 28: Details of Steering magnets and Beam Position Indicators (BPI) for FODO and Hybrid lattice.*

| **Parameter** | **FODO lattice** | **Hybrid lattice** |
|---|---|---|
| No. of horizontal steering magnets | 24 | 24 |
| No. of vertical steering magnets | 24 | 28 |
| Kick form steering magnets | 0.5 mrad | 0.6 mrad |
| Magnetic field of steering magnet (considering 300 mm of effective length) | ~50 G | ~60 G |
| Tolerance on the BL fluctuation of steering magnet (for orbit stability) | $< 1\times10^{-3}$ | $< 1\times10^{-3}$ |
| No. of BPI | 48 | 63 |
| Resolution of BPI | < 10 μm | < 10 μm |

## 10.2 Gradient errors and compensating their effects

Gradient errors in the lattice results in deviation of beta functions from the design values, and generate asymmetry in the optical functions. This asymmetry leads the reduction in the dynamic aperture and therefore requires correction. Each quadrupole magnet of the lattice will have a trim coil. Presently, study is carried out considering gradient error of 0.1% (1σ), truncated at 3σ. The results are shown in Tables 29 and 30 for FODO [24] and Hybrid lattices [25], respectively.

*Table 29: Effect of gradient error on the optical parameters of FODO lattice*

| **Tune point** | $\frac{\Delta\beta_x}{\beta_x}$ | $\frac{\Delta\beta_y}{\beta_y}$ | $\frac{\Delta D_x}{\sqrt{\beta_x}}$ **(m$^{-1/2}$)** | $\Delta\nu_x$ | $\Delta\nu_y$ |
|---|---|---|---|---|---|
| (7.20, 7.20) | 0.045 | 0.0523 | 0.0325 | 0.01 | 0.01 |
| (6.82, 6.82) | 0.0448 | 0.0523 | 0.0374 | 0.01 | 0.01 |
| (6.82, 6.20) | 0.0416 | 0.0422 | 0.0352 | 0.009 | 0.008 |
| (7.20, 6.20) | 0.0448 | 0.0414 | 0.0292 | 0.01 | 0.008 |
| (6.20, 6.20) | 0.4 | 0.0430 | 0.0372 | 0.008 | 0.008 |

*Table 30: Effect of gradient error on the optical parameters of Hybrid lattice*

| **Tune point** | $\frac{\Delta\beta_x}{\beta_x}$ | $\frac{\Delta\beta_y}{\beta_y}$ | $\frac{\Delta D_x}{\sqrt{\beta_x}}$ **(m$^{-1/2}$)** | $\Delta\nu_x$ | $\Delta\nu_y$ |
|---|---|---|---|---|---|
| (6.82, 6.82) | 0.0524 | 0.0875 | 0.0483 | 0.011 | 0.016 |
| (7.20, 7.20) | 0.0566 | 0.0840 | 0.0437 | 0.013 | 0.016 |
| (7.20, 6.82) | 0.0556 | 0.0828 | 0.0437 | 0.012 | 0.015 |
| (7.20, 6.20) | 0.0474 | 0.0797 | 0.038 | 0.011 | 0.016 |
| (6.82, 6.20) | 0.0592 | 0.0783 | 0.0478 | 0.012 | 0.015 |

Elementary analysis shows that the required strength of the trim coil is less than 1% of the strength of the quadrupole magnet. Response matrix method as well as resonance correction scheme, are explored to correct the effects due to gradient errors. In response matrix method, all the trim coils are used and effectively the beta function, tune and dispersion restores to its original value. In resonance correction scheme, the width of resonance line due to harmonics of beta functions and dispersion are reduced, such that the particles are in the resonance free zone on the tune diagram, even after including the tune shift due to space charge. For $\beta_x$, $\beta_y$ and $D_x$, harmonics are close to $2\nu_x$, $2y$, and $\nu_x$, respectively. Different strings of trim coils are excited to correct these resonances. Figs. 68, 69 and 70 show the results of beta and dispersion restoration with resonance correction scheme for FODO lattice at one of the working points.

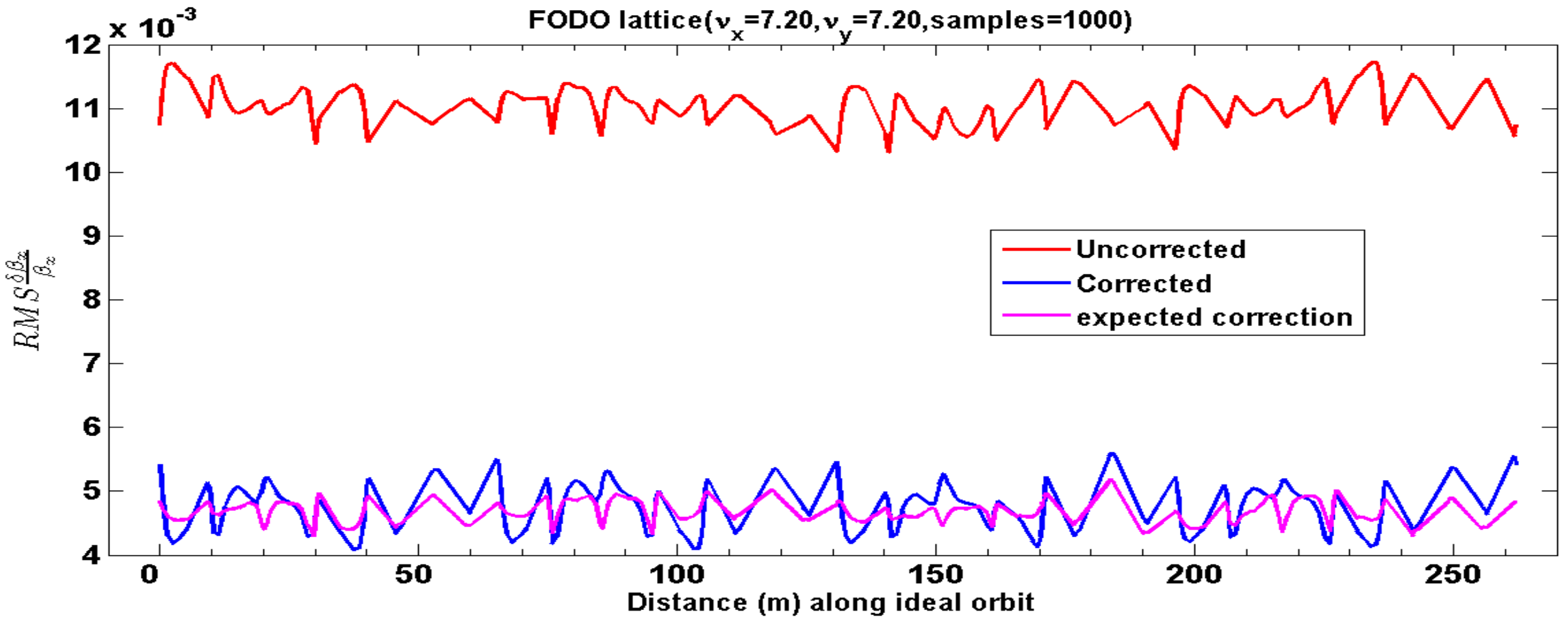

*Fig. 68: Resonance correction of horizontal beta function for optics1 ($\nu_x$=7.20, $\nu_y$=7.20).*

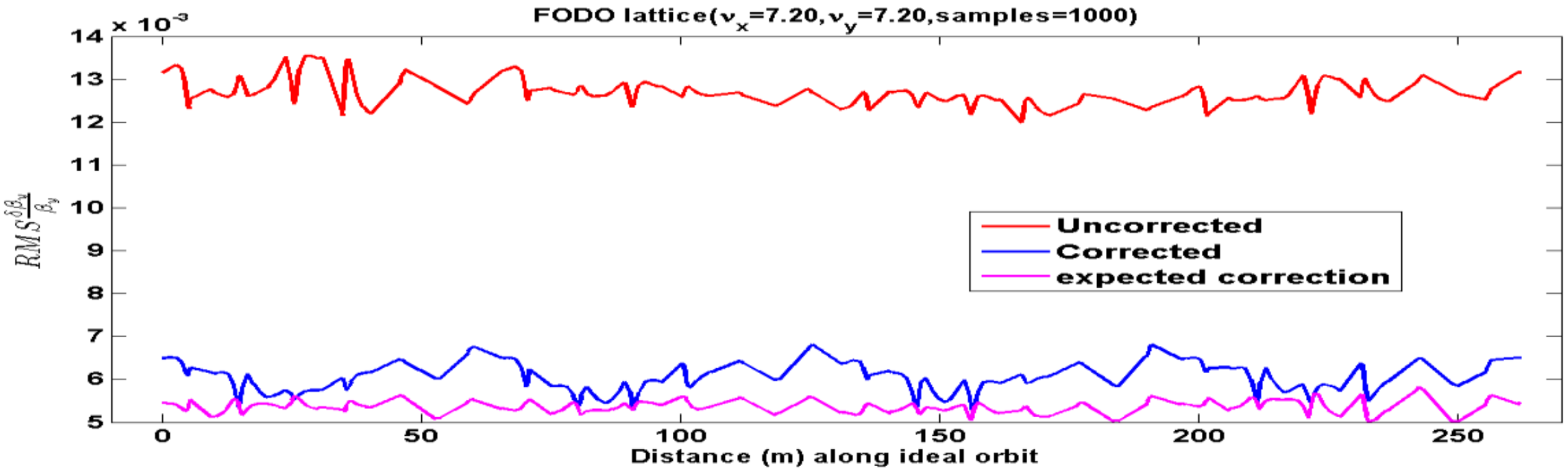

*Fig. 69: Resonance correction of vertical beta function for optics1 ($\nu_x$=7.20, $\nu_y$=7.20).*

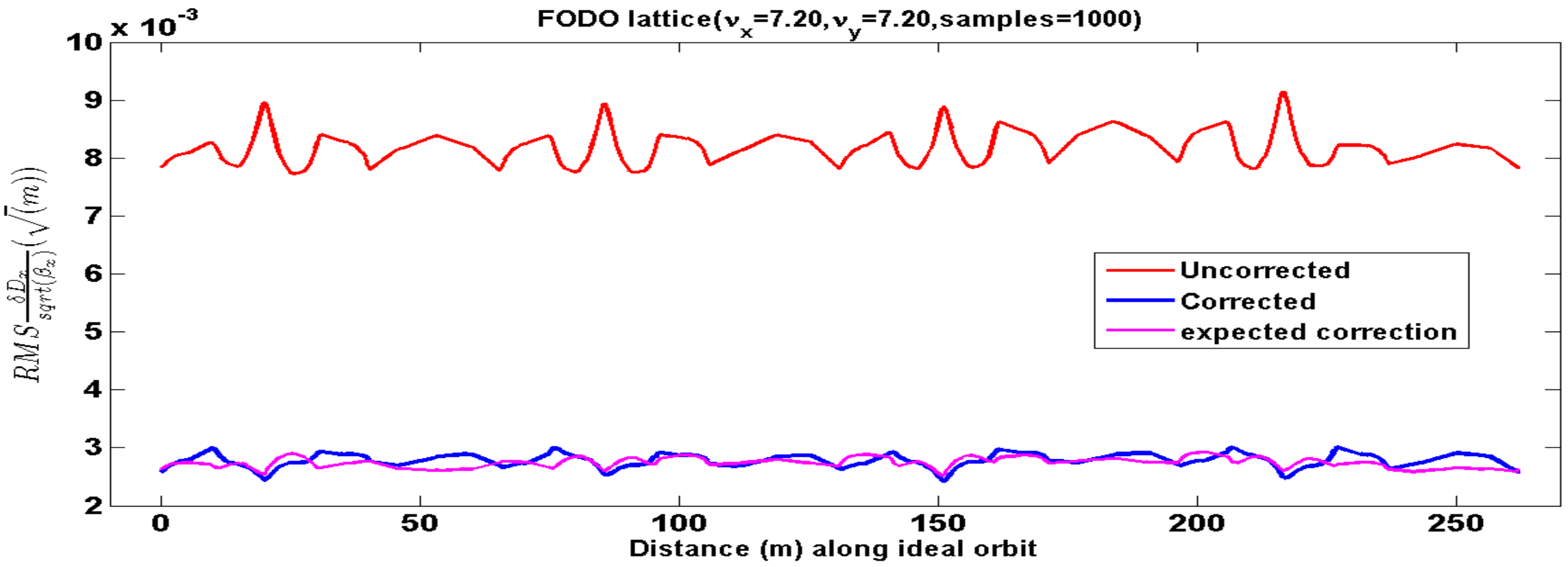

*Fig. 70: Resonance correction of horizontal dispersion function for optics1($\nu_x$=7.20, $\nu_y$=7.20).*

Similar to FODO lattice, results of resonance correction scheme for hybrid lattice at one of the working points are shown in Figs. 71, 72 and 73.

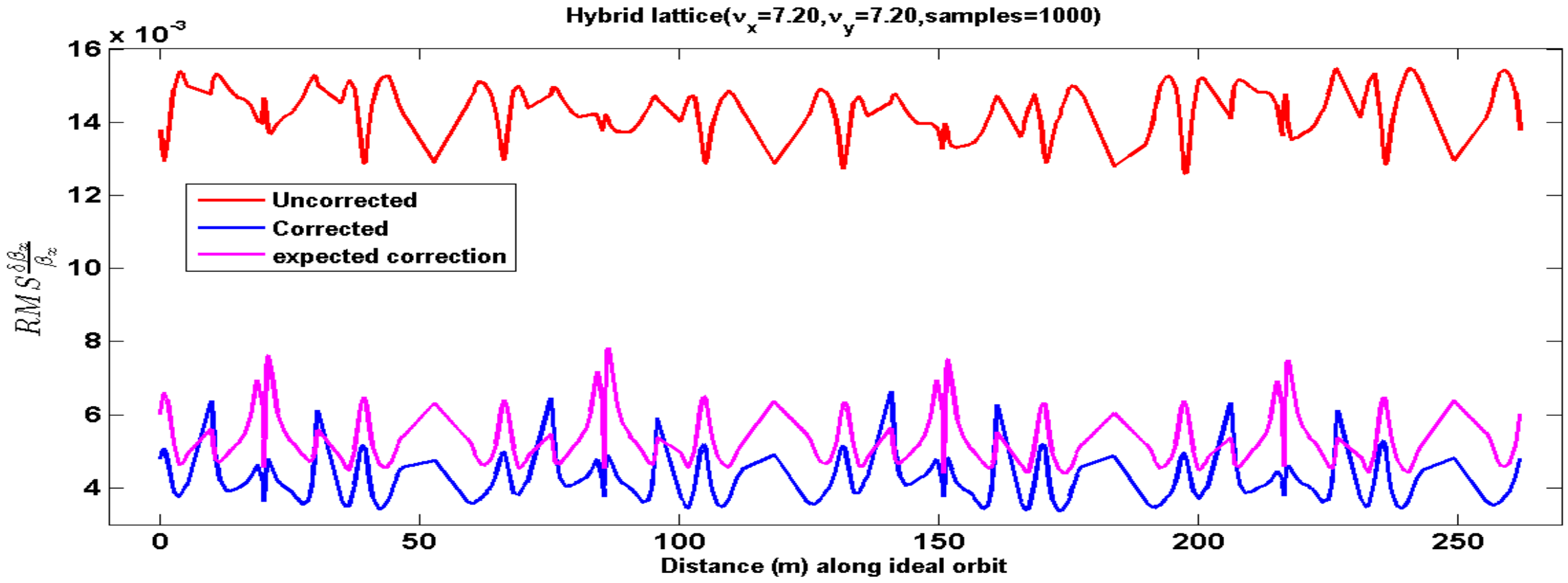

*Fig. 71: Resonance correction of $\beta_x$ for optics1 ($v_x = v_y = 7.20$).*

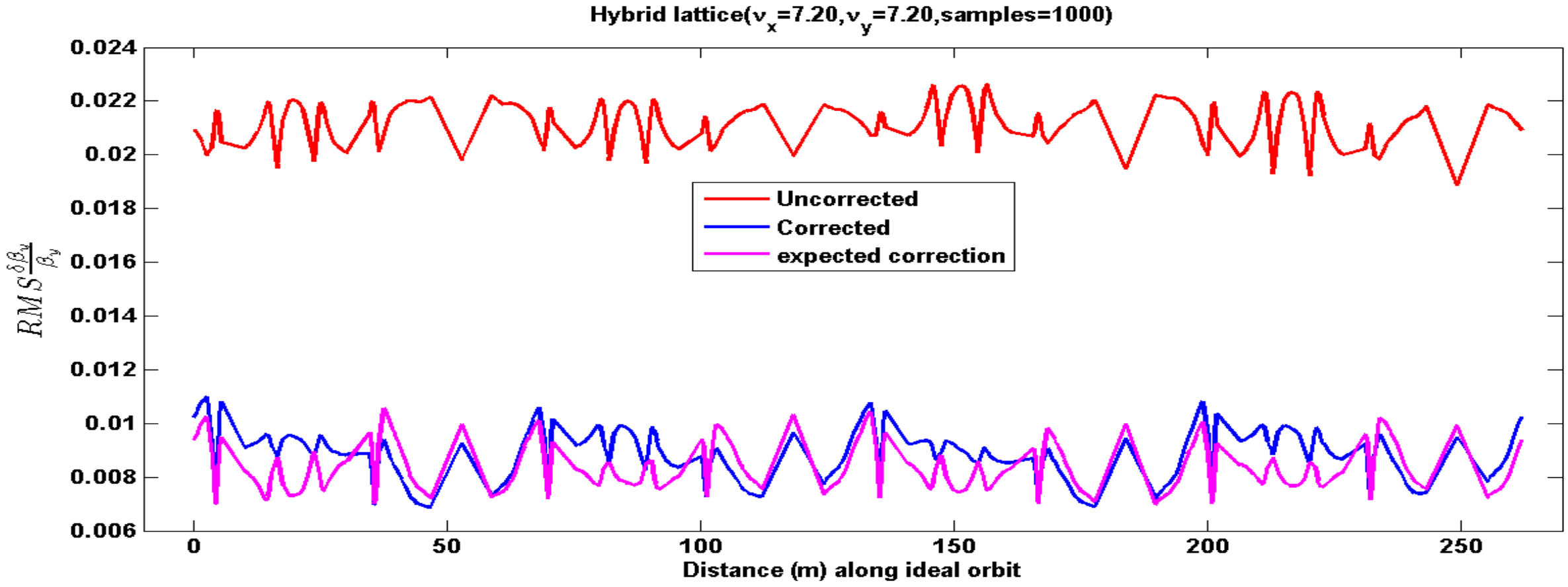

*Fig. 72: Resonance correction of $\beta_y$ for optics1 ($v_x = v_y = 7.20$).*

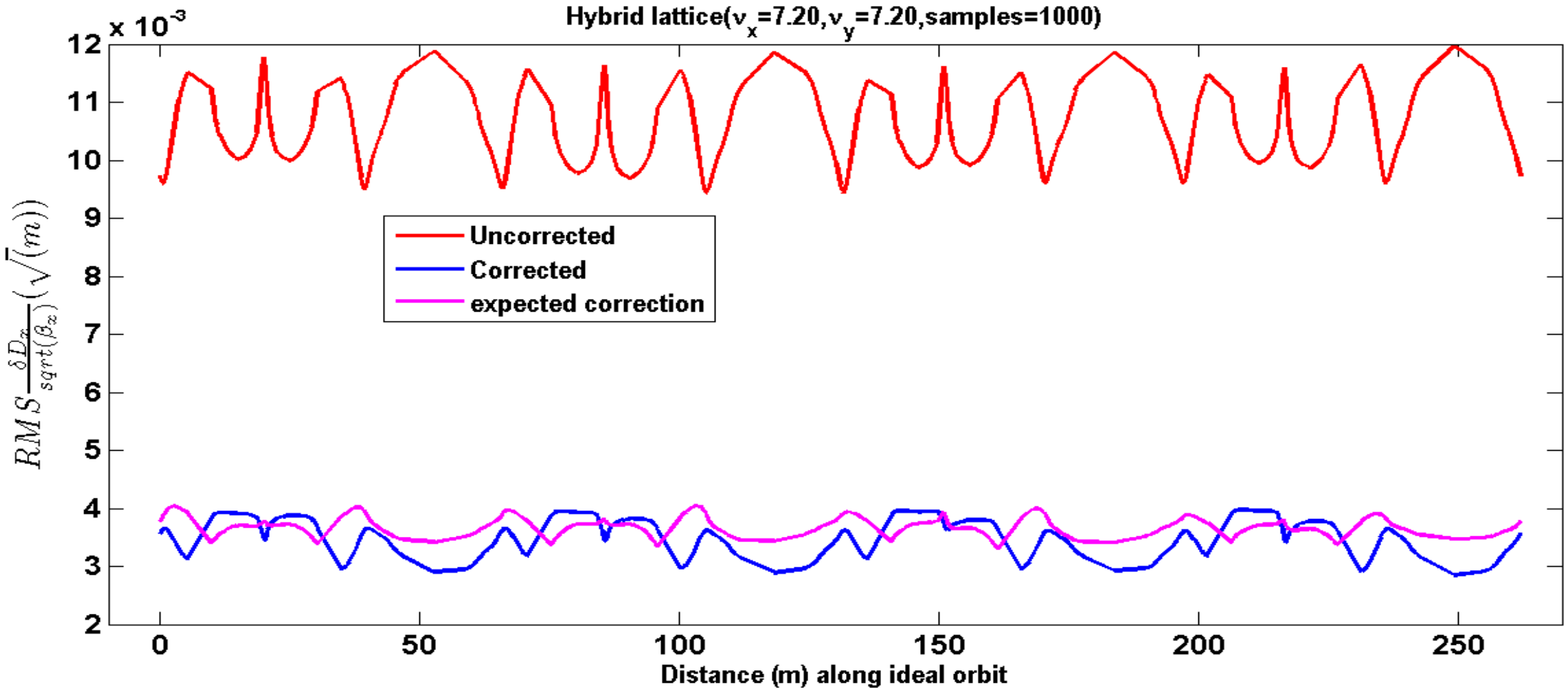

*Fig. 73: Resonance correction of $D_x$ for optics1 ($v_x = v_y = 7.20$).*

## 10.3 Skew quadrupole errors

Skew errors in quadrupole magnets introduce coupled betatron motion and also generate changes in beta functions and tunes, and produce spurious vertical dispersion. Studies have been carried out for minimizing the effects of these errors. An axial roll error of 1 mrad is assumed in the quadrupole magnets. For carrying out studies, these errors are distributed in quadrupole magnets independently. Such 10000 random configurations are studied to evolve the correction scheme. A maximum spurious dispersion of ~ 0.2 m is generated in FODO lattice. Change in beta function and tune is not much. Resonance correction scheme is used to counter the effects of skew errors. Prominent harmonics in the vertical dispersion are corrected, along with the sum and difference resonances, using suitable group (string) of skew quadrupoles. Figure 74 shows result for one of the working points in FODO lattice. A total number of 48 skew quadrupole magnets are proposed to be installed in the FODO lattice for this correction [26].

Similar result for one of the tune points for the Hybrid lattice is shown in Fig. 75, using 11 skew quadrupole magnets per period [27].

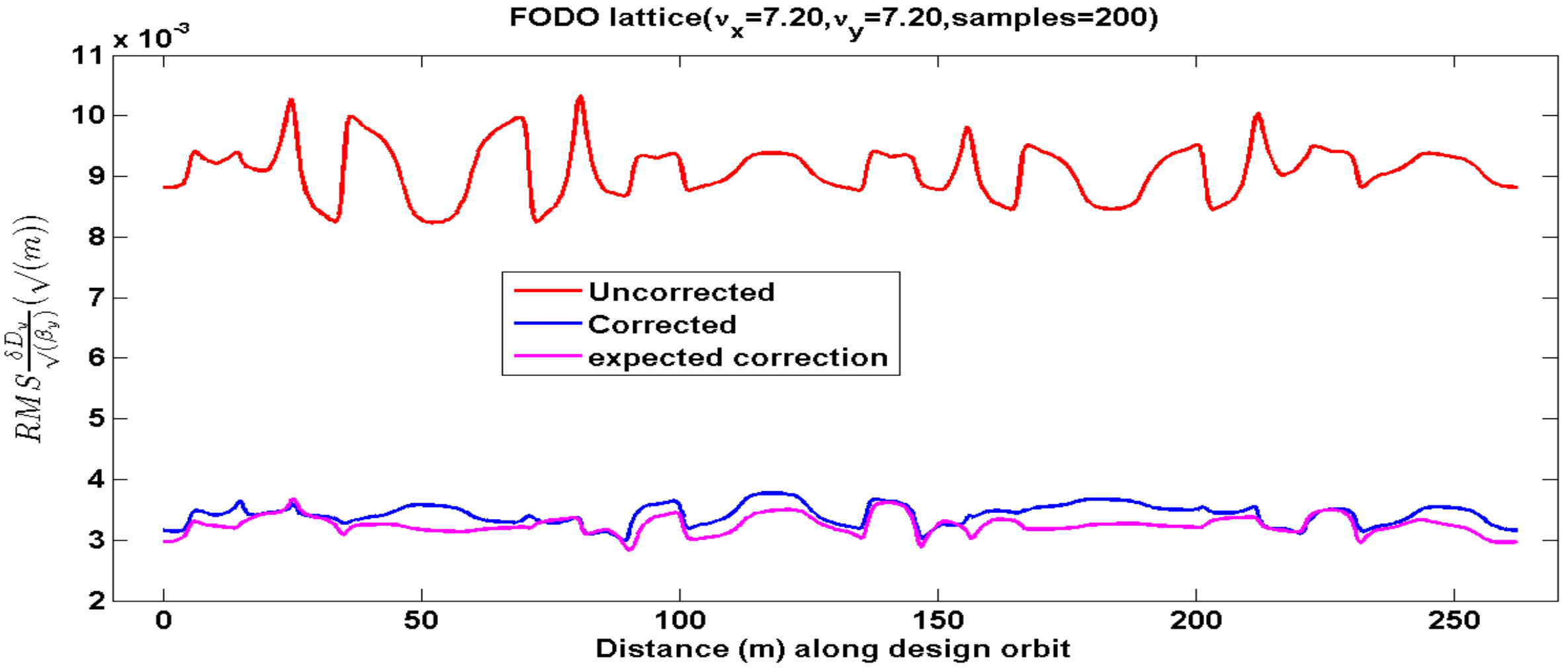

*Fig. 74: Resonance correction of $D_y$ function for optics1 ($\nu_x$= $\nu_y$=7.20).*

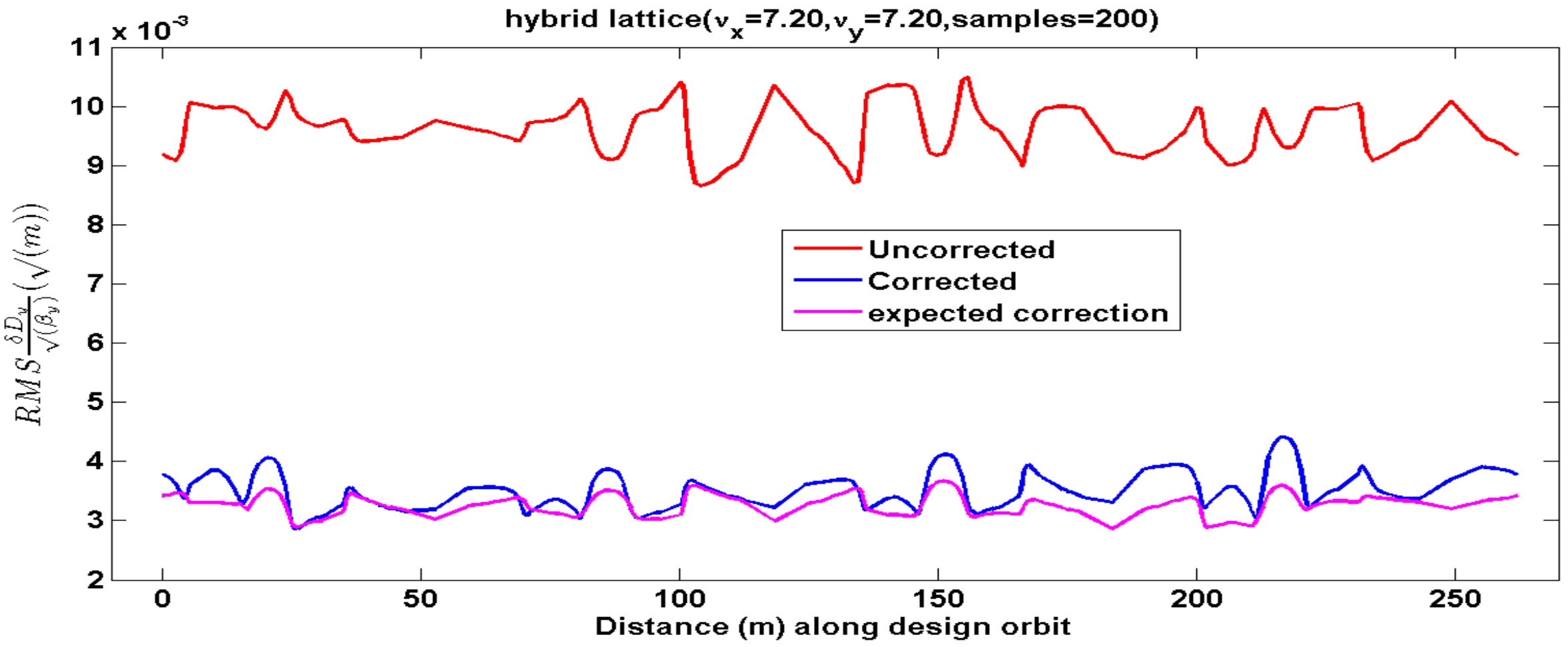

*Fig. 75: Resonance correction of $D_y$ function for optics1 ($\nu_x$= $\nu_y$=7.20).*

## 11. Sextupole scheme for the Accumulator Ring

Momentum dependent tune spread of the beam due to quadrupole magnets can lead to beam loss, which is required to be corrected in AR lattice. A natural chromaticity of the order of ~10 can lead a spread in tune up to 0.1 for a momentum spread of 1%. This is significant, and is comparable to tune spread generated by space charge. For correcting this chromaticity, there is a plan to incorporate sextupole magnets at finite dispersion region. Inclusion of sextupole magnets can deteriorate the dynamic aperture and therefore reduces the stable area in phase space for betatron oscillation. This reduction in stable area can lead to beam loss and in certain cases, even beam injection may be difficult. A study of optimization of the sextupole scheme is under progress. A schematic of chromaticity correcting sextupole magnets is shown in Fig. 76 for the AR lattice.

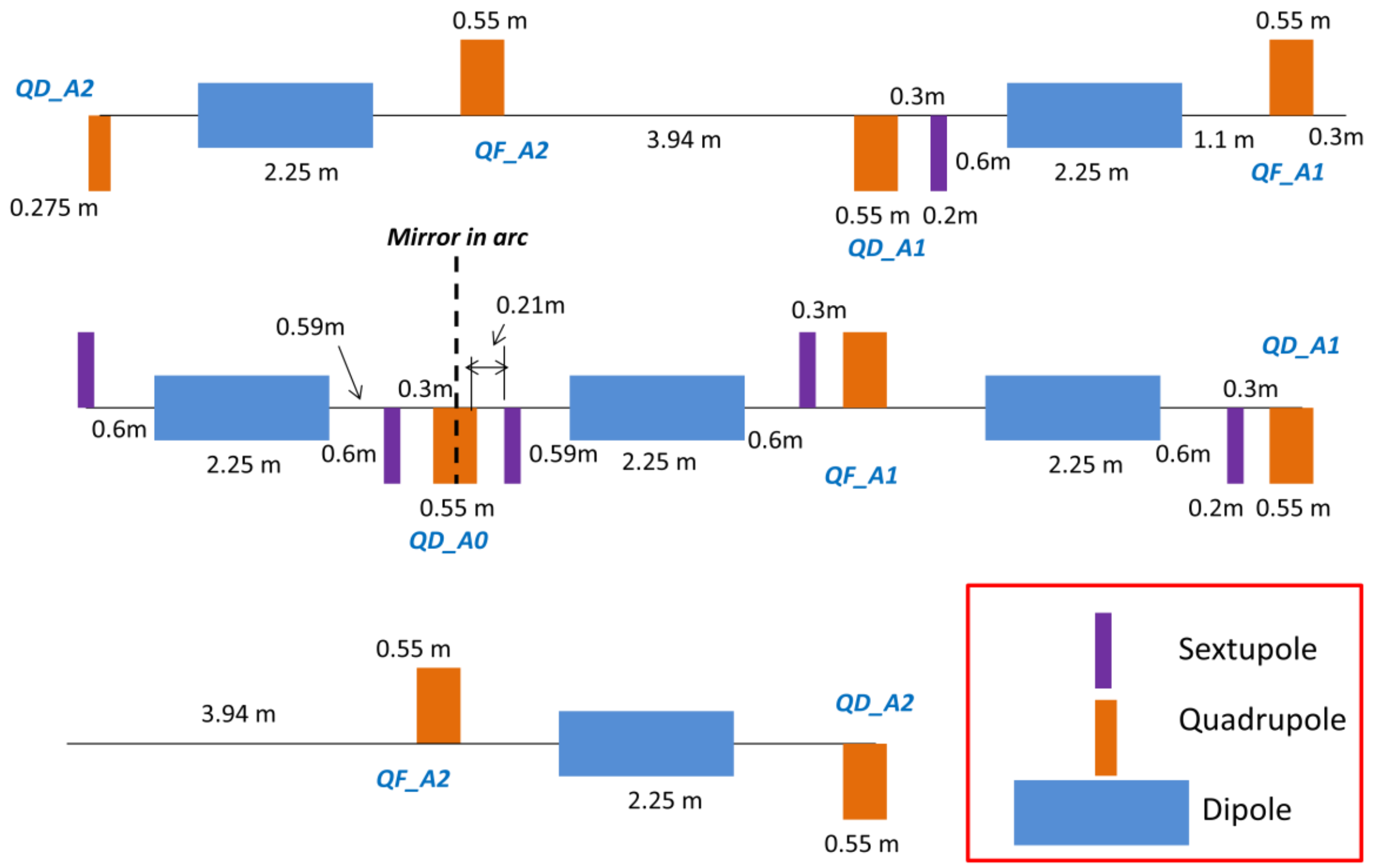

*Fig. 76: Schematic showing the location of sextupole magnets in the arc of AR lattice.*

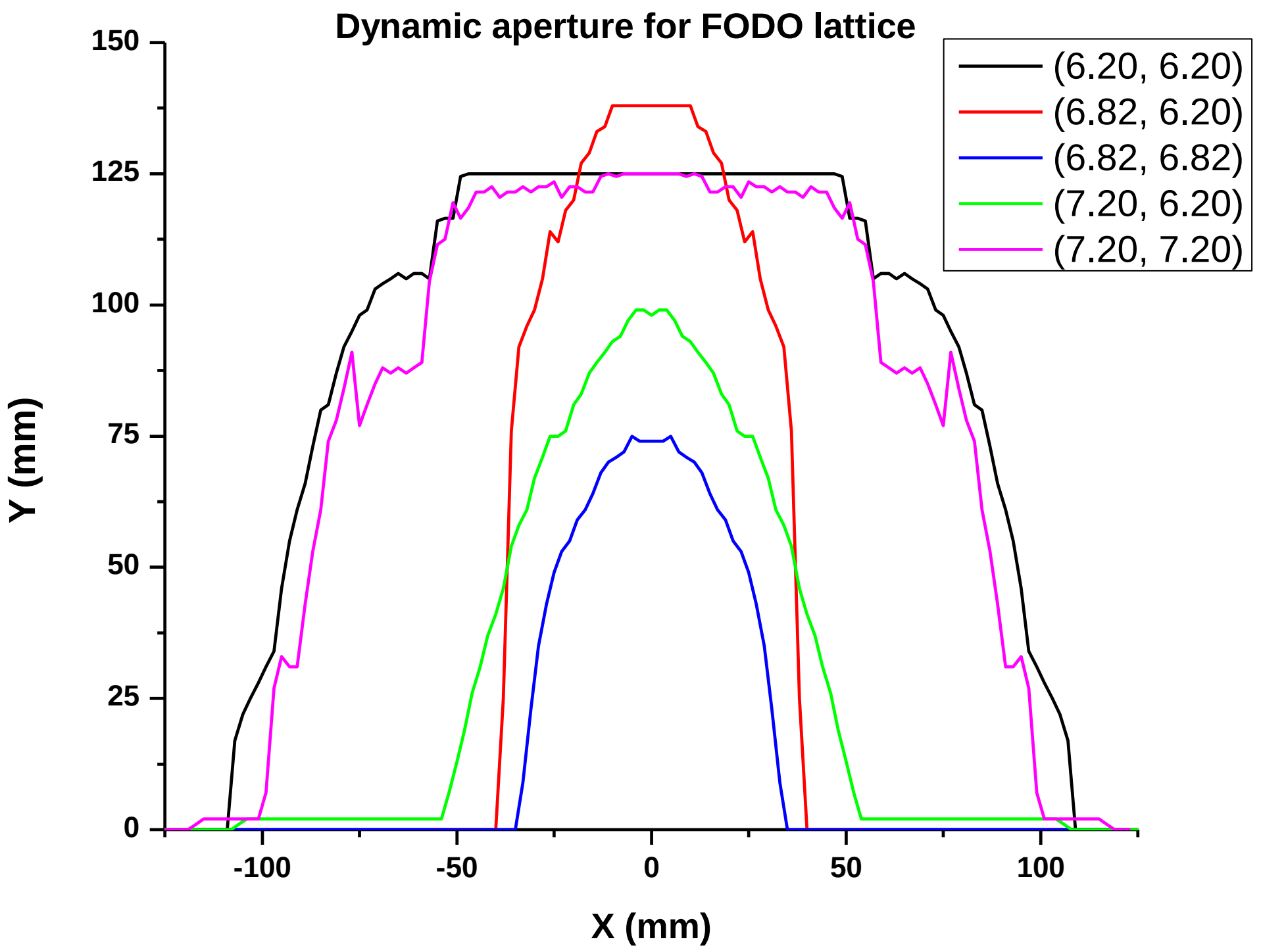

*Fig. 77: Dynamic aperture of FODO lattice with inclusion of chromaticity correcting sextupole magnets (number of turns = 1500 and chromaticity is set to zero).*

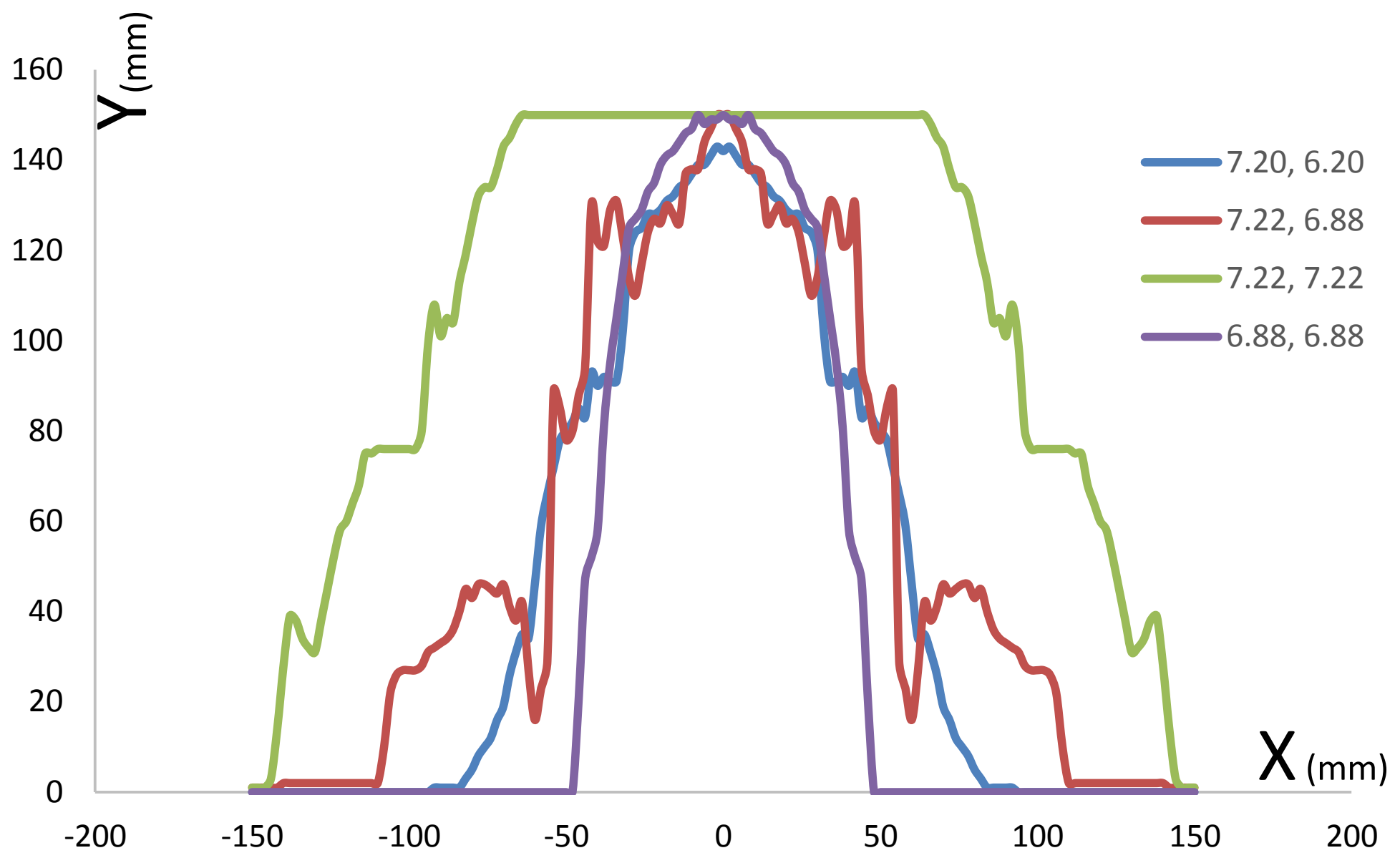

*Fig. 78: Dynamic aperture of Hybrid lattice with inclusion of chromaticity correcting sextupole magnets (number of turns = 1500 and chromaticity is set to zero).*

The chromaticity correcting sextupole magnets are six per superperiod, grouped in three families to reduce the resonances excited by these sextupoles. Considering 200 mm of effective length of sextupole magnets, the maximum strength ($\frac{\partial^2 B_y}{\partial x^2}$) is nearly 20 T/m$^2$. Initial estimation of dynamic aperture at zero chromaticity is shown in Fig. 77 and Fig. 78 for FODO and Hybrid lattice, respectively. This analysis shows that for some working points, dynamic aperture is less than the required, and inclusion of more sextupole families in zero dispersion region or further optimization of chromaticity correcting sextupole scheme and working points are required to enhance the dynamic aperture.

## 12. Beam injection into Accumulator Ring

Studies on multi-turn, charge exchange injection have been started. Presently, studies for FODO lattice are going on, considering injection through a carbon foil. Later, feasibility study on laser stripping scheme will also be explored.

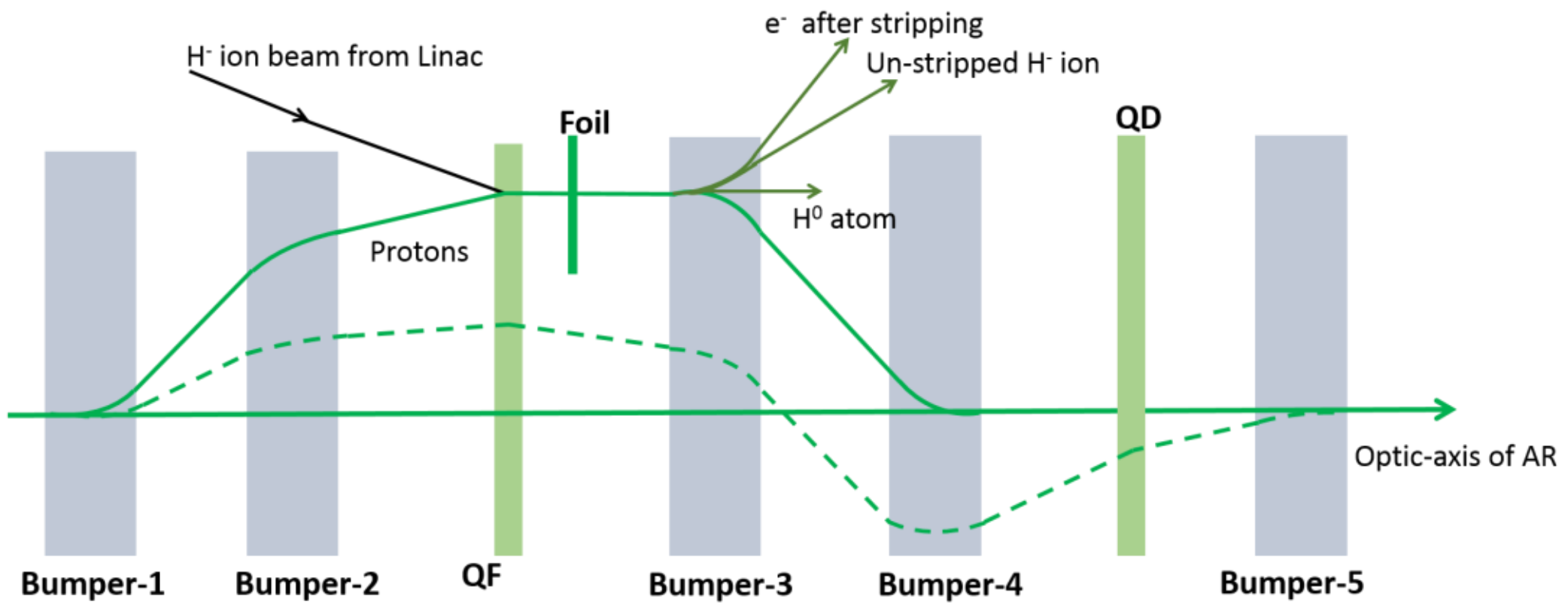

*Fig. 79: A possible configuration of injection chicane for FODO lattice*

A possible configuration of injection chicane for FODO lattice is shown schematically in Fig. 79. In FODO lattice, a maximum bump of ~100 mm using injection chicane is considered. Initial geometrical arrangement of bumper magnets of chicane and painting kickers are carried out.

This scheme presently has five horizontal injection kickers and four vertical kickers in the AR. The Option of having vertical kickers in HEBT has not been explored yet. Injected beam of $H^-$ ion is merged with the circulating proton beam, off-axis in a focusing quadrupole magnet of AR.

12.1 Kicker Locations and strengths

Painted emittance of 210 mm-mrad is required to achieve 1 MW of beam power with a space charge tune shift lower than 0.15. The horizontal beta function is ~ 20 m at the injection quadrupole magnet, resulting in a beam size of ~ 65 mm. Horizontal alpha parameter at this point is ~2, which gives $\alpha/\beta \cong 0.1$. Thus if a bump is generated up to 100 mm, the change in the angle required during the course of injection is 10 mrad (in case of mismatch injection, which is preferable from the point of view of number of hits on the foil). Figure 80 shows the locations of injection kickers. HK1 to HK4 are horizontal bumpers (kickers) and VK1 to VK4 are vertical injection kickers. Injection is carried out at QF_A3 quadrupole magnet.

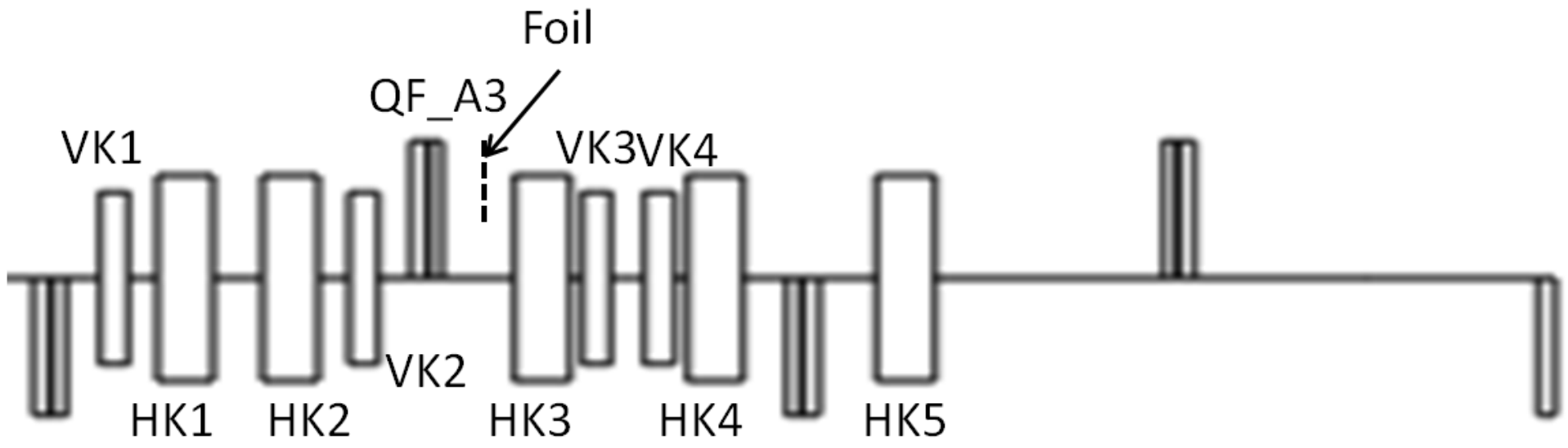

*Fig. 80: Injection system in straight section of FODO lattice.*

In order to keep the constant bend angles for stripped electrons towards the dump, with changing bump for painting, HK3 angle must not be changed. The kick angles and different bumps, required with time are shown in Fig. 81. These results are for the working point (7.186, 7.186).

These results show that the required strength for all the kickers is less than 50 mrad, which means that the required magnetic field is < 0.3 T for a 0.95 m long kicker. However, in order to generate a large angle without displacement at the foil, the orbit shifts by 45 mm at other places and it may demand larger aperture of quadrupole magnets, which are placed in the injection region. In the vertical plane, beta function at the foil is < 6 m and therefore, beam size is ~ 36 mm. Thus, the maximum bump in the vertical plane will be less than 40 mm. Figure 82 shows the vertical bump.

For the vertical steering, the required strength of kickers is higher, i.e. ~ 0.45 T for 0.5 m long kicker. Optimized locations of kickers are yet to be obtained for lowering the kicker strength. Since the option of vertical steering from the HEBT is not yet studied, finalization of locations for vertical kickers will be decided after exploring this option.

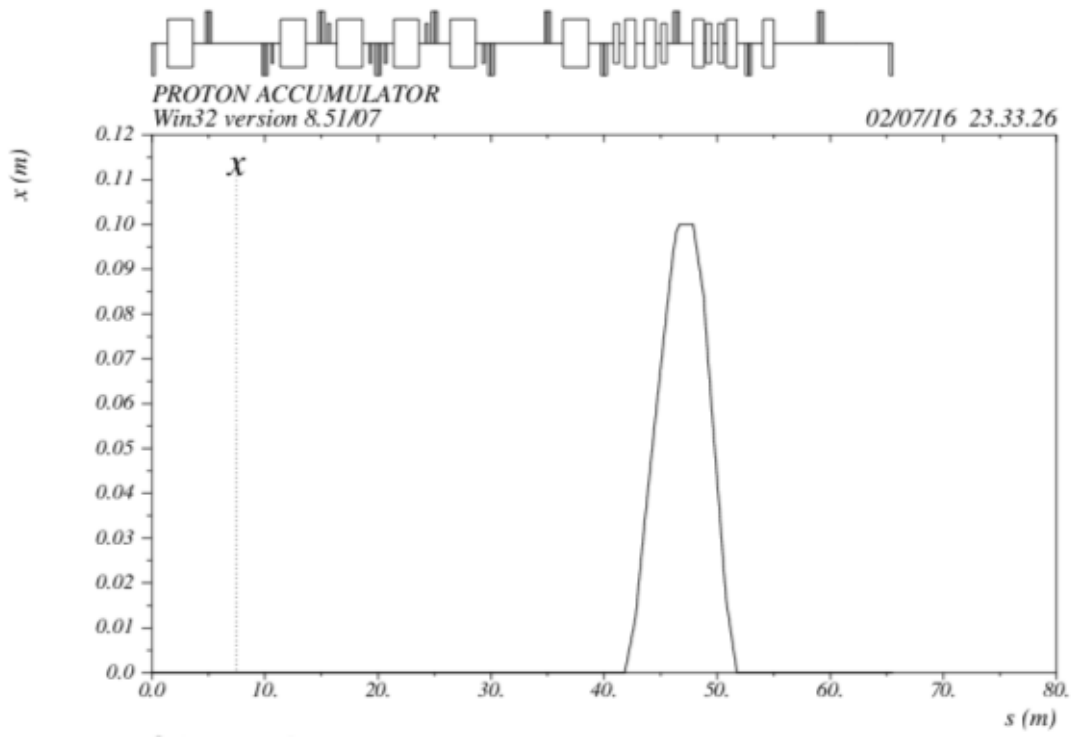

Bump of 100 mm (maximum) with zero angle
K1= 25.83151 mrad
K2= -2.369104 mrad
K3= -34.13668 mrad
K4= 34.13668 mrad
K5= OFF

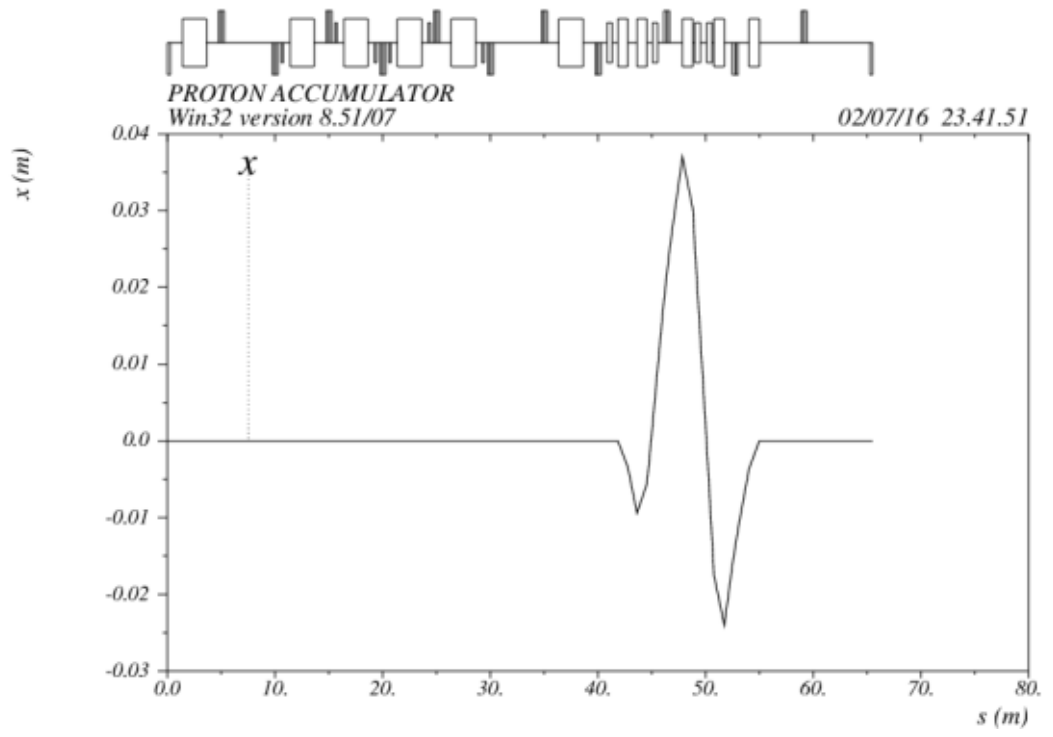

Bump with 30 mm and 10 mrad angle
K1= -7.203326 mrad
K2= 22.41652 mrad
K3= -34.13668 mrad
K4= 34.50035 mrad
K5= -7.688297 mrad

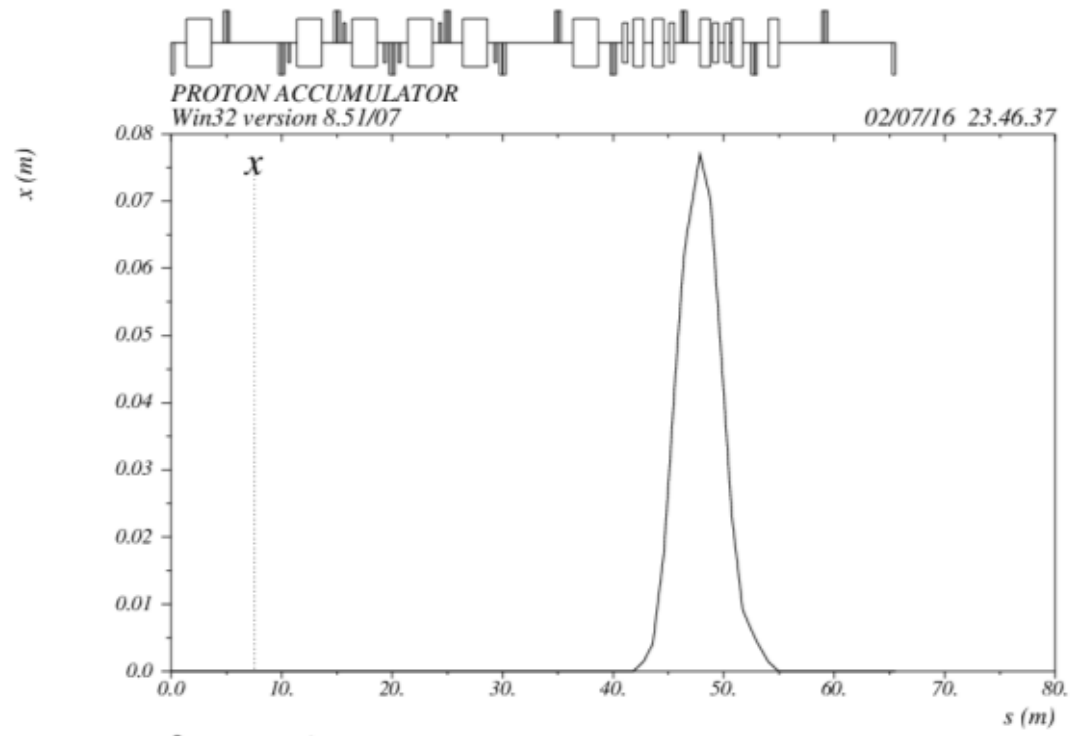

Bump of 70 mm with 10 mrad angle
K1= 3.129277 mrad
K2= 21.46888 mrad
K3= -34.13668 mrad
K4= 20.18391 mrad
K5= 2.932368 mrad

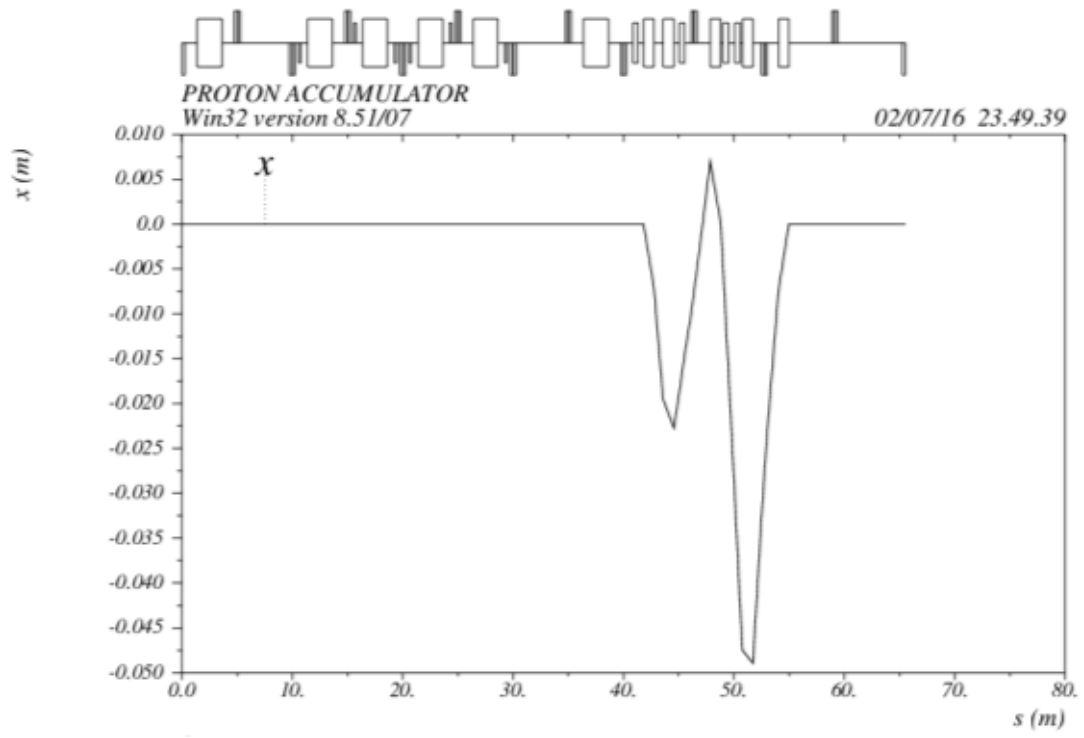

Bump with no displacement and 10 mrad angle
K1= -14.95278 mrad
K2= 23.12725 mrad
K3= -34.13668 mrad
K4= 45.23768 mrad
K5= -15.65380 mrad

*Fig. 81: Different bumps and kicker strengths for injecting the H- beam in AR.*

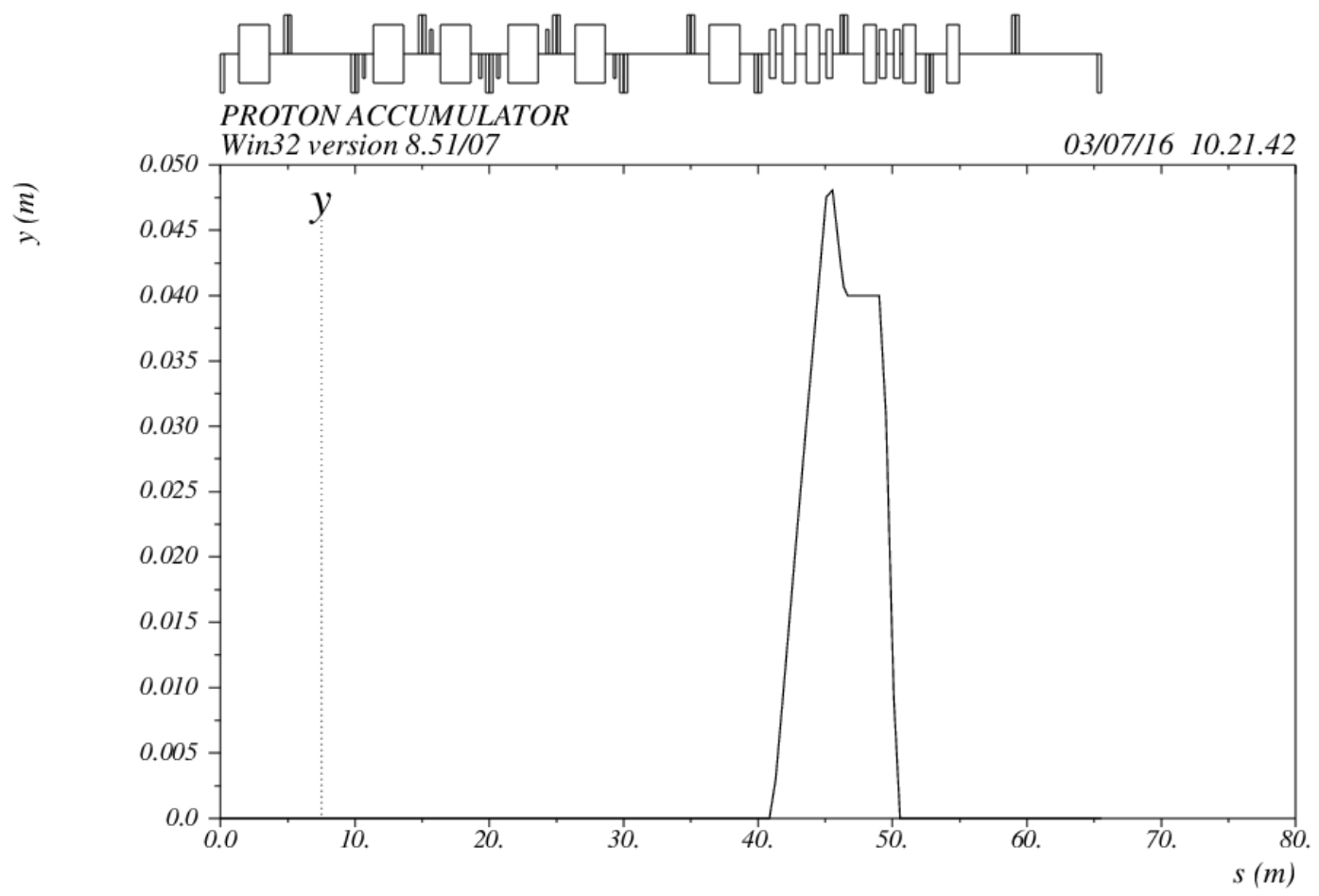

*Fig. 82: Vertical steering of beam for injection.* Here, K1= 11.94334 mrad, K2= -21.75023 mrad K3= -37.05762 mrad, and K4= 37.05762 mrad.

12.2 Bump Program

In order to mitigate the space charge, the phase space distribution of the beam after the painting is very important. The required phase space distribution can be achieved using the optimum programming of the profile of kicker power supplies. It is also important to perform a comparative study of correlated painting and anti-correlated painting to achieve this [28]. Another criteria for good injection scheme is the minimum average hit of the stored beam on the injection foil. Therefore, optimization of the injection scheme requires studies with different kicker profiles, and also a comparative study of matched and mismatched injection schemes [29], including the space charge. Currently, our studies are concentrated towards obtaining a uniform painted emittance in horizontal and vertical phase space, without the space charge. Space charge will be included in the next stage of the study. Different combinations of bumps, which are presently under study, are show in the following Table 31 [30,31].

*Table 31: Bump programs used in present injection studies*

| | **(A) Bump from maximum to centre** | **(B) Bump from centre to maximum** |
|---|---|---|
| **1** | $x_{bump} = x_0\sqrt{1-\frac{t}{T_{bump}}}$ | $x_{bump} = x_0\sqrt{\frac{t}{T_{bump}}}$ |
| **2** | $x_{bump} = -x_0\sqrt{1-\frac{t}{T_{bump}}}$ From negative side to zero | |
| **3** | $x_{bump} = x_0\sqrt{1-\left\{\frac{2t}{T_{bump}}-\left(\frac{t}{T_{bump}}\right)^2\right\}}$ | $x_{bump} = x_0\sqrt{\frac{2t}{T_{bump}}-\left(\frac{t}{T_{bump}}\right)^2}$ |

The simulation parameters taken in studies are following:

| | |
|---|---|
| Number of injected turns: | 1000 |
| Macro-particles per turn: | 500 |
| Total particles at end of injection: | $10^5$ |
| Emittance Painted in both planes: | 210 $\pi$ mm-mrad |

Selected working point is (7.186, 7.186), i.e. fully coupled tune. Studies for 2000 turn injection will be taken in near future. This working point (near to 7.20, 7.20) also has largest dynamic aperture amongst the five different working points of FODO lattice.

12.3 Tracking Results

Under the bump program 3 (Table 31), uniform painted emittance is obtained. These simulations are carried out using the code ORBIT [32]. Presently, simulations do not include space charge and sextupoles.

The distributions in phase space and in real space are shown in Figs. 83 and 84, respectively. Fig. 85 shows the number of hit on injection foil on each turn.

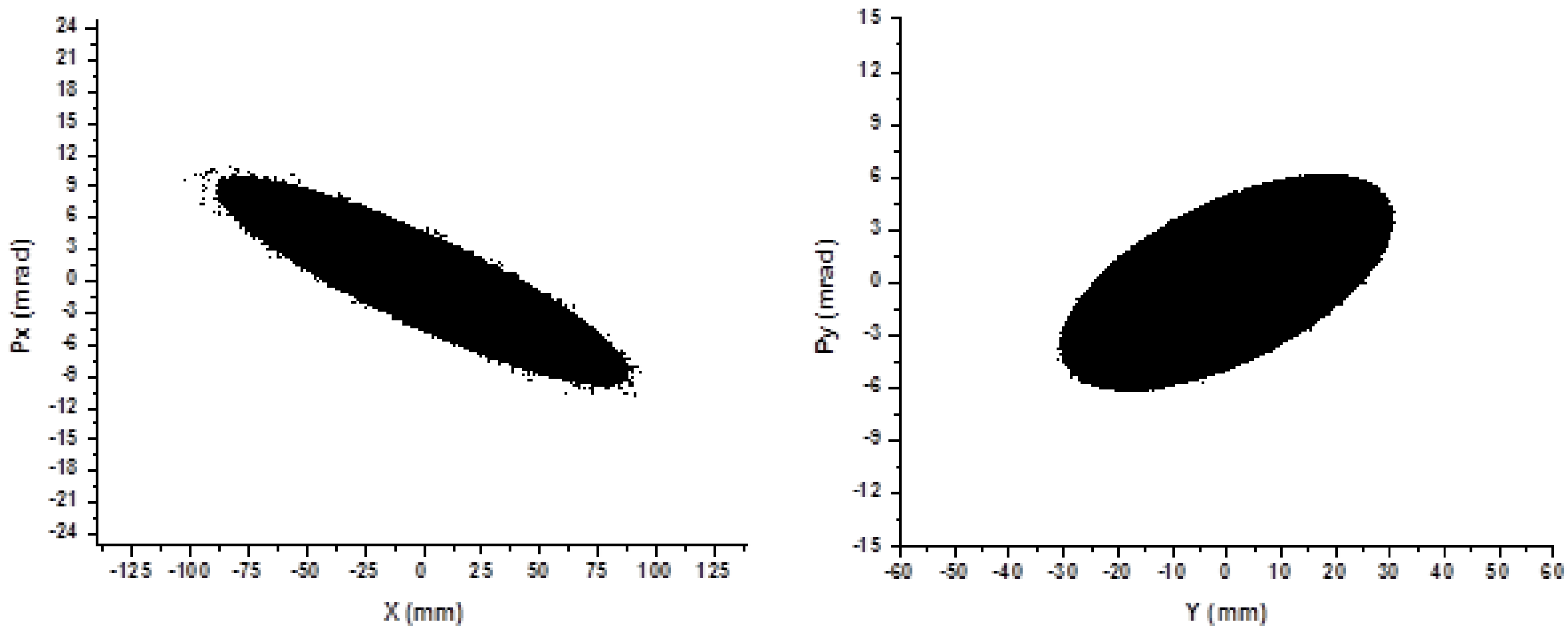

(a) Horizontal Plane (b) Vertical Plane

*Fig. 83: Distribution of the injected beam in phase space.*

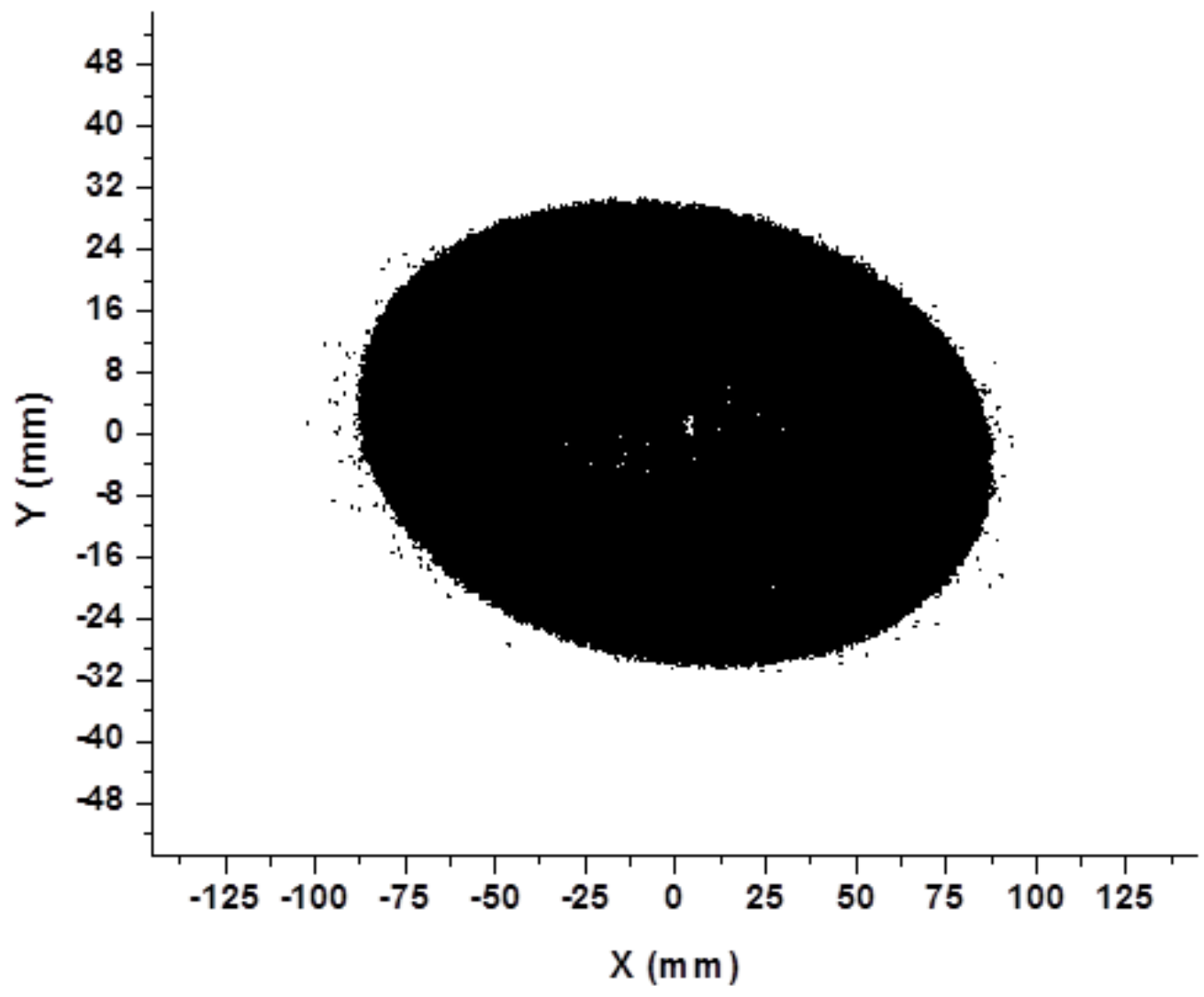

*Fig. 84: Injected beam distribution in real space.*

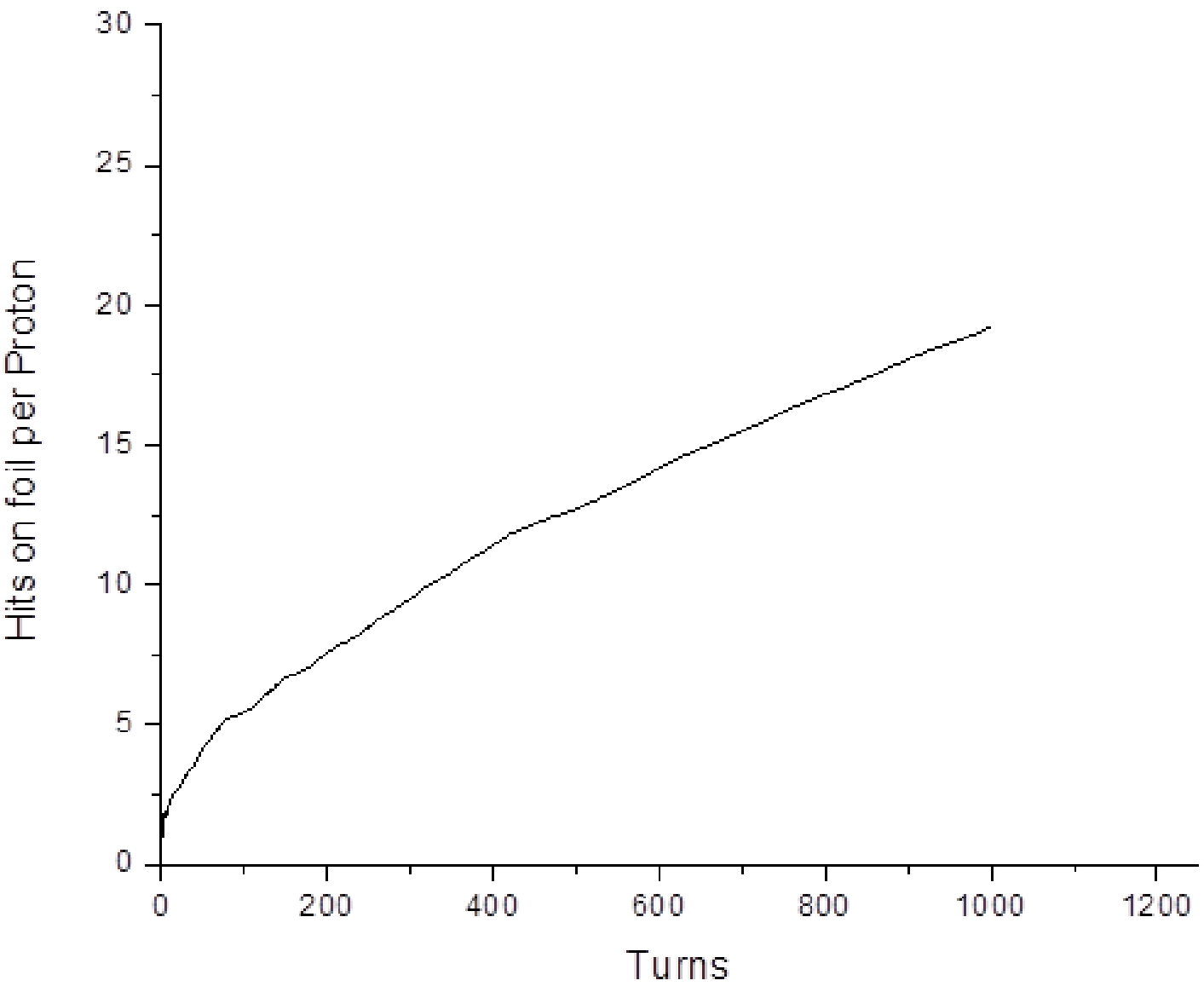

*Fig. 85: Number of hits on injection foil with each turn.*

Initial injection studies show that the considered scheme can be studied in detail and uniform painted emittance can be obtained using programmable kickers. Studies on injection requires the inclusion of space charge, optimization for minimum hit on foil, injecting a longer pulse length and finalizing the vertical kicker locations, i.e. AR or HEBT. Presently, while simulating the space charge using the ORBIT code, some difficulties are faced, and the troubleshooting is currently underway. After completing the studies on FODO lattice, similar studies will be carried out for the Hybrid lattice.

## 13. Beam extraction system

After accumulation and compression of beam in AR, beam has to be extracted safely from the ring and then will be sent to the target. Fast beam extraction scheme is developed for FODO and hybrid lattice. In this scheme, in the clear gap of the beam, magnetic field of kicker magnets will rise to the desired values, and then the beam will be kicked towards the extraction septum magnet. This extraction septum magnet will then send the beam towards the Ring to Target Beam Transfer Line (RTBT). The scheme is designed in a way so that even in the case of failure of two kickers, beam will be extracted safely [33]. Figures 86 and 87 show the schematic layout of extraction optics of FODO and Hybrid lattice, respectively.

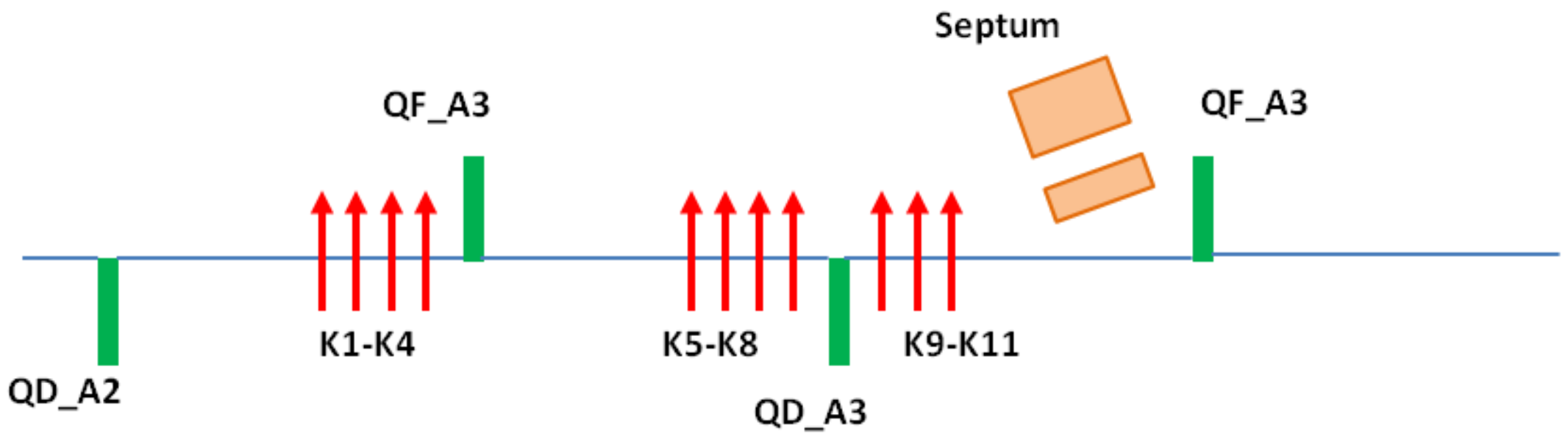

*Fig. 86: Schematic layout of extraction optics of FODO lattice (red arrows show the kicker magnets and green blocks represent quadrupole magnet).*

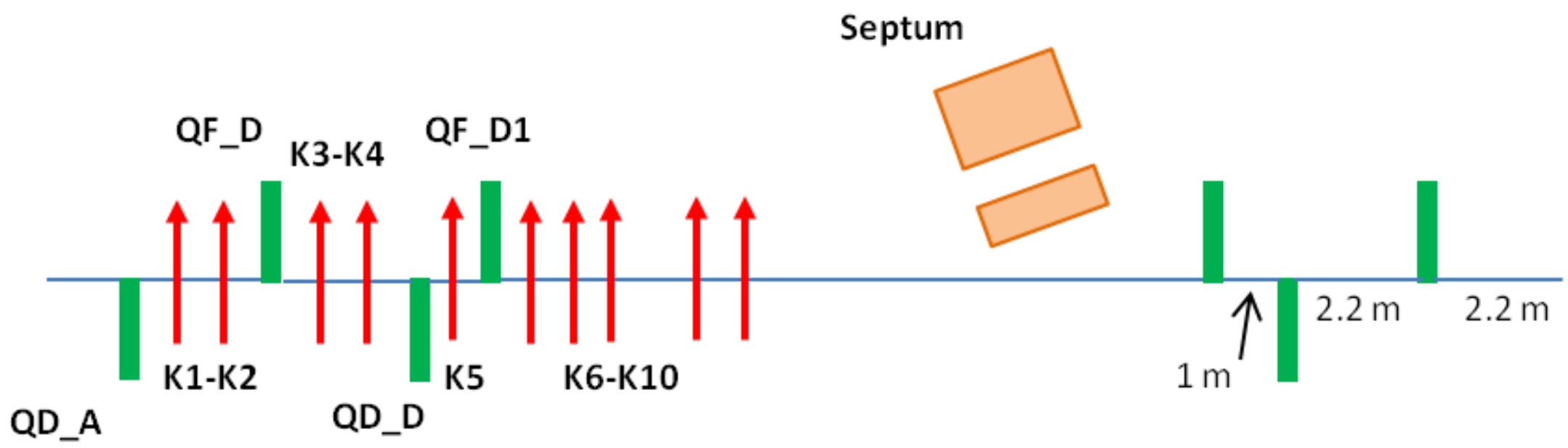

*Fig. 87: Schematic layout of extraction optics of Hybrid lattice (red arrows show the kicker magnets and green blocks represent quadrupole magnets).*

There are total 11 nos. of kicker magnets in the FODO lattice, while there are 10 kicker magnets in Hybrid lattice [34]. Table 32 and 33 show the required kick angles of different kicker magnets to produce a vertical bump of 150 mm. Extraction scheme is able to produce a bump from 140 mm to 175 mm with a bump angle at the septum from 5 to 15 mrad. The main parameters related to extraction kicker are provided in Table 34.

*Table 32: Kick angles required by different kicker magnet for proper extraction for all the five tune points in case of FODO lattice.*

| **Tune points → Kick angles (mrad)↓** | **7.20, 6.20** | **6.82, 6.20** | **7.20, 7.20** | **6.82, 6.82** | **6.20, 6.20** |
|---|---|---|---|---|---|
| K1 | 2.56 | 2.50 | 2.50 | 2.70 | 2.70 |
| K2 | 2.60 | 2.66 | 2.70 | 2.70 | 2.70 |
| K3 | 2.41 | 2.37 | 2.41 | 2.40 | 2.41 |
| K4 | 2.66 | 2.62 | 2.48 | 2.47 | 2.57 |
| K5 | 2.79 | 2.76 | 2.79 | 2.49 | 2.50 |
| K6 | 2.79 | 2.89 | 2.89 | 2.50 | 2.50 |
| K7 | 3.09 | 3.06 | 3.09 | 2.52 | 2.50 |
| K8 | 2.93 | 2.91 | 2.59 | 2.87 | 2.50 |
| K9 | 3.12 | 3.10 | 3.09 | 2.48 | 2.41 |
| K10 | 3.13 | 3.13 | 3.09 | 2.48 | 2.44 |
| K11 | 3.15 | 2.45 | 3.11 | 2.49 | 2.48 |

*Table 33: Kick angles required by different kicker magnet for proper extraction for all the five tune points in case of Hybrid lattice.*

| **Tune points → Kick angles (mrad)↓** | **6.82, 6.82** | **7.20, 7.20** | **7.20, 6.82** | **7.20, 6.20** | **6.82, 6.20** |
|---|---|---|---|---|---|
| KK1 | 2.55 | 2.72 | 2.6 | 2.60 | 2.64 |
| KK2 | 2.82 | 3.10 | 2.90 | 2.86 | 2.81 |
| KK3 | 2.82 | 2.83 | 2.93 | 2.87 | 2.81 |
| KK4 | 3.10 | 3.57 | 3.14 | 3.14 | 3.11 |
| KK5 | 2.78 | 3.15 | 2.75 | 2.83 | 2.76 |
| KK6 | 2.45 | 2.28 | 2.43 | 2.49 | 2.43 |
| KK7 | 2.43 | 2.59 | 2.44 | 2.45 | 2.40 |
| KK8 | -3.13 | -2.94 | -2.16 | -3.04 | -3.06 |
| KK9 | -3.18 | -3.29 | -3.70 | -3.50 | -3.33 |
| KK10 | -3.23 | -2.98 | -3.23 | -2.96 | -2.71 |
| KK1 | 2.55 | 2.72 | 2.6 | 2.60 | 2.64 |

*Table 34: Main parameters of extraction kicker magnet*

| **Parameter** | **Value** |
|---|---|
| Length | 500 mm |
| Maximum field | < 550 G |
| Rise time | 200 ns |
| Flat top time | ~800 ns |
| Fall time | Not specified, it should become zero well before the next injected beam pulse arrival |
| Field error | < 0.5% |

## 14. Ring to Target Beam Transport (RTBT)

Extracted beam from AR will be transferred to the spallation target via Ring to Target Beam Transport Line (RTBT). RTBT has beam expander optics towards the target, to produce the desired density of proton beam on the target and it has collimators for proton beam, and also for back scattered neutrons. RTBT should have provision for branching the beam for two or three target stations in future. RTBT may also incorporate vertical dipole magnets to cancel the vertical dispersion and vertical angle of beam orbit generated at the septum magnet during beam extraction. In case of extraction kicker failures, RTBT optics must ensure the beam transport up to the dump. These requirements again increase the length of RTBT. For the SNS at Oak Ridge, the length of RTBT is ~160 m [35]. J-PARC facility at Japan has even longer beam line, i.e. 280 m [3]. This is however for the case where the extracted energy is higher (3 GeV), and it also has a Muon target in addition to the spallation Neutron target. Proposed Chinese SNS has a design of RTBT with a length of 144 m [36]. Design studies for RTBT for the ISNS project is currently undergoing.

The initial studies for RTBT have been carried out considering an achromat for bending the beam to the desired angle, in accordance to the available topographical layout. After the achromat, a matching optics is required, which matches the optical parameters at the exit of achromat to the parameters of FODO cells, which are to be used for transporting the beam to the target. After the FODO cells (which provides sufficient space for collimator and diagnostics), a suitable optics is required for beam size manipulation at the target.

A typical magnetic arrangement for RTBT is presented in Fig. 88. In the current study, the last section for the beam size manipulation optics is not included. This optics section will be included after the beam size requirements at the target are finalized.

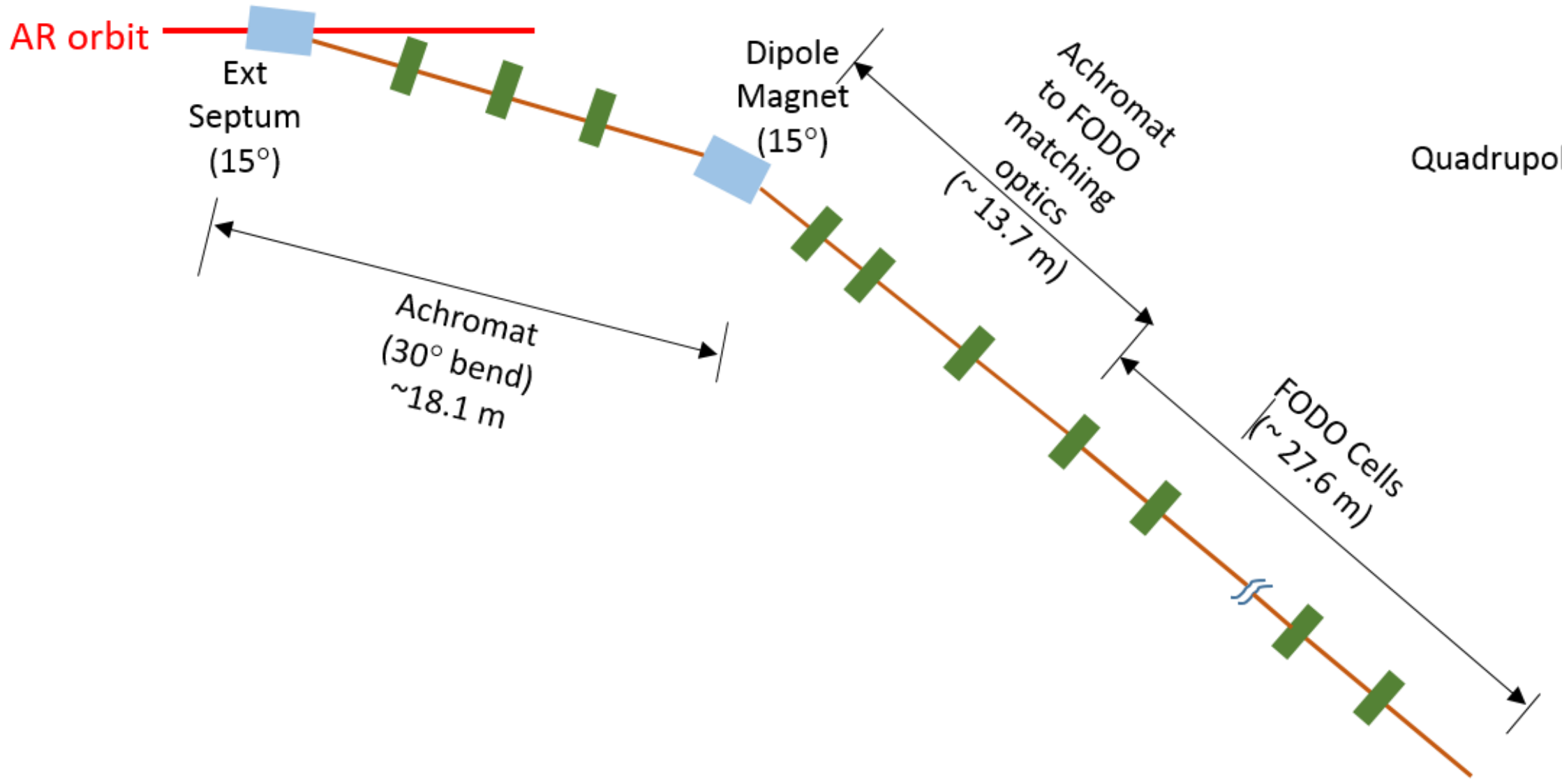

*Fig. 88: Layout of Ring to Target Beam Transport Line (RTBT).*

Twiss parameters of this line are shown in Fig. 89. Initial Twiss parameters at the extraction septum were chosen for working point (7.20, 7.20) of AR.

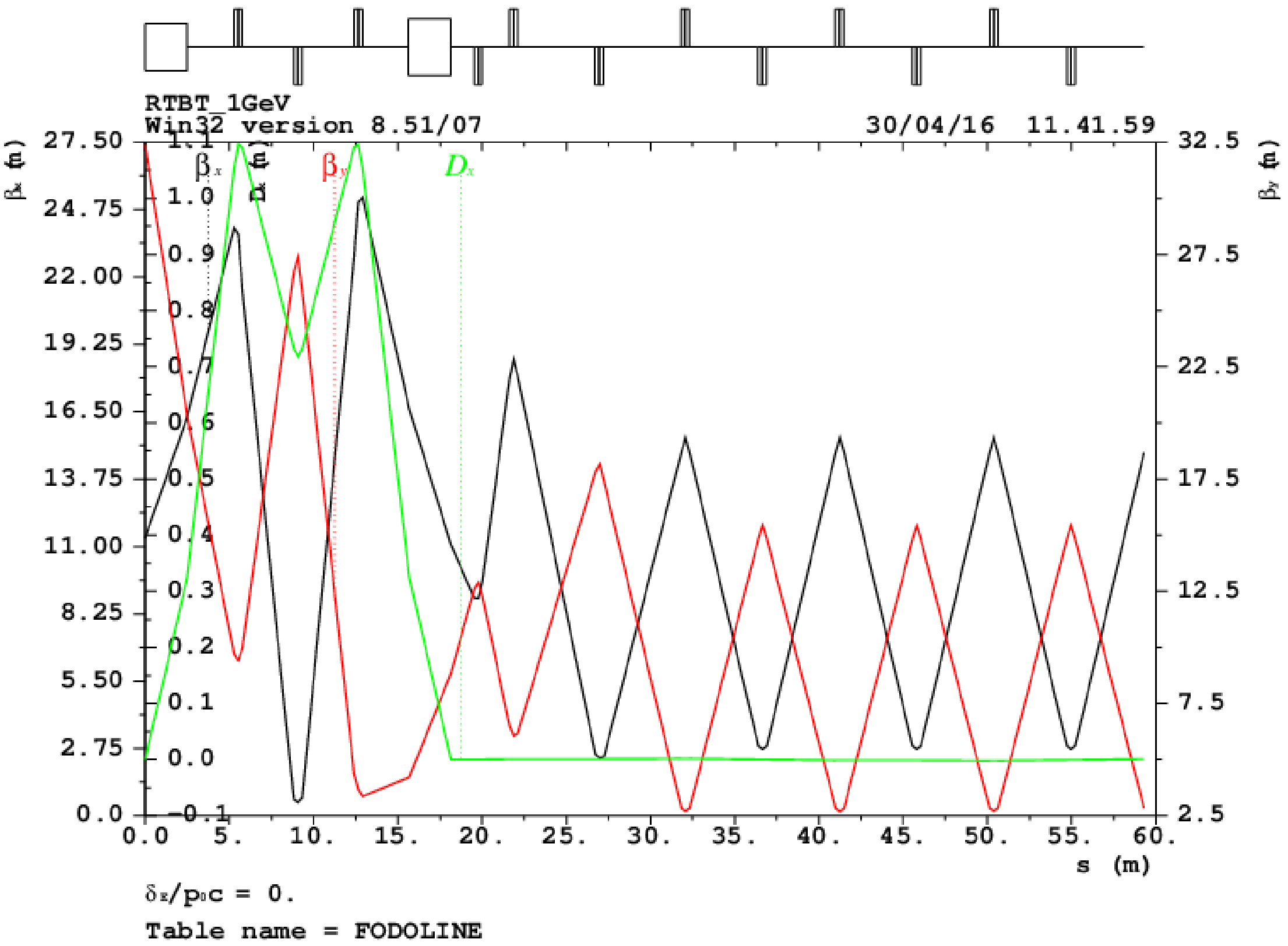

Fig. 89: Twiss parameters of RTBT.

The maximum value beta function in either plane is below 35 m, which results in the maximum beam size below ~ 85 mm. The inner radius of the vacuum chamber is ~140 mm. This gives an acceptance, which is ~ 2.7 times the beam emittance (210 mm-mrad), and a margin of ~ 50 mm for the centroid shift.

Studies of RTBT beam dynamics under various field errors and space charge are currently in progress. Collimator studies will be taken up shortly. After these studies and after finalizing the building geometry, complete layout and specifications of RTBT will be evolved.

## 15. Beam Halo and Collimation studies for accumulator ring and transport lines

Beam loss is a major concern in such machines. Space charge studies are required to estimate the behaviour of beam in the optics of the ring during accumulation as well as in transfer lines. Space charge can generate a beam halo, which will be lost eventually in the ring or transfer line and can activate the components. To prevent the activation from such beam loss, collimators will be used in transfer lines and in ring [3,37,38,39]. Presently, a space charge study during beam injection is started. Detailed studies about beam loss will be started very soon.

In a collimator system, primary collimator provides the minimum aperture to the beam in a ring. However, this aperture is large enough, so that the core proton beam traverses from this location without hitting the aperture, even in case of allowed practical errors. The halo particles, which have larger amplitude of betatron oscillations scatter through this primary collimator and their amplitude of oscillation increase further. At a suitable location, secondary collimator (absorber) is placed to collect these scattered particles. In this way a collimation scheme provides a controlled loss of particles.

In FODO lattice, the phase advance in straight section is ~150° , and therefore, aperture ratio of ~0.85 between the primary and secondary collimator can be accommodated. Presently, the phase advance in Hybrid lattice is ~120°, and therefore needs to be increased for efficient collimation. Collimator studies (i.e. location, aperture and orientation optimization etc.) have been recently started.

## Summary and Conclusions

To summarize, significant progress has been made on the physics design of various sub-systems of 1 GeV injector linac for the proposed Indian Spallation Neutron Source at RRCAT. Start to end beam dynamics studies for the full injector linac, after integrating various sub-systems in the design has been completed. Physics design has also been performed for various subsystems of accumulator ring and transport lines. The scheduled date of completion of physics design of the injector linac, and the accumulator ring with the transport lines is March 2017. This document has been prepared in order to obtain comments on various aspects of the design before the scheduled completion date of reference physics design.

## Acknowledgements

We thank Dr. S. B. Roy, Head, MAASD for several useful discussions on issues related to design of superconducting cavities, and mentoring the accelerator physics design team for the work presented in this report. We sincerely acknowledge his important contributions. We are grateful to Director, RRCAT for constant support and encouragement. We sincerely acknowledge Mr. A. D. Ghodke, Head, BDL/IAPDD for useful discussions related to the design of accumulator ring.